\documentclass[aps,prb,twocolumn,superscriptaddress,a4paper,]{revtex4-2}

\usepackage[dvipdfmx]{graphicx}

\usepackage{hyperref}
\usepackage{color}
\usepackage{amsmath}
\usepackage{amssymb}
\usepackage{lipsum}
\usepackage{mathtools}
\usepackage{microtype}
\usepackage{mathtools}
\usepackage{braket}
\usepackage[caption=false]{subfig} 
\usepackage{overpic}

\usepackage{xcolor}

\usepackage[normalem]{ulem}

\allowdisplaybreaks
\newcommand{\Wzv}{W_z^{(v)}}
\newcommand{\Wzh}{W_z^{(h)}}
\newcommand{\Wxv}{W_x^{(v)}}
\newcommand{\Wxh}{W_x^{(h)}}

\begin{document}

\title{The toric code under antiferromagnetic isotropic Heisenberg interactions}

\author{Won Jang}
\affiliation{Department of Physics, Kyoto University, Kyoto 606-8502, Japan}
\email[]{jangwon001129@gmail.com}
\author{Robert Peters}
\affiliation{Department of Physics, Kyoto University, Kyoto 606-8502, Japan}
\author{Thore Posske}
\affiliation{I. Institute for Theoretical Physics, Universit\"at Hamburg, Notkestraße 9, 22607 Hamburg, Germany}
\affiliation{The Hamburg Centre for Ultrafast Imaging, Luruper Chaussee 149, 22761 Hamburg, Germany}
\begin{abstract}
 We investigate the impact of an isotropic antiferromagnetic Heisenberg perturbation on the toric code, focusing on the resulting quantum phase transition and the nature of the phase that emerges beyond topological order. Using neural-network quantum states (NQS), we compute ground states over a wide range of Heisenberg couplings while fully respecting the exact symmetries of the model. In the weak-coupling regime, the numerical results are in excellent agreement with an effective low-energy description derived from a Schrieffer-Wolff (SW) transformation, providing analytic control over the perturbative breakdown of topological order. We show that the Heisenberg perturbation only renormalizes local operators at low orders, whereas mixing
between topological sectors occurs only at a perturbative order proportional to the system size.
At intermediate values of the Heisenberg interaction, the topological phase breaks down.
We estimate the critical point through a combination of the fidelity susceptibility and the logarithmic susceptibility of non-contractible Wilson loops for various system sizes. Furthermore, we utilize the topological entanglement entropy to provide a comprehensive 
characterization of the phase transition. Beyond the transition, 
an antiferromagnetic $\pm X/\pm Z$ N\'eel phase emerges, characterized by a 
fourfold-degenerate symmetry-broken manifold, which is explicitly 
probed using staggered-magnetization-based diagnostics.
Our results show how local two-spin interactions, which naturally arise in realistic implementations of the toric code, drive the breakdown of topological order. Moreover, we establish the SW approach as a systematic framework for analyzing such perturbations in combination with variational many-body methods.
\end{abstract}

\maketitle

\section{Introduction}
The toric code is the paradigmatic exactly solvable lattice model exhibiting topological order. As a two-dimensional Hamiltonian built from mutually commuting local projectors, it realizes long-range entanglement characterized by a universal topological entanglement entropy $\gamma_{\text{TEE}}=\ln 2$ and hosts abelian anyonic quasiparticles $e$ and $m$ with mutual semionic statistics~\cite{kitaev_toric_code,hamma2005,Nayak2008}. 
When placed on manifolds of nontrivial topology, the model displays a ground state degeneracy that depends solely on the global topology, fourfold on the torus, reflecting its robust, locally indistinguishable subspace protected by a finite energy gap~\cite{TQOstability}.
These degenerate states are characterized by non-contractible loop operators,
that are generated by non-contractible Wilson loops, making the toric code both the canonical example of a stabilizer code and a foundational model for fault-tolerant quantum memories~\cite{quantum_memory}.
Beyond its role as a minimal model, the toric code serves as the canonical fixed-point representative of a broad class of 
\(\mathbb{Z}_2\)-topologically ordered phases~\cite{Lee2006,senthil2000}. It realizes the \(\mathrm{D}(\mathbb{Z}_2)\) quantum double, the fixed-point representative of its universality class within the Levin–Wen string net framework, and other exactly solvable \(\mathbb{Z}_2\) lattice gauge theories flow to the same topological order~\cite{levin_wen_stringnet}.  Toric-code physics emerges as a low-energy effective description in various microscopic systems, including gapped phases of the Kitaev honeycomb model and frustrated quantum dimer models near Rokhsar–Kivelson points, providing concrete routes from spin models to gauge-theory topological order~\cite{kitaev_honeycomb,qdm_rk,moessner_sondhi_qdm,kagome_qdm_z2}.
Moreover, recent experiments highlight the relevance of the toric code~\cite{GoogleQuantumAI2023,Yao2012,Lu2009}. In superconducting-qubit processors, prepared toric-code ground states display topological entanglement entropy close to \(\ln 2\) and interferometric signatures of anyon braiding, while surface-code (toric-code) stabilizer measurements demonstrate logical-error suppression with increasing code distance~\cite{realizing_toric_code,google_surface}. However, realistic platforms cannot realize the ideal toric code Hamiltonian in isolation: stray fields, cross-talk, and residual two-body couplings inevitably introduce perturbations that break exact solvability.  The effect of uniform magnetic fields has been studied extensively. Fields applied in the $x$- or $z$-direction each mobilize a single species of anyon, driving its condensation and confining the dual species~\cite{parallelmag_toric_code,parallelmag_toric_code2,random_mag_toric_code}.
The resulting $(h_x,h_z)$ phase diagram features two second-order Ising-type transition lines meeting a first-order line at a topological multicritical point~\cite{parallelmag_toric_code2}. By contrast, a transverse $y$-field is self-dual and drives a first-order transition to a polarized phase accompanied by multi-anyon bound states~\cite{transversemag_toric_code,random_mag_toric_code}. Similar Ising-type couplings trigger analogous confinement transitions
~\cite{ferro_ising1,Trebst2007}.

In this work, we investigate how the robustness of the $\mathbb{Z}_2$ deconfined phase is affected by perturbations that simultaneously excite all sectors of the code. Among such perturbations, the isotropic antiferromagnetic Heisenberg exchange 
is both physically natural and qualitatively distinct. Arising ubiquitously from superexchange in Mott insulators and crosstalk in engineered qubit arrays, the Heisenberg interaction couples all spin components ($\sigma^x, \sigma^y, \text{ and } \sigma^z$). It thus drives a mixture of diagonal and off-diagonal perturbation terms that lack the orientation-dependent selectivity of simple magnetic fields. Consequently, Heisenberg perturbations generate a competition between local stabilizer renormalizations and nonlocal multi spin loop processes, potentially altering the character of the emergent gauge structure. This makes the Heisenberg perturbation a rigorous and physically motivated probe of robustness beyond the well-studied Ising case.

To investigate the toric code under isotropic antiferromagnetic Heisenberg exchange, we employ neural-network quantum states (NQS). Properly designed NQS can encode non-perturbative correlations, long-range entanglement, and nonlocal constraints beyond the reach of small-cluster or low-order expansions; with sufficient depth and parameters, they have demonstrated strong expressivity in interacting 2D systems, including volume-law capacity~\cite{carleo2017,nomura2017,Lange_2024,GaoDuan2017NatCommun,capture_ent,deepCNN_RNN_capture_ent,Allah2020,sharir2020,NQS_volume_law,joshi2023,Joshi2024}. Our objective is to use an NQS to track how topological order is renormalized or destroyed by Heisenberg perturbations and to test whether the ansatz cleanly discriminates the deconfined $\mathbb{Z}_2$ phase from magnetically ordered phases. 
From a methods standpoint, NQS complements existing numerics. Exact diagonalization provides unbiased benchmarks but is restricted to very small lattices; tensor-network methods (e.g., Projected Entangled Pair States) capture the toric-code fixed point exactly yet can become numerically costly as bond dimensions grow away from the stabilizer limit and can be less expressive than NQS~\cite{glasser2018,chen2018,Huang2021}. As a flexible alternative at intermediate sizes, NQS gives direct access to the variational wavefunction, enabling overlaps/fidelity, sector-resolved observables, and nonlocal matrix elements, and remains applicable when non-stoquastic deformations are introduced. 
While its optimization remains a challenge, we note several stabilization strategies: (i) adiabatic annealing from guided Hamiltonians~\cite{annealing}, (ii) gentle sector-bias terms when needed~\cite{corrRBM}, and (iii) reusing pretrained parameters across couplings~\cite{fine-tuning}.

While these NQS can probe the full microscopic Hamiltonian, it is equally valuable to obtain a controlled analytic expansion of the emergent low-energy theory to gain physical insight into the underlying processes. To this end, we employ the Schrieffer-Wolff transformation (SWT)~\cite{bravyiSW} and obtain a low-energy description of the toric code under isotropic AFM Heisenberg exchange. The SWT  block-diagonalizes the Hamiltonian into the toric-code subspace, yielding a gauge-invariant effective Hamiltonian together with dressed observables. 
Remarkably, the SWT enables us to show that the topological sector mixing occurs at the order of system size and higher orders, which is in agreement with earlier works regarding the robustness of the topological phase to local perturbations. On top of that, local perturbation only renormalizes the Hamiltonian and operators at lower orders.

The rest of this paper is organized as follows: In Sec.~\ref{model and method}, we describe the model and the NQS used in our simulations. We also present several techniques and observables that we utilize to analyze the occurring phases. In Sec.~\ref{Schrieffer–Wolff transformation}, we derive effective operators via the SWT. In Sec.~\ref{ground_state}, we show our results and discuss the different ground states. Finally, we summarize our work in Sec.~\ref{summary}.

\section{Model and method}
\label{model and method}

In this section, we present the model and its symmetries, the neural-network variational ansatz, and the observables used to characterize the quantum phase transition.
Subsection~\ref{Toric code with Heisenberg interactions} reviews the unperturbed toric code and its mutually commuting stabilizer operators, which ensure exact solvability and define the encoded qubit structure. We then examine how isotropic Heisenberg interactions break this structure by violating the stabilizer algebra and, in particular, by failing to commute with non-contractible Wilson loop operators.
Subsection~\ref{Marshall Gauge} discusses how the non-stoquastic sign structure arising in the antiferromagnetic (AFM) case can be removed via a Marshall gauge transformation and how restoring stoquasticity improves numerical stability by eliminating sign-related instabilities.
Subsection~\ref{Global Parity Symmetries and Logical Structure} shows that the global parity operator remains conserved even in the presence of the Heisenberg interaction and clarifies the relation between global parity sectors and the ground-state manifold of the unperturbed model.
Subsection~\ref{Variational Neural-Network Ansatz} introduces the neural-network architecture and outlines the implementation details relevant for the optimization.
Subsection~\ref{Parity on Marshall Gauge and Duality Constraints} analyzes which global parity sector contains the finite-size ground state and explains how this structure becomes transparent in the effective low-energy description derived via the SWT. We also discuss the duality symmetry that persists under perturbation.
Finally, Subsection~\ref{Detecting of quantum phase transition} introduces the observables used to identify the quantum phase transition: the fidelity susceptibility, whose finite-size scaling yields the critical coupling and correlation-length exponent; a composite non-contractible Wilson loop diagnostic that exhibits a smooth crossover with increasing coupling, and whose peak in the
logarithmic susceptibility yields a nonlocal estimate of the critical point consistent with the fidelity-susceptibility analysis; the topological entanglement entropy, which signals the breakdown of long-range entanglement; and staggered magnetization, which characterizes the emergent symmetry-breaking phase.

\subsection{Toric code with Heisenberg interactions}
\label{Toric code with Heisenberg interactions}

\begin{figure}[t]
  \centering

\begin{minipage}{0.48\columnwidth}
  \centering
  \begin{overpic}[width=\linewidth,height=0.28\textheight,keepaspectratio]{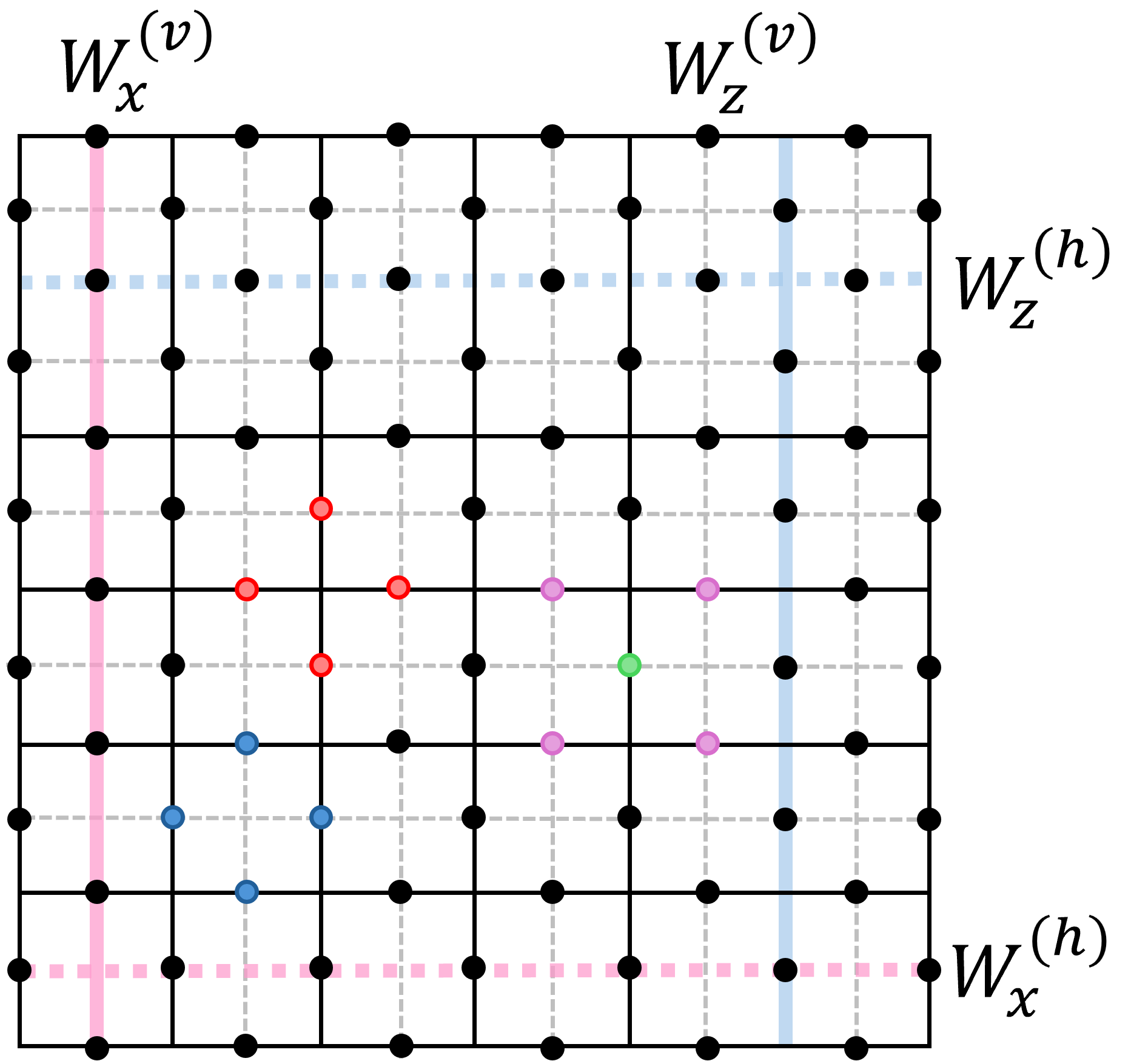}
    \put(-2,96){(a)}
  \end{overpic}
\end{minipage}\hfill
\begin{minipage}{0.48\columnwidth}
  \centering
  \begin{overpic}[width=\linewidth,height=0.28\textheight,keepaspectratio]{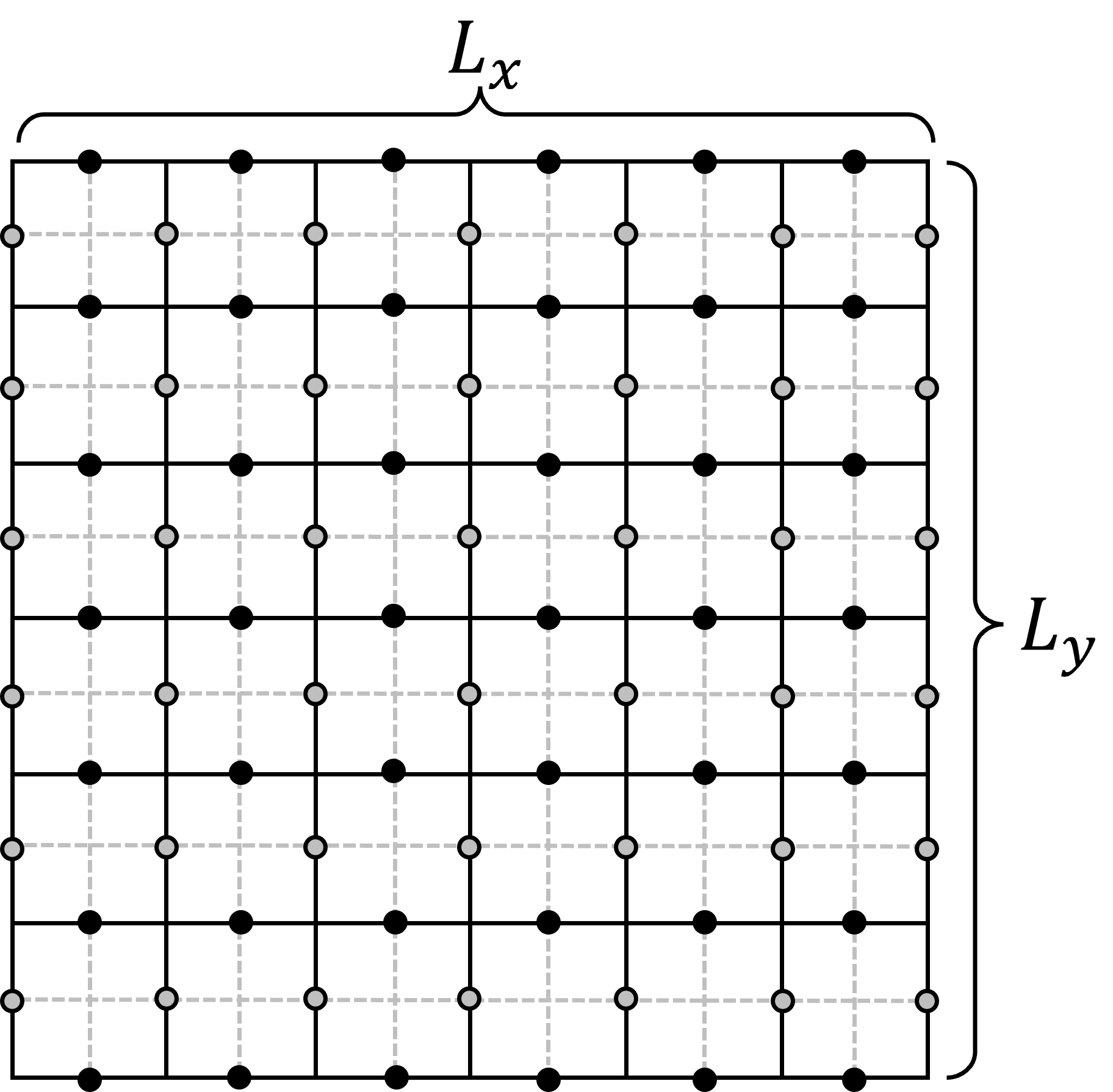}
    \put(-2,102){(b)}
  \end{overpic}
\end{minipage}

  \caption{\label{fig:tc}
Structure of the toric-code lattice.
  (a) The four red sites indicate the star operator
  $A_v=\prod_{e\in +_v}\sigma^x_e$; the four blue sites indicate the plaquette operator
  $B_p=\prod_{e\in \partial p}\sigma^z_e$.
  The four purple sites show the nearest neighbors of the green site used in the Heisenberg interaction. Moreover, the gray dashed lines describe the dual lattice.
  We show examples of non-contractible Wilson loops under Periodic Boundary Condition, where $W_z^{(v)}$ and $W_z^{(h)}$ are products of $\sigma^z$ along the light-blue vertical solid and horizontal dashed lines. In the same manner, the pink solid vertical and horizontal dashed lines represent the $\sigma^x$ loops $W_x^{(v)}$ and $W_x^{(h)}$~\cite{kitaev_toric_code}.
  (b) The Marshall-gauge pattern used in this work, where gray sites belong to the $A$ sublattice (flipped by the Marshall transformation) and black sites belong to the $B$ sublattice. We specifically use $L_x=L_y=L$}
\end{figure}

We study the spin-$\tfrac12$ toric code with antiferromagnetic, isotropic Heisenberg interactions on a two-dimensional square lattice \(\Lambda\) with periodic boundary conditions. Spin-$\tfrac12$ degrees of freedom (qubits) reside on the edges \(e\in \mathcal{E}(\Lambda)\). The Heisenberg interaction is introduced as a generic residual two-spin interaction acting within an effective low-energy theory whose dominant terms are described by the toric-code Hamiltonian. 
Such interactions naturally arise from perturbations, virtual processes, or imperfect fine-tuning in realistic implementations of topologically ordered phases, and provide a controlled way to study the stability and breakdown of toric-code topological order.

For a vertex \(v\), let \(+_v\subset \mathcal{E}(\Lambda)\) be the four edges incident on \(v\); for a plaquette \(p\), let \(\partial p\subset \mathcal{E}(\Lambda)\) be the four edges on the boundary of \(p\).
The toric-code stabilizers (see Fig.~\ref{fig:tc}\,(a) for star and plaquette conventions) are
\[
A_v \equiv \prod_{e\in +_v} \sigma^x_e,\qquad
B_p \equiv \prod_{e\in \partial p} \sigma^z_e,
\]
where \(\sigma^\alpha_e\) is the Pauli matrix \(\sigma^\alpha\) acting on edge \(e\). 
The stabilizers generate an abelian group; the basic relations are
\begin{gather}
A_v^2=B_p^2=\mathbf 1,\\
[A_v,A_{v'}]=0,\qquad [B_p,B_{p'}]=0,\\
[A_v,B_p]=0 .
\end{gather}
The last line follows because, on the square lattice, the overlap
\(\bigl|(+_v)\cap(\partial p)\bigr|\in\{0,2\}\) is always even,
where \(\bigl|(+_v)\cap(\partial p)\bigr|\) denotes the number of edges
shared by the star \(+_v\) and the plaquette boundary \(\partial p\).
The unperturbed toric-code Hamiltonian is
\[
H_0=-\sum_v A_v-\sum_p B_p,
\]
whose ground state space is the common \(+1\) eigenspace of all \(A_v\) and \(B_p\).
Let \(C_{v}\), \(C_{h}\) be the vertical/horizontal non-contractible loops on the direct lattice,
and \(\tilde C_{v}\), \(\tilde C_{h}\) the corresponding dual-lattice cycles (see Fig.~\ref{fig:tc}\,(a)).
We consider the Wilson loop operators
\begin{gather}
\Wzv = \prod_{e\in \mathcal{E}(C_v)}\sigma^z_e,\qquad
\Wzh = \prod_{e\in \mathcal{E}(C_h)}\sigma^z_e,\\
\Wxv = \prod_{e\in \mathcal{E}(\tilde C_v)}\sigma^x_e,\qquad
\Wxh = \prod_{e\in \mathcal{E}(\tilde C_h)}\sigma^x_e.
\end{gather}
These operators square to the identity and commute with all stabilizers:
\begin{gather}
\bigl(W_z^{(\alpha)}\bigr)^2=\mathbf 1,\quad 
\bigl(W_x^{(\alpha)}\bigr)^2=\mathbf 1,\quad 
\alpha\in\{v,h\},\\
[W_z^{(\alpha)},A_v]=[W_z^{(\alpha)},B_p]=0,\quad \forall\, v,p,\\
[W_x^{(\alpha)},A_v]=[W_x^{(\alpha)},B_p]=0,\quad \forall\, v,p.
\end{gather}

Operators of the same type commute, and operators from different types that cross once anticommute:
\begin{gather}
[\Wzv,\Wzh]=0,\quad [\Wxv,\Wxh]=0,\\
\Wzv\Wxh=-\,\Wxh\Wzv,\\
\Wzh\Wxv=-\,\Wxv\Wzh .
\end{gather}

On the torus, the commuting Wilson loop operators provide a natural set of nonlocal observables whose eigenvalues distinguish the four topological sectors in the ground state space. Since mutually anticommuting loop operators act as Pauli operators, the ground space may be viewed as encoding two logical qubits.
A convenient logical choice is
\begin{equation}\label{eq:logical_ops}
\begin{aligned}
\bar Z_1&=\Wzh,\quad \bar X_1=\Wxv,\\
\bar Z_2&=\Wzv,\quad \bar X_2=\Wxh .
\end{aligned}
\end{equation}
which satisfy the two-qubit Pauli algebra.

We next introduce an isotropic Heisenberg interaction with non-negative coupling $J$, 
\[
J \sum_{\langle i,j\rangle}\!\bigl(\sigma_i^x\sigma_j^x+\sigma_i^y\sigma_j^y+\sigma_i^z\sigma_j^z\bigr),
\]
defined on a chosen set of nearest-neighbour edge pairs \(\langle i,j\rangle\), see Fig.~\ref{fig:tc}(a). 
In general, this interaction does not commute with the stabilizers \(A_v\) and \(B_p\) and therefore induces virtual transitions out of the stabilizer ground state space. We arrive at the full Hamiltonian
\[
H
= - \sum_{v} A_v \;-\;  \sum_{p} B_p
\;+\; J \sum_{\langle i,j\rangle}
\bigl(\sigma_i^x\sigma_j^x+\sigma_i^y\sigma_j^y+\sigma_i^z\sigma_j^z\bigr).
\]

For a vertex \(v\) with star \(+_v\) and a plaquette \(p\) with boundary \(\partial p\), the parity of the overlap of the bond \(\{i,j\}\) with these sets determines the anticommutation relations:
\begin{gather}
\begin{aligned}
&\{\sigma_i^x\sigma_j^x,\; B_p\}=0 &&\text{if }|\{i,j\}\!\cap\!\partial p|=1,
\end{aligned}\\[2pt]
\begin{aligned}
&\{\sigma_i^z\sigma_j^z,\; A_v\}=0 &&\text{if }|\{i,j\}\!\cap\!+_v|=1,
\end{aligned}\\[2pt]
\begin{aligned}
&\{\sigma_i^y\sigma_j^y,\; A_v\}=0 &&\text{if }|\{i,j\}\!\cap\!+_v|=1,\\
&\{\sigma_i^y\sigma_j^y,\; B_p\}=0 &&\text{if }|\{i,j\}\!\cap\!\partial p|=1,
\end{aligned}
\end{gather}
and otherwise the corresponding commutators vanish.  Here \(\langle i,j\rangle \) denotes the bond on which the two-spin operator
\(\sigma_i^\alpha\sigma_j^\alpha\) acts, and
\(|\langle i,j \rangle\cap X|\) counts how many of its endpoints belong to the edge set \(X\).
In words: a bond operator \(\sigma_i^x\sigma_j^x\) always commutes with all star operators but flips a plaquette eigenvalue if it overlaps with exactly one edge of that plaquette; a bond operator \(\sigma_i^z\sigma_j^z\) satisfies the dual statement with stars and plaquettes interchanged; and a bond operator \(\sigma_i^y\sigma_j^y\) flips a star (plaquette) eigenvalue whenever it touches the corresponding star (plaquette) on an odd number of edges.
These local rules will be the building blocks in our SWT analysis.

\subsection{Marshall Gauge}
\label{Marshall Gauge}
In the computational (\(\sigma^z\)) basis, the full Hamiltonian has mixed-sign off-diagonal elements: 
the star term \(-\sum_v A_v\) contributes negative off-diagonals, 
whereas the Heisenberg exchange interaction  \(J\sum_{\langle i,j\rangle}(\sigma_i^x\sigma_j^x+\sigma_i^y\sigma_j^y)\) contributes positive off-diagonal elements.
Consequently, the Hamiltonian is non-stoquastic , i.e., some off-diagonal matrix elements are positive, in this basis, implying that the ground-state wave function generally exhibits a nontrivial sign structure \cite{Bravyi2008,Troyer2005}.

However, the Heisenberg bonds \(\langle i,j\rangle\) form a bipartite edge graph (Fig.~\ref{fig:tc}\,(b)) and the coupling is antiferromagnetic (\(J\ge 0\)). 
Under these conditions, a Marshall sign transformation renders all off-diagonal elements non-positive, i.e., making the Hamiltonian stoquastic in this basis. 
As a result, the Perron-Frobenius theorem guarantees that the ground state can be chosen with non-negative amplitudes~\cite{Bravyi2008}. This substantially simplifies variational optimization: we may employ a positive real-valued NQS without learning a complex sign structure.

Let $\mathcal{E}=A\cup B$ be a partition of edges, $A\cap B=\varnothing$, chosen such that
$A$ contains only dual-lattice edges and $B$ contains the direct-lattice edges, as illustrated in Fig.~\ref{fig:tc}(b).
We then define the Marshall transformation~\cite{marshall1955} as
\begin{equation}\label{eq:Marshall_transformation}
U_M \coloneqq \prod_{i\in A} \sigma_i^z .
\end{equation}
For any bond \(\langle i,j\rangle\) with \(i\in A,\ j\in B\), one finds
\begin{align}
U_M(\sigma_i^x\sigma_j^x)\,U_M^\dagger &= -\sigma_i^x\sigma_j^x, \\
U_M(\sigma_i^y\sigma_j^y)\,U_M^\dagger &= -\sigma_i^y\sigma_j^y, \\
U_M(\sigma_i^z\sigma_j^z)\,U_M^\dagger &= \phantom{-}\sigma_i^z\sigma_j^z,
\end{align}
while the stabilizers are invariant by construction,
\begin{equation}
U_M A_v U_M^\dagger = A_v,\qquad
U_M B_p U_M^\dagger = B_p .
\end{equation}
Thus, the gauge-transformed Hamiltonian
\begin{align}
H' &\coloneqq U_M H U_M^\dagger \nonumber\\
   &= -\sum_v A_v - \sum_p B_p \nonumber\\
   &\phantom{=}\; + J \sum_{\langle i,j\rangle}
      \bigl[-(\sigma_i^x\sigma_j^x+\sigma_i^y\sigma_j^y)
            + \sigma_i^z\sigma_j^z\bigr]
\end{align}
is stoquastic in the \(\sigma^z\) basis. 
Then, the ground state of \(H'\) can be chosen non-negative, and the ground state of the original \(H\) has the Marshall sign structure
\begin{equation}
\psi(\sigma)=(-1)^{N_\downarrow(A;\sigma)}\,\phi(\sigma),\qquad \phi(\sigma)\ge 0,
\end{equation}
where \(N_\downarrow(A;\sigma)\) counts the number of down-spins on \(A\) in configuration \(\sigma\).

We next explain how to obtain the expectation value of observables in the physical gauge from the ground state in the Marshall gauge. 
Let the ground state of $H$ be $\ket{\psi_{phys}}$. Then 
the ground state in Marshall gauge can be expressed by \(\ket{\psi_M}=U_M\ket{\psi_{\rm phys}}\)  Since \(U_M\) is unitary, any expectation value in the physical gauge
 follows from the Marshall gauge via
\begin{equation}\label{eq:OM_definition}
\langle O\rangle_{\rm phys}=\langle \psi_M|\,O_M\,|\psi_M\rangle,\qquad
O_M \equiv U_M O U_M^\dagger .
\end{equation}
With \(U_M=\prod_{e\in A}\sigma^z_e\), one only needs an edgewise sign flip for \(x/y\) on the \(A\) sublattice:
\begin{gather}
U_M \sigma^{x,y}_e U_M^\dagger=\eta_e\,\sigma^{x,y}_e,\quad
U_M \sigma^{z}_e U_M^\dagger=\sigma^{z}_e,\\
\eta_e=\begin{cases}-1,& e\in A,\\ +1,& e\in B.\end{cases}
\end{gather}

\subsection{Global Parity Symmetries and Logical Structure}
\label{Global Parity Symmetries and Logical Structure}

We define the global spin-parity operators
\begin{equation}
P_x=\prod_i \sigma_i^x, \qquad
P_z=\prod_i \sigma_i^z ,
\end{equation}
where the product runs over all spins residing on the edges of the square lattice.
Because the star, plaquette, and bond operators include an even number of Pauli spin operators, one readily verifies 
\begin{equation}
[H,P_x]=0, \qquad [H,P_z]=0.
\end{equation}
Furthermore, for a two-atomic unit cell on an $L\times L$ torus, the total number of spins is $N=2L^2$, which is even.
Thus, parity operators commute among themselves.
\begin{equation}
P_x P_z = (-1)^N P_z P_x = P_z P_x
\quad \Rightarrow \quad
[P_x,P_z]=0 .
\end{equation}
These symmetries allow one to organize the Hilbert space into fixed $(P_x,P_z)$ sectors,
which our neural-network ansatz explicitly targets.
We note that the Heisenberg interaction preserves these global symmetries for all $J$, ensuring that total parity remains a robust quantum number. In contrast, the interaction does not commute with the non-contractible Wilson loops, meaning topological sectors are no longer conserved and their eigenvalues are hybridized by anyonic fluctuations. 
However, fixing the total parity has important consequences for the structure of the ground state space of the unperturbed toric code, which depends crucially on the lattice geometry.
Since a non-contractible Wilson loop of Pauli type $\alpha\in\{x,z\}$ along direction
$\gamma\in\{h,v\}$ is a product of $L$ single-spin operators, conjugation by the global
parities gives
\begin{equation}
\begin{aligned}
P_x\, W_z^{(\gamma)}\, P_x &= (-1)^{L}\, W_z^{(\gamma)},\\
P_z\, W_x^{(\gamma)}\, P_z &= (-1)^{L}\, W_x^{(\gamma)}.
\end{aligned}
\end{equation}
In contrast, Wilson loops constructed from the same Pauli component as the parity operator commute trivially,
\begin{equation}
[P_x, W_x^{(\gamma)}] = [P_z, W_z^{(\gamma)}] = 0,
\qquad \gamma\in\{h,v\}.
\end{equation}
For lattices with even $L$, the factor $(-1)^L=+1$.
Therefore, all non-contractible Wilson loops commute with both parity operators,
\begin{equation}
[P_x, W_\alpha^{(\gamma)}] = [P_z, W_\alpha^{(\gamma)}] = 0,
\qquad (\text{even } L),
\end{equation}
for all $\alpha\in\{x,z\}$ and $\gamma\in\{h,v\}$.
Consequently, the ground state can be chosen as a simultaneous eigenstate of the Hamiltonian,
all four non-contractible Wilson loops, and both total parity operators.

For lattices with odd $L$, the situation is qualitatively different.
In this case, the cross-type pairs anticommute,
\begin{equation}
\{P_x, W_z^{(\gamma)}\} = \{P_z, W_x^{(\gamma)}\} = 0,
\qquad (\text{odd } L),
\end{equation}
while $[P_x, W_x^{(\gamma)}]=0$ and $[P_z, W_z^{(\gamma)}]=0$ still hold.
This algebraic structure implies that no state can be a simultaneous eigenstate of both
the global parity operators and all logical Wilson loops.
For example, let $\ket{w_h,w_v}$ denote the common eigenbasis of the two $z$-type logical Wilson loops,
\begin{equation}
\begin{gathered}
W_z^{(\gamma)}\ket{w_h,w_v} = w_\gamma\,\ket{w_h,w_v},\\
\gamma \in \{h,v\},\quad w_h,w_v = \pm 1.
\end{gathered}
\end{equation}
Then, the eigenstate with $(p_x,p_z)=(-1,-1)$ can be chosen as
\begin{equation}
\begin{aligned}
\ket{\Psi_{p_x=-1,\;p_z=-1}}
&=
\frac{1}{\sqrt{2}}
\Bigl(
\ket{w_h=+1,\;w_v=-1}
\\[-2pt]
&\hspace{2.6em}
-
\ket{w_h=-1,\;w_v=+1}
\Bigr).
\end{aligned}
\label{eq:Pxpz_minusminus}
\end{equation}

\subsection{Variational Neural-Network Ansatz}
\label{Variational Neural-Network Ansatz}

To approximate the ground state of the Hamiltonian $H$, we employ a variational
neural-network quantum state (NQS) that represents the many-body wave function
$\psi(\sigma)$ in the computational ($\sigma^z$) basis.
Since the Hamiltonian can be rendered stoquastic by a Marshall sign transformation, see Eq.~\eqref{eq:Marshall_transformation}, the ground-state amplitudes can be chosen real and non-negative.
This allows us to focus exclusively on learning the amplitude structure of the wave
function using a real-valued neural-network ansatz.

The network architecture is designed to exploit the physical symmetries of the lattice \cite{Liang2018,Liang2021,choo2019,Szabo2020}.
We utilize translational symmetry via convolutional weight sharing, by mapping the edge-spin
configuration onto a two-dimensional grid.
A naive encoding, however, leads to unstable training, as star and plaquette couplings
are distributed across distant grid locations \cite{corrRBM}.
We resolve this issue by introducing a geometry-aware positional embedding, which maps the four bonds incident on each vertex into local
$2\times4$ blocks.
This construction, detailed in Appendix~\ref{appendix_A}, condenses the off-diagonal
star operators into local features that can be efficiently processed by the network.
To further restrict the variational space to physically relevant states, we explicitly enforce invariance under discrete $C_4$ lattice rotations.
For a network that outputs the log-amplitude
$\ell_\theta(\sigma)=\log\psi_\theta(\sigma)$,
we define the $C_4$-symmetrized model as
\begin{equation}
\ell_\theta^{(C_4)}(\sigma)
=
\log\!\left[
\frac{1}{4}
\sum_{k=0}^{3}
\exp\!\left(\ell_\theta(\mathcal{R}^k\sigma)\right)
\right],
\end{equation}
where $\mathcal{R}$ denotes a $90^\circ$ rotation acting on the spin configuration.
This symmetrization increases the computational cost but significantly
reduces the effective variational search space and improves convergence toward the
symmetric ground state~\cite{Reh2023,Vieijra2020,Vieijra2021,nomura2021,Roth2021GCNN,Mezera2023ShastrySutherland}.

In practice, the network utilizes modern CNN techniques such as skip connections and increased cardinality to enhance expressivity while maintaining numerical stability, which are specified in Appendix~\ref{appendix_A}. To prevent divergence in the log-amplitudes as the system size increases, we employ normalized global aggregation layers. By construction, the resulting output is invariant under lattice translations by integer units of stars or plaquettes.
To train the NQS, we employ a variational Monte Carlo (VMC) approach, the details of which are provided in Appendix~\ref{appendix_B}.

\subsection{Parity on Marshall Gauge and Duality Constraints}
\label{Parity on Marshall Gauge and Duality Constraints}

Our variational ansatz is chosen to be even under a global spin flip in the $\sigma^z$
basis,
\begin{equation}
\psi(\{\sigma^z\})=\psi(\{-\sigma^z\})
\quad \Longleftrightarrow \quad
P_x|\psi\rangle = +|\psi\rangle .
\end{equation}
The calculation is performed in the Marshall gauge,
$H' = U_M H U_M^\dagger$, under which the variational ground state is restricted
to the $P_x=+1$ eigenspace of $H'$.
The corresponding physical expectation value follows from
\begin{equation}
U_M P_x U_M^\dagger = (-1)^{L^2} P_x ,
\end{equation}
so that
\begin{equation}\label{x-parity}
\langle P_x\rangle_{\mathrm{phys}}
=
(-1)^{L^2}\,\langle P_x\rangle ,
\end{equation}
where the expectation value on the right-hand side is evaluated in the Marshall gauge.

For even $L$, the operator $P_z$ can be expressed as a product of plaquette operators,
which fixes $P_z=+1$.
For odd $L$, no such identity exists.
Nevertheless, the Hamiltonian possesses an exact $x\!\leftrightarrow\!z$ duality.
Let $H_{\rm ad}$ denote the single-qubit Hadamard operator,
\begin{equation}
H_{\rm ad}\sigma^x H_{\rm ad}=\sigma^z, \quad
H_{\rm ad}\sigma^z H_{\rm ad}=\sigma^x, \quad
H_{\rm ad}\sigma^y H_{\rm ad}=-\sigma^y .
\end{equation}
Let $\mathcal T$ be the unitary operator implementing a one-site lattice translation,
\begin{equation}
\mathcal T\,\sigma_e^\mu\,\mathcal T^\dagger
=
\sigma_{\mathcal T(e)}^\mu ,
\end{equation}
and define the duality operator
\begin{equation}
\mathcal D_{\rm dual}
=
\mathcal T\,(H_{\rm ad})^{\otimes N}.
\end{equation}
Its action exchanges stars and plaquettes and maps
\begin{equation}
\mathcal D_{\rm dual} P_z \mathcal D_{\rm dual}^\dagger = P_x .
\end{equation}
Since the Heisenberg interaction is invariant under this transformation,
\begin{equation}
[\mathcal D_{\rm dual},H]=0 ,
\end{equation}
one expects $\langle P_x\rangle = \langle P_z\rangle$ in any finite-size system
with a nondegenerate ground state.
Combining this with the Marshall-Gauge relation yields
\begin{equation}\label{z-parity}
\langle P_z\rangle_{\mathrm{phys}} = (-1)^{L^2} ,
\end{equation}
i.e., $\langle P_z\rangle_{\mathrm{phys}}=+1$ for even $L$ and
$\langle P_z\rangle_{\mathrm{phys}}=-1$ for odd $L$.
Since $P_z$ is not explicitly fixed within our variational ansatz,
we introduce a weak guiding term that biases the optimization toward the
appropriate parity sector, as detailed in Appendix~\ref{appendix_B}.
We have verified this relation by exact diagonalization on $2\times2$ and
$3\times3$ tori. Already for a very small Heisenberg coupling,
$J=10^{-4}$, the finite-size ground state lies in the parity sector consistent
with $\langle P_x\rangle_{\mathrm{phys}}=\langle P_z\rangle_{\mathrm{phys}} = (-1)^{L^2}$.

\subsection{Observables for the detection of the quantum phase transition}
\label{Detecting of quantum phase transition}

In this subsection, we explain the observables used to detect phase transitions and determine the phase after the transition.

We evaluate the ground-state fidelity between the states at $J$ and $J+\delta J$,
\begin{equation}
F(J,J+\delta J)\equiv
\bigl|\langle \psi_M(J)\mid \psi_M(J+\delta J)\rangle\bigr|^2,
\label{eq:fidelity_def}
\end{equation}
and estimate the fidelity susceptibility from the small-$\delta J$ expansion
\begin{equation}
F(J,J+\delta J)=1-\chi_F(J)(\delta J)^2+\mathcal{O}\!\left((\delta J)^3\right).
\label{eq:fidelity_expansion}
\end{equation}
In practice, we assign the estimate to the midpoint
\begin{equation}
J_{\rm mid}\equiv J+\frac{\delta J}{2},
\end{equation}
and use the finite-difference formula
\begin{equation}
\chi_F(J_{\rm mid})\approx \frac{\,[1-F(J,J+\delta J)]}{(\delta J)^2}.
\label{eq:chiF_fdiff}
\end{equation}
For a continuous quantum phase transition, $\chi_F$ obeys the finite-size scaling form
\begin{equation}
\chi_F(J,L)=L^{2/\nu}\,\Phi\!\bigl((J-J_c)L^{1/\nu}\bigr),
\label{eq:chiF_fss}
\end{equation}
where  $\nu$ is the correlation-length exponent~\cite{zanardi2006,zanardi2007,albuquerque2010,you2007,gu2010}.
{Equation~\eqref{eq:chiF_fss} implies that the peak height scales as
\begin{equation}
\chi_F^{\max}(L)\propto L^{2/\nu},
\label{eq:chiF_peak_height}
\end{equation}
while the peak position $J^{*}(L)$ 
drifts towards the thermodynamic critical point $J_c$ as
\begin{equation}
J^{*}(L)-J_c \propto L^{-1/\nu}.
\label{eq:chiF_peak_shift}
\end{equation}
In Sec.~\ref{ground_state}, we utilize these relations to extract $\nu$ from the peak height and estimate $J_c$ via the peak shift using Eq.~(\ref{eq:chiF_peak_shift}).

To complement the fidelity-based analysis with a diagnostic that directly probes the nonlocal topological content of the ground state, we introduce
\begin{equation}
\begin{split}
\bar{W}_d(J,L) &= \frac{1}{2}\Bigl(\bigl|\langle W_x^{(h)}W_x^{(v)}\rangle_{\mathrm{phys}}\bigr| \\
&\quad + \bigl|\langle W_z^{(h)}W_z^{(v)}\rangle_{\mathrm{phys}}\bigr|\Bigr).
\end{split}
\label{eq:Wd_bar}
\end{equation}
Deep in the topological phase, $\bar{W}_d$ remains of order unity, although its baseline value differs between odd and even $L$: for odd $L$, both loop products approach $-1$ in magnitude (see, Eq.~\eqref{odd_double}), giving $\bar{W}_d \simeq 1$, whereas for even $L$ they take values slightly larger than $1/2$, giving $\bar{W}_d \simeq 0.5$--$0.6$ (see, Eq.~\eqref{even_double}). This parity-dependent baseline is a finite-size effect originating from the interplay between global parity constraints and the non-contractible Wilson loop algebra, and is discussed in Sec.~\ref{sec:methods_sw_logical_summary}. Beyond the transition, the non-contractible loop expectation values are exponentially suppressed in $L$, so $\bar{W}_d \to 0$ as $L\to\infty$ in the trivial phase. To locate the steepest decay of topological order, we monitor the logarithmic susceptibility
\begin{equation}
\chi_W^{(\ln)}(J,L) \;=\; -\frac{d \ln \bar{W}_d}{dJ}\,,
\label{eq:chi_W_ln}
\end{equation}
which we evaluate numerically via finite differences. We adopt the logarithmic form to measure the relative rate of change, which is insensitive to the parity-dependent baseline.
This yields peak positions providing a robust nonlocal estimate of the critical coupling that complements the fidelity-susceptibility analysis.

In addition, we compute the second R\'enyi entropy \(S_2(R)\) of subsystems \(R\)
using a replica (SWAP) estimator and extracting the topological constant via the
Kitaev--Preskill (KP) construction~\cite{KPTEE,swap}.
For three adjacent regions \(A,B,C\) each forming a simply connected patch, the KP
combination
\begin{equation}
\begin{split}
\gamma_{\mathrm{KP}}
\equiv{}& S_2(A) + S_2(B) + S_2(C) \\
       &- S_2(AB) - S_2(BC) - S_2(CA) \\
       &+ S_2(ABC)
\end{split}
\label{eq:KP_def}
\end{equation}
is designed to cancel all boundary-law contributions \(\alpha|\partial R|\) as well as corner-dependent and other local short-range terms, leaving only the
universal \(O(1)\) constant associated with long-range entanglement.
With this convention, the $\mathbb{Z}_2$ topologically ordered phase is 
characterized by a constant topological entanglement entropy $\gamma_{\mathrm{KP}} = -\ln 2$ in the thermodynamic limit. 
Beyond the topological transition, conventional order may emerge.
In a finite system without an explicit symmetry-breaking field, the ground state can remain symmetry-preserving and effectively form a superposition over symmetry-related configurations \cite{Beekman2019SSB}.
For broken discrete symmetries with degeneracy $d$, this sector mixing contributes an additive Shannon-like term $+\ln d$ to the entanglement entropy (opposite in sign to the topological reduction), which does not necessarily cancel in the Kitaev--Preskill subtraction and can therefore yield an apparent $\gamma_{\mathrm{KP}}>0$ even in a topologically trivial ordered regime~\cite{Jiang2012SpinLiquidJ1J2,stephan2009,Kiely2022}.
Moreover, at intermediate lattice sizes, the entanglement-entropy extraction can be contaminated by non-universal finite-size effects (e.g., when the correlation length is comparable to the linear size of the subregions).
Hence, conventional order in finite size system can lead to an intrinsically ambiguous residual nonzero $\gamma_{\mathrm{KP}}$ in the non-topological regime~\cite{You2022EntanglementLens,Laflorencie2016EntanglementReview,stephan2009,Kallin2011}.}
Furthermore, we note that fidelity and topological entanglement entropy(TEE) are invariant under the Marshall gauge, \(U_M\); inner products are preserved, and Rényi-2 entropies from the swap estimator are unchanged because local unitaries on a region commute with the swap operator.
Finally, to analyze conventional antiferromagnetic order, we use the Marshall-transformation adapted staggered magnetization:
\begin{equation}
\label{eq:mstag}
\begin{aligned}
m_\alpha^{\mathrm{stag}} &= \frac{1}{N}\sum_i \eta_i\,\sigma_i^\alpha, \qquad \alpha\in\{x,y,z\},\\
\eta_i &= \begin{cases}+1,& i\in B,\\ -1,& i\in A.\end{cases}
\end{aligned}
\end{equation}

\section{Schrieffer-Wolff transformation}
\label{Schrieffer–Wolff transformation}
The SW transformation is a canonical unitary approach for deriving effective low-energy
descriptions of quantum many-body systems, which have been successfully applied for condensed matter physics~\cite{Voleti2021,Kale2022,Hu2023,Manning-Coe2023} and quantum information theory~\cite{Benito2019,Mutter2020,Mutter2021,Bosco2021,Liu2022,Hetenyi2022,Kawakami2023,Zhang2023,Delvecchio2023,Zhang2022,Marecat2023}.
Given a Hamiltonian with a clear separation of energy
scales, the SW procedure constructs a unitary rotation $e^{S}$ 
that block-diagonalizes
the perturbed Hamiltonian with respect to a chosen low-energy subspace, thereby integrating out virtual transitions to high-energy
sectors and producing an effective Hamiltonian acting within the low-energy manifold \cite{bravyiSW,schrieffer1966,Kim2021,Hillmann2022,Reascos2024,Winkler2003,Landi2024}.

In our setting, the Heisenberg exchange does not commute with the toric-code stabilizers and thus couples the
ground-state manifold to sectors with virtual $e$ and $m$ anyons. We apply the SW transformation in the projector formalism~\cite{bravyiSW} to derive the effective Hamiltonian on the stabilizer ground state space, integrating out these virtual anyon processes order by order.
This yields a controlled expansion of the effective Hamiltonian and explicitly identifies the operator contributions arising at each perturbative order. In the following, we present the resulting effective Hamiltonian and show that the leading orders merely renormalize the star and plaquette terms, while mixing of topological sectors appears only at order $L$, i.e., at a perturbative order proportional to the system size. Our results are consistent
with earlier work that finds topological-sector mixing exponentially small in $L$, and further clarify how a local
perturbation dress stabilizers and Wilson loops in the effective theory. In addition, we analyze how the parity of the lattice size $L$ influences the finite-size ground-state selection and lifts the degeneracy differently for even and odd systems. Detailed derivations are provided in Appendix~\ref{appendix:SW-details}.

\subsection{Model and SW setup}

We split the Hamiltonian as
\begin{equation}
\begin{aligned}
H_0^{SW} &= -J_e\sum_v A_v - J_m\sum_p B_p,
\end{aligned}
\end{equation}
introducing independent positive coupling constants $J_e$ and $J_p$ for the star and plaquette operators.
The isotropic antiferromagnetic  Heisenberg perturbation on the edge graph  with coupling $J\geq0$  is given by

\begin{equation}
\begin{aligned}
V &= J\sum_{b=\langle i,j\rangle}\bigl(X_b+Y_b+Z_b\bigr),\\
X_b &= \sigma^x_i\sigma^x_j,\quad
Y_b = \sigma^y_i\sigma^y_j,\quad
Z_b = \sigma^z_i\sigma^z_j .
\end{aligned}
\label{hei_SW}
\end{equation}
Acting once with these bond operators creates two-anyon excitations with gaps 
\begin{equation}
\begin{aligned}
\Delta_x &= 4J_m,\\
\Delta_z &= 4J_e,\\
\Delta_y &= 4(J_e+J_m).
\end{aligned}
\end{equation}
We thus identify the controlled-expansion regime as $J \ll \min(4J_m,4J_e)$.

\subsection{Projector--resolvent form of the Schrieffer--Wolff generator}
\label{sec:sw_generator_resolvent}

We briefly summarize the standard SW construction in the
projector-resolvent language~\cite{bravyiSW}.

Let $P$ project onto the unperturbed code space and $Q=\mathbf 1-P$ onto its orthogonal complement.  We seek an anti-Hermitian generator $S^\dagger=-S$ such
that the rotated Hamiltonian $\tilde H \equiv e^{S}H^{\rm SW}e^{-S}$ is block
diagonal, effectively decoupling the $P$ and $Q$ subspaces order by order in the perturbation strength.
Working in the off-diagonal gauge
\begin{equation}
P S P = Q S Q = 0 ,
\end{equation}
and expanding $S=S_1+\mathcal O(J^2)$, the first-order generator can be written
compactly in terms of the reduced resolvent
\begin{equation}
R \equiv Q\,(E_P-H_0^{\rm SW})^{-1}Q, \qquad PR=RP=0,
\end{equation}
as
\begin{equation}
S_1 = P V R - R V P ,
\label{eq:S1_projector_resolvent}
\end{equation}
which cancels the $\mathcal O(J)$ mixing between $P$ and $Q$.
Projecting onto $P$ gives
\begin{equation}
\begin{split}
H_{\rm eff}
&\equiv P\,e^{S}(H_0^{\rm SW}+V)e^{-S}P \\
&= P(H_0^{\rm SW}+V)P + \mathcal{O}(J^2).
\end{split}
\end{equation}
For the present Heisenberg perturbation, a single bond creates stabilizer defects, hence $PVP=0$ and the first-order correction vanishes:
\begin{equation}
H_{\rm eff} = P H_0^{\rm SW} P + \mathcal O(J^2).
\end{equation}

\subsection{Effective Hamiltonian to \texorpdfstring{$\mathcal O(J^2)$}{O(J²)}}
\label{sec:sw_Heff_J2}

The leading nontrivial contribution to $H_{\rm eff}$ arises at second order.
Using the SW expansion with the generator in Eq.\eqref{eq:S1_projector_resolvent}, we
find an $\mathcal O(J^2)$ correction controlled by two-bond virtual processes. The projected effective Hamiltonian through $\mathcal O(J^2)$ is
\begin{equation}
\begin{split}
H_{\rm eff} ={} & P\biggl[ -\Bigl(J_e+\frac{J^2}{J_m}\Bigr)\sum_v A_v -\Bigl(J_m+\frac{J^2}{J_e}\Bigr)\sum_p B_p \\
&\quad + E_{\rm const}^{(2)} \biggr]P +\mathcal{O}(J^3)
\end{split}
\end{equation}
with a second-order constant shift
\begin{equation}
E_{\mathrm{const}}^{(2)}
=
-\,L^2 J^2
\left(
\frac{1}{J_m} + \frac{1}{J_e+J_m} + \frac{1}{J_e}
\right).
\label{eq:Econst_OJ2}
\end{equation}
Since $P$ projects onto the code space where $A_v=B_p=+1$, $P H_{\rm eff}P$
reduces to an overall energy shift within the ground space sector up to second order.
Specifically, on an $L\times L$ torus, there are $L^2$ vertex stabilizers $A_v$ and $L^2$ plaquette stabilizers $B_p$, while the total number of spins is $N=2L^2$.
Hence, the energy per spin becomes
\begin{equation}
\begin{aligned}
\frac{E_0(J)}{N}
&=
-\frac{J_e+J_m}{2}
-\frac{J^2}{2}\left[
\frac{2}{J_m}+\frac{2}{J_e}\right. \\
&\qquad\left.
+\frac{1}{J_e+J_m}
\right]
+\mathcal O(J^3),
\end{aligned}
\label{eq:E0_over_N_OJ2}
\end{equation}
which is independent of $L$ up to second order.
For the choice $J_e=J_m=1$, Eq.~\eqref{eq:E0_over_N_OJ2} reduces to
\begin{equation}
\label{energyperspin}
\frac{E_0(J)}{N} = -1 - \frac{9}{4}J^2 + \mathcal O(J^3).
\end{equation}

\subsection{Higher-order SW terms and logical-operator effective theory}
\label{sec:methods_sw_logical_summary}

Beyond second order, the SW expansion generates a hierarchy of
gauge-invariant operator structures built from connected strings of perturbations
and reduced resolvents.  After projection onto the toric-code ground space, products of local stabilizers—equivalently, any contractible loop operators—act as constants and can therefore be absorbed into an overall energy shift.
Consequently, the operator contents that remain qualitatively relevant for topological-sector splitting on the torus are those involving non-contractible Wilson loops and firstly appear at \(L\)\text{-th order}.

While an $L \times L$ torus admits $L$ distinct straight paths for each 
winding direction $\gamma \in \{h,v\}$ and Pauli type $\alpha \in \{x,z\}$, 
these operators are not independent. One can define $L$ representatives 
$W_\alpha^{(\gamma)}(r)$ by translating a loop transversally by $r$ lattice 
units; however, all these loops operators are the same up to multiplication by contractible strips generated 
by $\{A_v\}$ and $\{B_p\}$. Since the projector $P$ enforces $A_v=B_p=+1$ 
in the toric-code ground space, such contractible factors act trivially after 
projection, and all translated representatives act identically within 
the code space:
\begin{equation}
P\,W_\alpha^{(\gamma)}(r)\,P = P\,W_\alpha^{(\gamma)}(r')\,P, \qquad \forall\,r,r'.
\label{eq:W_r_equivalence_main}
\end{equation}
We may therefore drop the transverse label and define the corresponding
code-space winding operator as
\begin{equation}
W_\alpha^{(\gamma)}
\;\equiv\;
P\,W_\alpha^{(\gamma)}(r_0)\,P
\;=\;
\frac{1}{L}\,
P\Big(\sum_{r=1}^{L} W_\alpha^{(\gamma)}(r)\Big)P,
\label{eq:W_def_projected_mean_main}
\end{equation}
for any fixed representative $r_0$.  The normalization ensures eigenvalues
$\pm1$ in the ground space (matching a logical Pauli operator), while the
translation-average on the right-hand side makes the underlying symmetry
manifest and is often convenient in numerical measurements.
Importantly, loops winding along $h$ and $v$ remain independent logical
operators on the torus.

At the dual point $J_e=J_m=1$, the requirements of hermiticity, $C_4$ 
rotational symmetry, and duality restrict the most general effective 
Hamiltonian to the simpler form.
The resulting operator content is subject to an even--odd selection rule: 
for odd $L$, the global parities $P_x$ and $P_z$ forbid all terms linear 
in a single winding loop. For even $L$, however, such single-loop terms 
are allowed and appear at the \(L\)\text{-th order}.
Using Eq.~\eqref{eq:logical_ops}, the general effective Hamiltonian can be written as
\begin{equation}
\begin{aligned}[b]
H_{\rm eff}^{\rm odd}
&=
c_0\,P
+P\Bigl[
u\bigl(\bar X_1\bar X_2+\bar Z_1\bar Z_2\bigr)
+w\,\bar Y_1\bar Y_2
\Bigr]P,
\\[2pt]
H_{\rm eff}^{\rm even}
&=
c_0\,P
+P\Bigl[
t\bigl(\bar X_1+\bar X_2+\bar Z_1+\bar Z_2\bigr)
\\
&\qquad
+u\bigl(\bar X_1\bar X_2+\bar Z_1\bar Z_2\bigr)
\\
&\qquad
+v\bigl(\bar X_1\bar Z_2+\bar Z_1\bar X_2\bigr)
+w\,\bar Y_1\bar Y_2
\Bigr]P,
\end{aligned}
\label{eq:methods_Heff_odd_even_main}
\end{equation}
where $\bar Y_j \equiv i\bar X_j\bar Z_j$. All logical couplings are generated by winding (and higher-order connected)
virtual processes and scale parametrically as
$t,u,v,w=\mathcal{O}\!\left(J^{L}/\Delta^{L-1}\right)$ in the topological regime.

For odd $L$, the single winding loops are odd under the conserved global
parities $P_x$ and $P_z$, hence all single-loop expectation values vanish in any
fixed $(P_x,P_z)$ block,
\begin{equation}
\langle \bar X_1\rangle=\langle \bar X_2\rangle=
\langle \bar Z_1\rangle=\langle \bar Z_2\rangle=0
\qquad (L\ \text{odd}).
\label{eq:methods_odd_singleloops_vanish}
\end{equation}
The leading dynamics in $H_{\rm eff}^{\rm odd}$ is therefore governed by the
commuting two-qubit operators
$\bar X_1\bar X_2$, $\bar Z_1\bar Z_2$, and $\bar Y_1\bar Y_2$. 
Moreover, it turns out that
$u>0$ and $w>0$ while  $|w|\ll |u|$  for odd L (Appendix~\ref{app:sw_higher_logical}).
This selects a Bell-type ground state in the ground space, which implies the characteristic loop pattern \begin{equation} \begin{aligned} \langle W_x^{(h)}W_x^{(v)}\rangle &\simeq -1,\\ \langle W_z^{(h)}W_z^{(v)}\rangle &\simeq -1, \end{aligned} \qquad (L\ \text{odd},\ J\ll J_c) \label{odd_double} \end{equation}
together with Eq.~\eqref{eq:methods_odd_singleloops_vanish}.  
Thus, odd-$L$ systems naturally realize a cat-state structure with strong 
sector mixing even deep within the topological regime: Expectation values of 
single winding loops vanish, while parity-even loop products attain 
degenerate, maximally negative values.

In the intermediate size of even-$L$ case, the hierarchy $|t| \gtrsim |u| \gg |v|, |w|$ is maintained, with all four coefficients taking negative values (Appendix~\ref{app:sw_higher_logical}). 
When the coefficient $t$ dominates, the leading contribution is a single-qubit
logical field proportional to $\bar X+\bar Z$ on each logical qubit.  The corresponding ground state is approximately polarized along the bisector of the
$\bar X$ and $\bar Z$ axes, giving
\begin{equation}
\begin{aligned}
\langle \bar X_1\rangle \simeq \langle \bar X_2\rangle &\simeq \frac{1}{\sqrt2},\\
\langle \bar Z_1\rangle \simeq \langle \bar Z_2\rangle &\simeq \frac{1}{\sqrt2},
\end{aligned}
\qquad (L\ \text{even},\ J\ll J_c).
\label{eq:methods_even_bisector_expect}
\end{equation}

i.e.,
\begin{equation}
\langle W_x^{(h)}\rangle\simeq \langle W_x^{(v)}\rangle\simeq
\langle W_z^{(h)}\rangle\simeq \langle W_z^{(v)}\rangle\simeq \frac{1}{\sqrt2}.
\label{eq:methods_even_wilson_bisector}
\end{equation}
The remaining couplings $u,v,w$ act as subleading two-qubit interactions that
correlate the logical qubits and shift these expectations away from the ideal
bisector values. 
Specifically, the same dominant-bisector picture predicts the baseline
$\langle W_x^{(h)}W_x^{(v)}\rangle \approx \langle W_x^{(h)}\rangle\langle W_x^{(v)}\rangle
\simeq 1/2$ (and similarly for $z$) if the two logical qubits were independent.
In practice, however, the two-qubit logical interactions---most importantly the
$u(\bar X_1\bar X_2+\bar Z_1\bar Z_2)$ term---favor correlated responses of the two
winding directions and thus enhance the products above this uncorrelated value,
\begin{equation}\label{even_double}
\langle W_x^{(h)}W_x^{(v)}\rangle,\ \langle W_z^{(h)}W_z^{(v)}\rangle \;\gtrsim\; \frac12
\qquad (L\ \text{even},\ J\ll J_c).
\end{equation} 
Furthermore, since all logical couplings are tunneling-generated at order $L$, the
overall topological splitting scale
$\Delta E_{\rm topo}(L,J)\sim \mathcal{O}(J^{L}/\Delta^{L-1})$ becomes exponentially
small at weak $J$. Strictly speaking, at any fixed finite system size the ground state is unique and must
respect all symmetries of the Hamiltonian; it is therefore a symmetric superposition of
the four logically polarized torus sectors rather than a symmetry-broken sector state.
In the thermodynamic limit, the tunneling-induced splittings vanish and this
near-degeneracy becomes an exact degeneracy such that any one of the symmetry-broken
(sector-polarized) states can be selected as a legitimate ground state.

\subsection{Dressed stabilizers \texorpdfstring{$B_p$,$A_v$}{Bp,Av} and non-contractible Wilson loops}
\label{sec:sw_dressed_stabilizers}
We can obtain dressed
observables by applying the same SW unitary transformation,
\begin{equation}
O^{\rm eff} \equiv P\,e^{S} O\,e^{-S}P .
\end{equation}
At order $\mathcal O(J^2)$, the plaquette and star operators remain purely
multiplicatively renormalized (no operator mixing within $P$ at this order).
Using the excitation gaps $\Delta_x,\Delta_y,\Delta_z$, the dressed stabilizers read
\begin{align}
B_p^{\rm eff}
&=
\Bigl[1-32\bigl(\tfrac{J}{\Delta_x}\bigr)^2
        -16\bigl(\tfrac{J}{\Delta_y}\bigr)^2\Bigr]\,
P B_p P + \mathcal O(J^3)
\nonumber\\
&=
\Bigl[1-2\Bigl(\tfrac{J}{J_m}\Bigr)^2
       -\Bigl(\tfrac{J}{J_e+J_m}\Bigr)^2\Bigr]\,
P B_p P + \mathcal O(J^3),
\\[0.5em]
A_v^{\rm eff}
&=
\Bigl[1-32\bigl(\tfrac{J}{\Delta_z}\bigr)^2
        -16\bigl(\tfrac{J}{\Delta_y}\bigr)^2\Bigr]\,
P A_v P + \mathcal O(J^3)
\nonumber\\
&=
\Bigl[1-2\Bigl(\tfrac{J}{J_e}\Bigr)^2
       -\Bigl(\tfrac{J}{J_e+J_m}\Bigr)^2\Bigr]\,
P A_v P + \mathcal O(J^3).
\end{align}
Furthermore,
We define dressed shortest non-contractible Wilson loop operators by
\begin{equation}
W^{\rm eff}_{z/x}(C) \equiv P\,e^{S} W_{z/x}(C)\,e^{-S}P ,
\end{equation}
where $C$ is a straight winding cycle of length $L$.  At $\mathcal O(J^2)$ the
projection again yields a purely multiplicative renormalization.
For $W_z(C)=\prod_{e\in C}\sigma^z_e$ one finds
\begin{equation}
\begin{aligned}
P W^{\rm eff}_z(C) P
&=
\Bigl[
1
-16L\bigl(\tfrac{J}{\Delta_x}\bigr)^2
- 8L\bigl(\tfrac{J}{\Delta_y}\bigr)^2
\Bigr]
\\
&\qquad \times P W_z(C) P
+\mathcal O(J^3),
\end{aligned}
\end{equation}
while for $W_x(C)=\prod_{e\in C}\sigma^x_e$,
\begin{equation}
\begin{aligned}
P W^{\rm eff}_x(C) P
&=
\Bigl[
1
-16L\bigl(\tfrac{J}{\Delta_z}\bigr)^2
- 8L\bigl(\tfrac{J}{\Delta_y}\bigr)^2
\Bigr]
\\
&\qquad \times P W_x(C) P
+\mathcal O(J^3).
\end{aligned}
\end{equation}

\section{Numerical Results for the Ground State}
\label{ground_state}

\begin{figure}[t]
  \centering
  \includegraphics[width=0.95\linewidth]{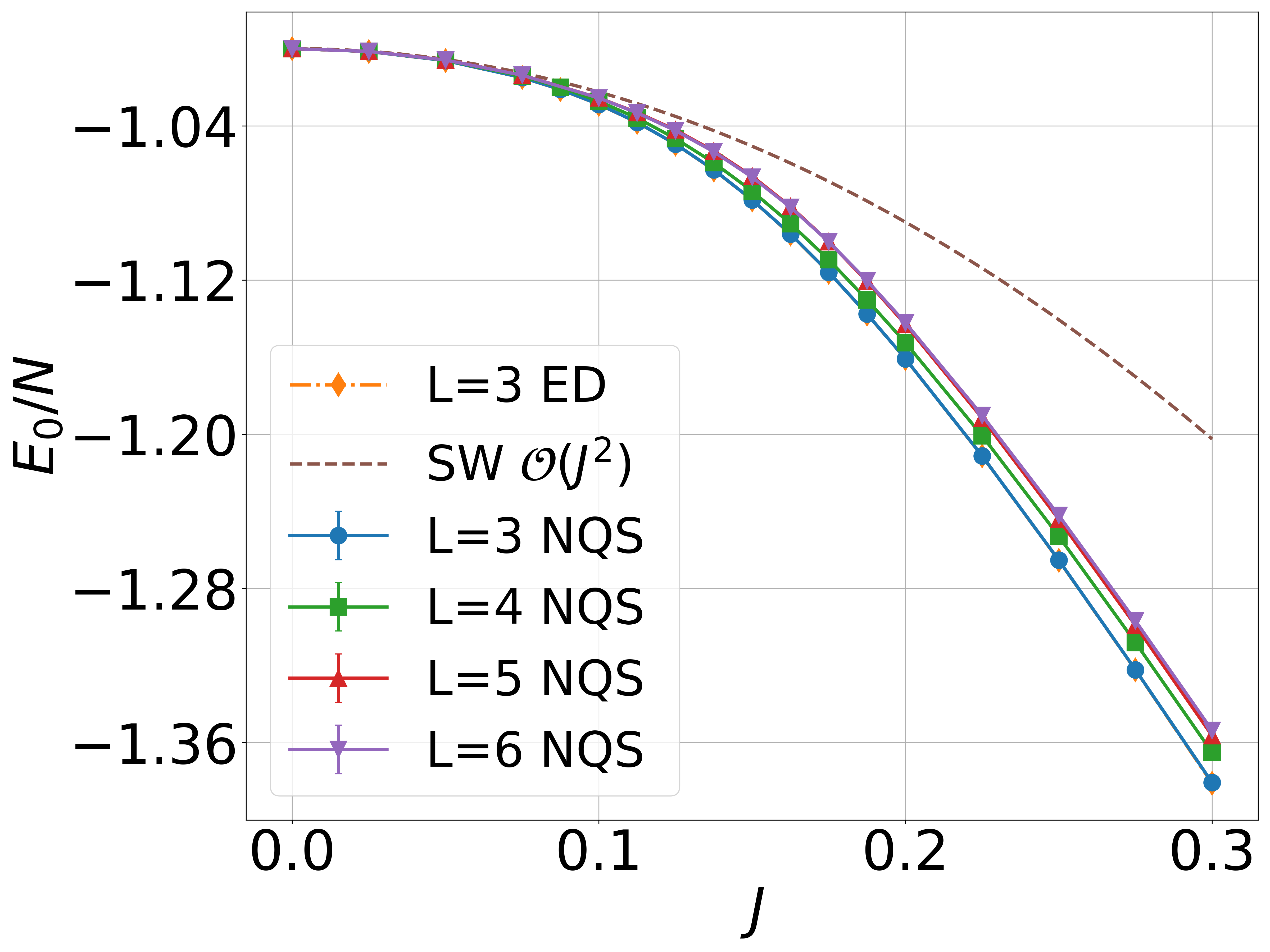}  \caption{\label{fig:energy}Ground state energy per spin $E_0/N$ (with $N=2L^2$) versus the AFM Heisenberg coupling $J$ for $L=3,4,5,6$.
Symbols show NQS results and error bars represent stochastic uncertainty; for $L=3$ we also plot the exact-diagonalization (ED) benchmark.
The dashed curve is the SW prediction truncated at $\mathcal{O}(J^2)$,
see Eq.~\eqref{energyperspin}.
Note that this approximation coincides with the result up to third order for $L \geq 4$; see Appendix~\ref{sec:methods_sw_third}.
In the perturbative regime $J\lesssim 0.1$, SW is in quantitative agreement with NQS/ED.
For $J\gtrsim 0.13$--$0.16$, systematic deviations from the $\mathcal{O}(J^2)$ curve appear and the $L$-dependence increases,
consistent with the breakdown of low-order perturbation theory and the approach to the critical region.  }

\end{figure}

\begin{figure}[t]
  \centering
 \includegraphics[width=0.95\linewidth]  
  {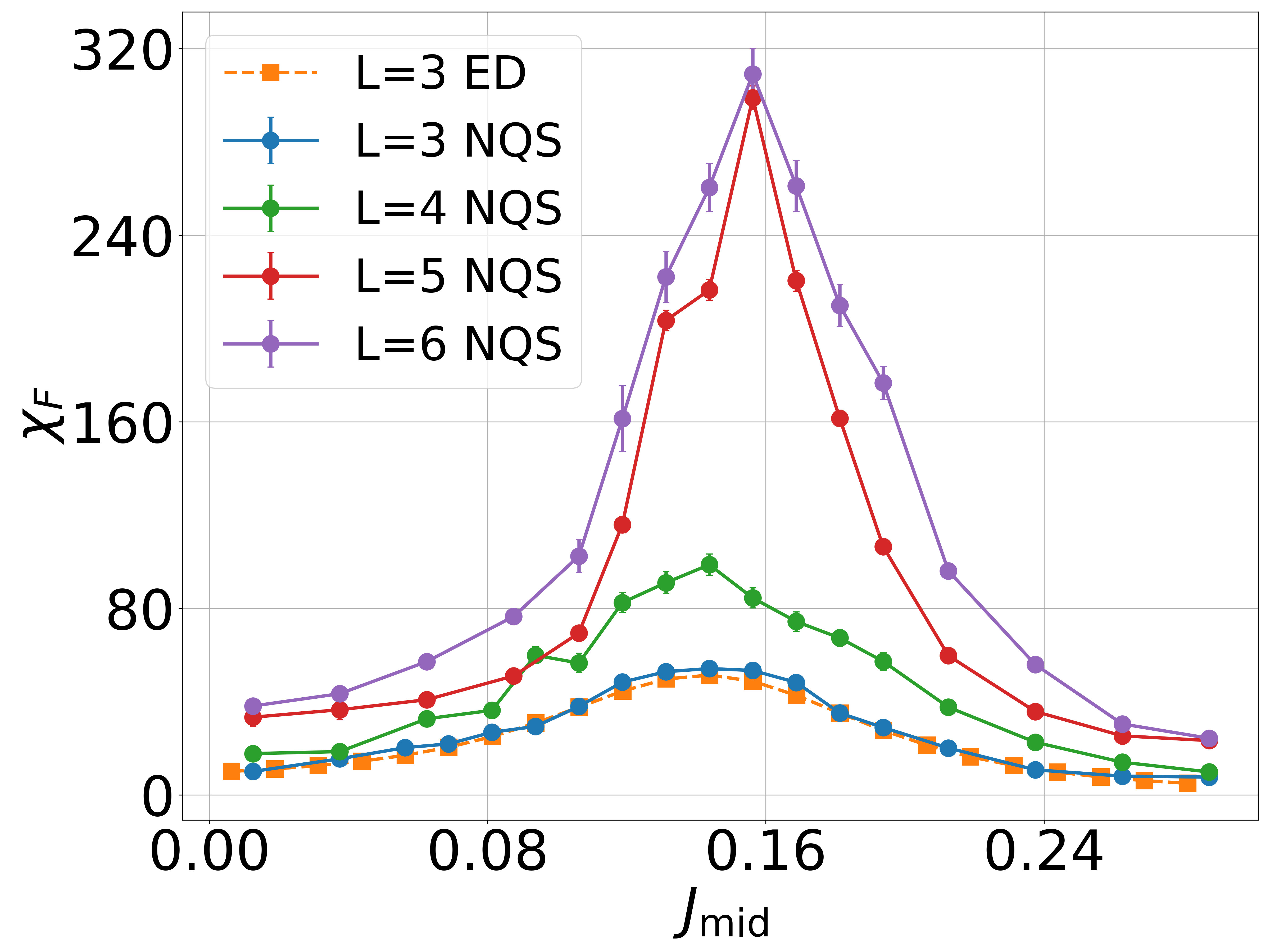}
  \caption{\label{fig:chiF_all}Fidelity susceptibility $\chi_F$ obtained from NQS for $L=3,4,5,6$ and from ED for $L=3$.
Prominent peaks indicate a phase crossover in this range of $J$.
The peak height increases with system size 
$L$, while the peak position exhibits a slight finite-size drift.
}

\end{figure}

\begin{figure}[t]
  \centering
  \begin{minipage}{0.48\columnwidth}
    \centering
    \begin{overpic}[width=\linewidth,height=0.28\textheight,keepaspectratio]{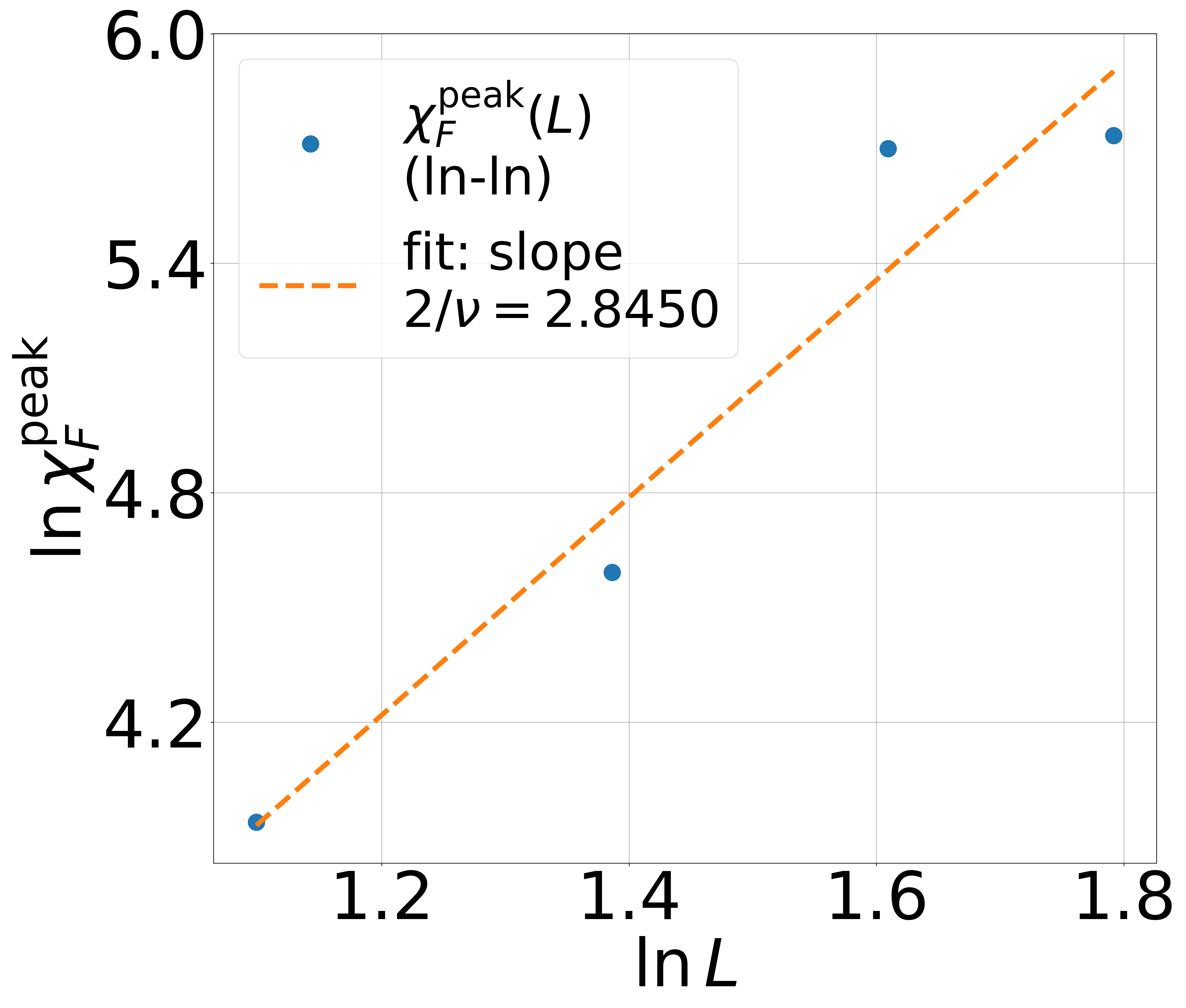}
      \put(2,87){(a)}
    \end{overpic}
  \end{minipage}\hfill
  \begin{minipage}{0.48\columnwidth}
    \centering
    \begin{overpic}[width=\linewidth,height=0.28\textheight,keepaspectratio]{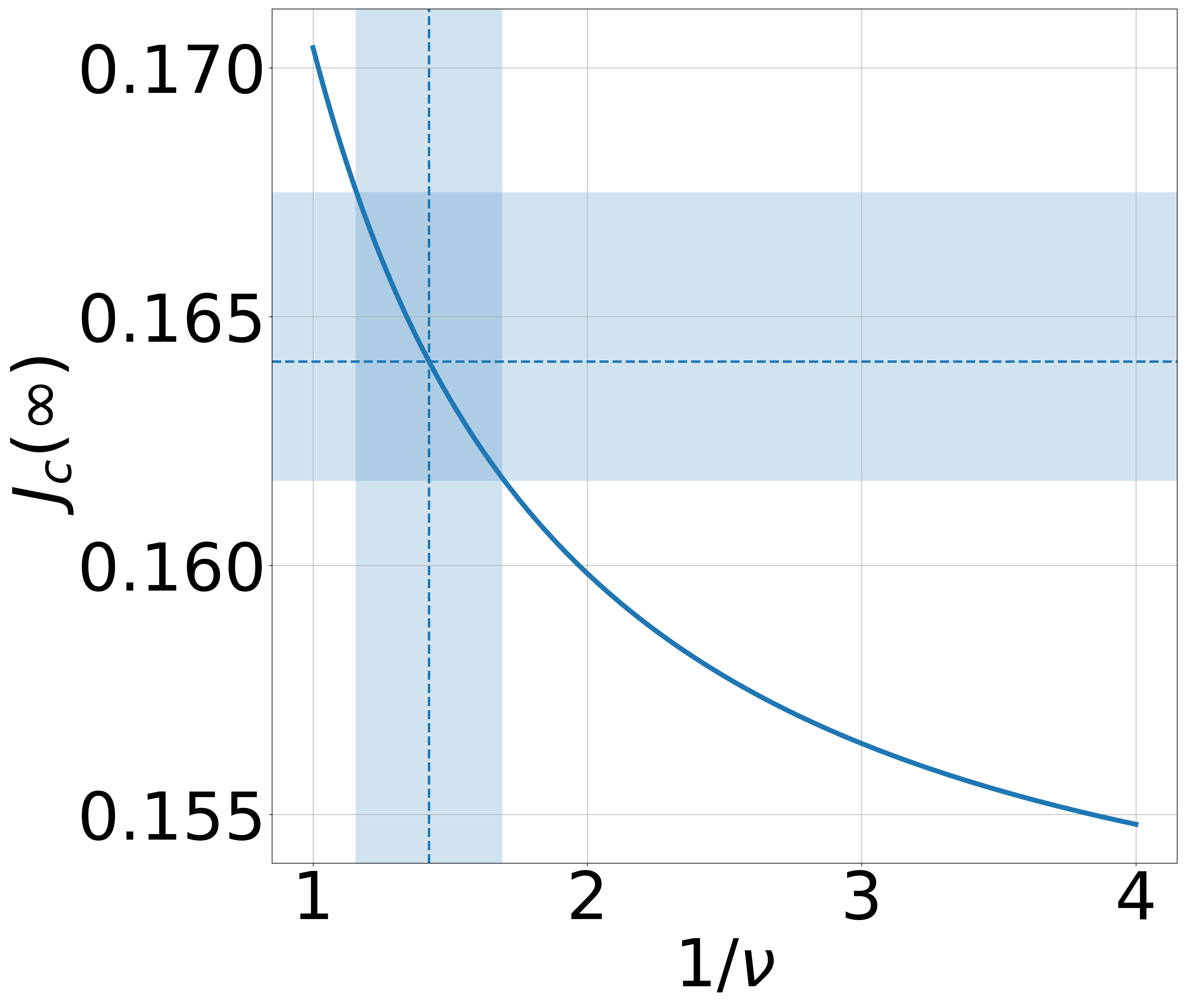}
      \put(2,87){(b)}
    \end{overpic}
  \end{minipage}

  \caption{\label{fig:fidsus_scaling}
  Finite-size scaling analysis of the fidelity-susceptibility peak.
(a) Log--log plot of the peak height, $\chi_F^{\mathrm{peak}}(L)$, versus $L$.
From a linear fit, we obtain the critical exponent from $2/\nu = 2.85 \pm 0.54$.
(b) Extrapolated critical coupling $J_c(\infty)$, obtained from the drift of the peak position by fitting
$J_{\mathrm{peak}}(L)=J_c(\infty)+a/L^{1/\nu}$,
shown as a function of the assumed shift exponent $1/\nu$. we obtained $a=-0.086$ in our fitting.
The vertical dashed lines indicate the value of $1/\nu$ from (a), and the horizontal dashed line and band marks the corresponding $J_c = 0.1641^{+0.0034}_{-0.0024}$.
The blue vertical band indicates the uncertainty range of $1/\nu$ from (a).}
  
\end{figure}

\begin{figure}[t]
  \centering

  \begin{minipage}{0.48\columnwidth}
    \centering
    \begin{overpic}[width=\linewidth,height=0.28\textheight,keepaspectratio]{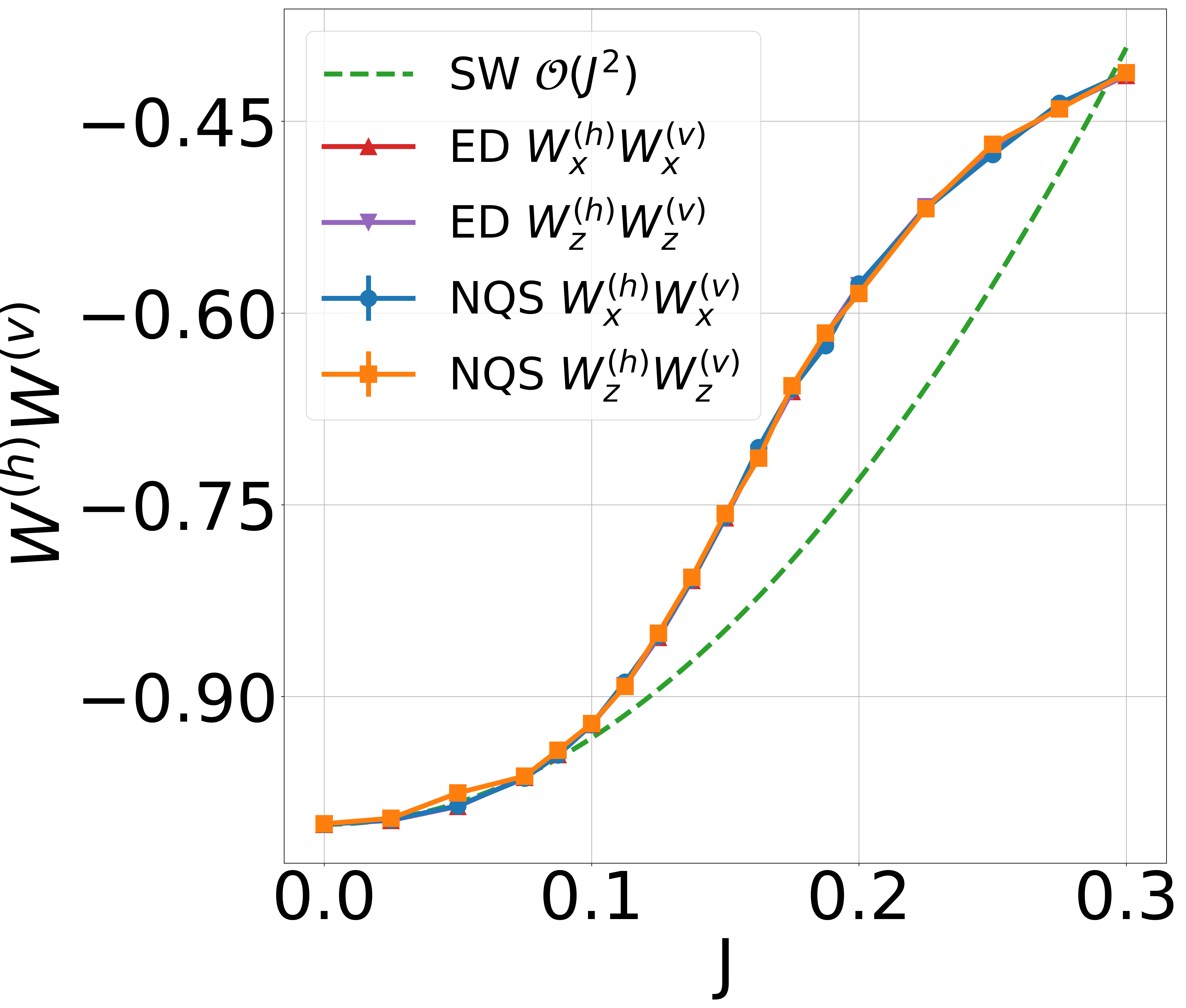}
      \put(0,85){(a)}
    \end{overpic}
  \end{minipage}\hfill
  \begin{minipage}{0.48\columnwidth}
    \centering
    \begin{overpic}[width=\linewidth,height=0.28\textheight,keepaspectratio]{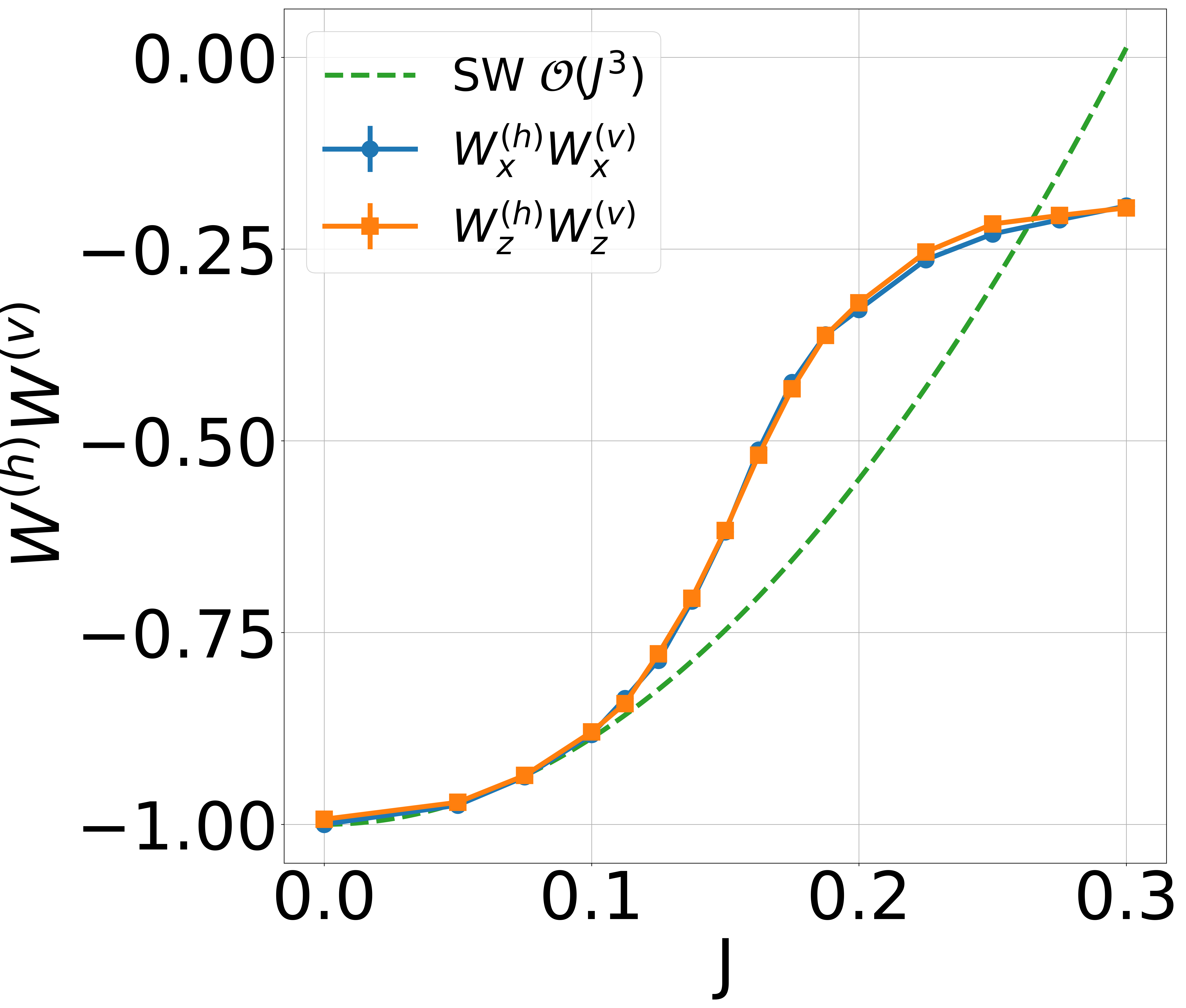}
      \put(0,85){(b)}
    \end{overpic}
  \end{minipage}

  \caption{\label{fig:oddwilson}
  Logical-qubit diagnostics from non-contractible Wilson loops for odd $L$.  
Shown are the parity-even double-winding correlators
$\langle W_x^{(h)}W_x^{(v)}\rangle_{\rm phys}$ and
$\langle W_z^{(h)}W_z^{(v)}\rangle_{\rm phys}$ for (a) $L=3$ and (b) $L=5$.
Their deviation from the ideal toric-code values directly quantifies correlated logical errors induced by the isotropic Heisenberg interaction.
Markers show NQS results for $L=3,5$, including the sampling/optimization uncertainty, as well as ED results for $L=3$.
Dashed curves correspond to the $\mathcal{O}(J^2)$ SW dressing of the logical correlators; good agreement at small $J$ validates the perturbative error model, while the rapid change and strong deviations near the transition region indicate the onset of the breakdown of the topological protection as the system crosses into the N\'eel-ordered phase.}
\end{figure}

\begin{figure*}[t]
\centering
\begin{overpic}[width=0.24\textwidth]{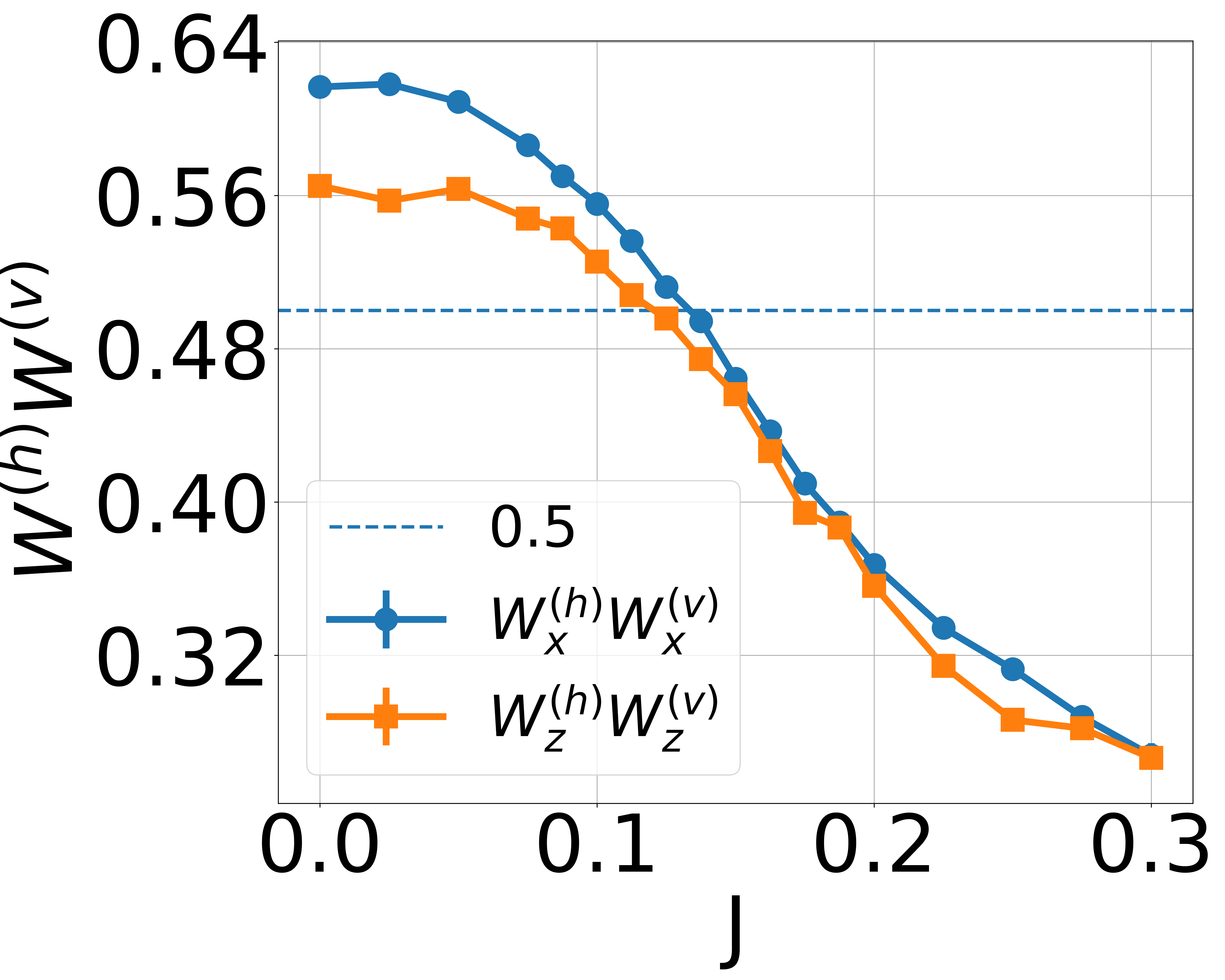}
 \put(0,82){\small(a)}
\end{overpic}
\begin{overpic}[width=0.24\textwidth]{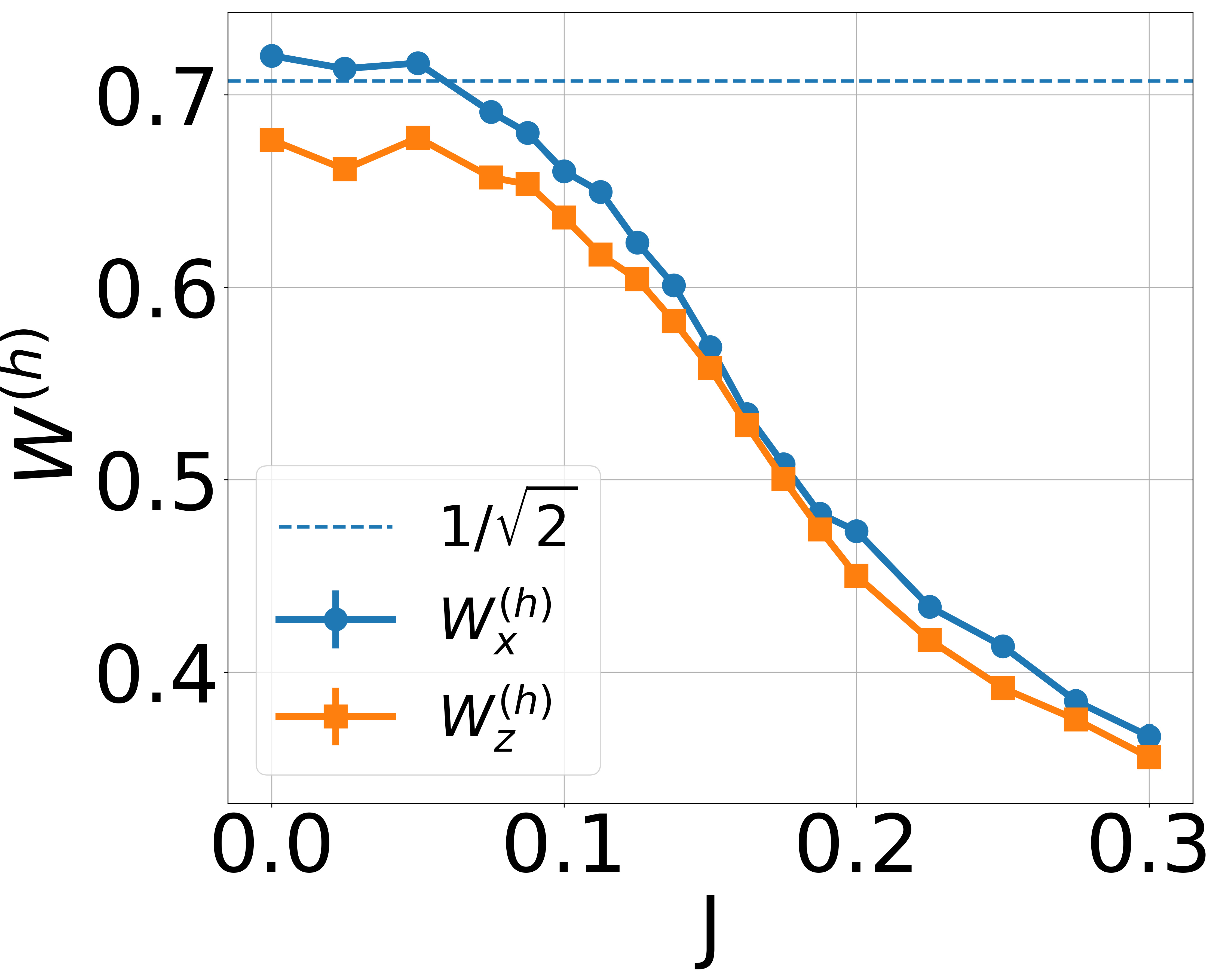}
 \put(0,82){\small(b)}
\end{overpic}
\begin{overpic}[width=0.24\textwidth]{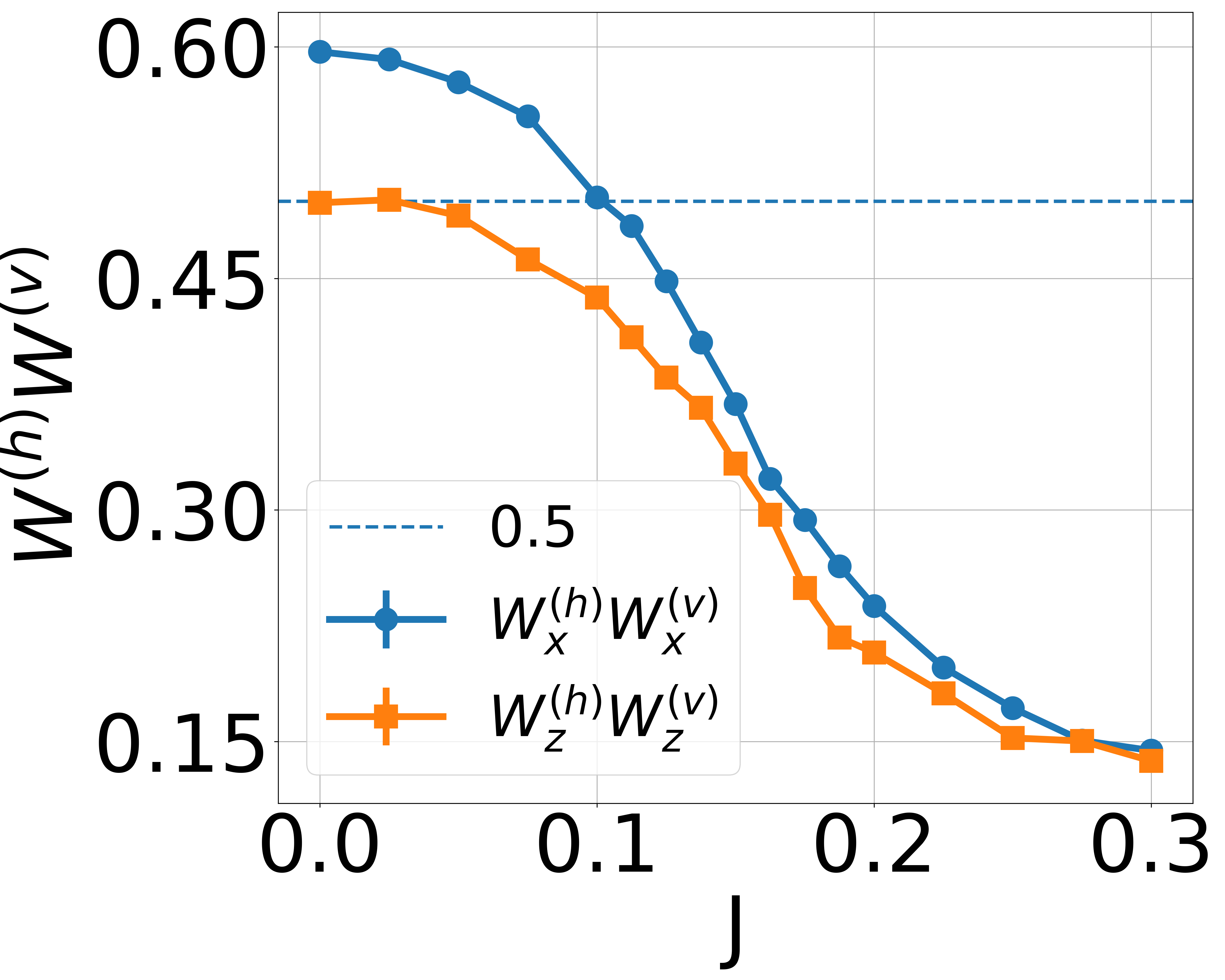}
 \put(0,82){\small(c)}
\end{overpic}
\begin{overpic}[width=0.24\textwidth]{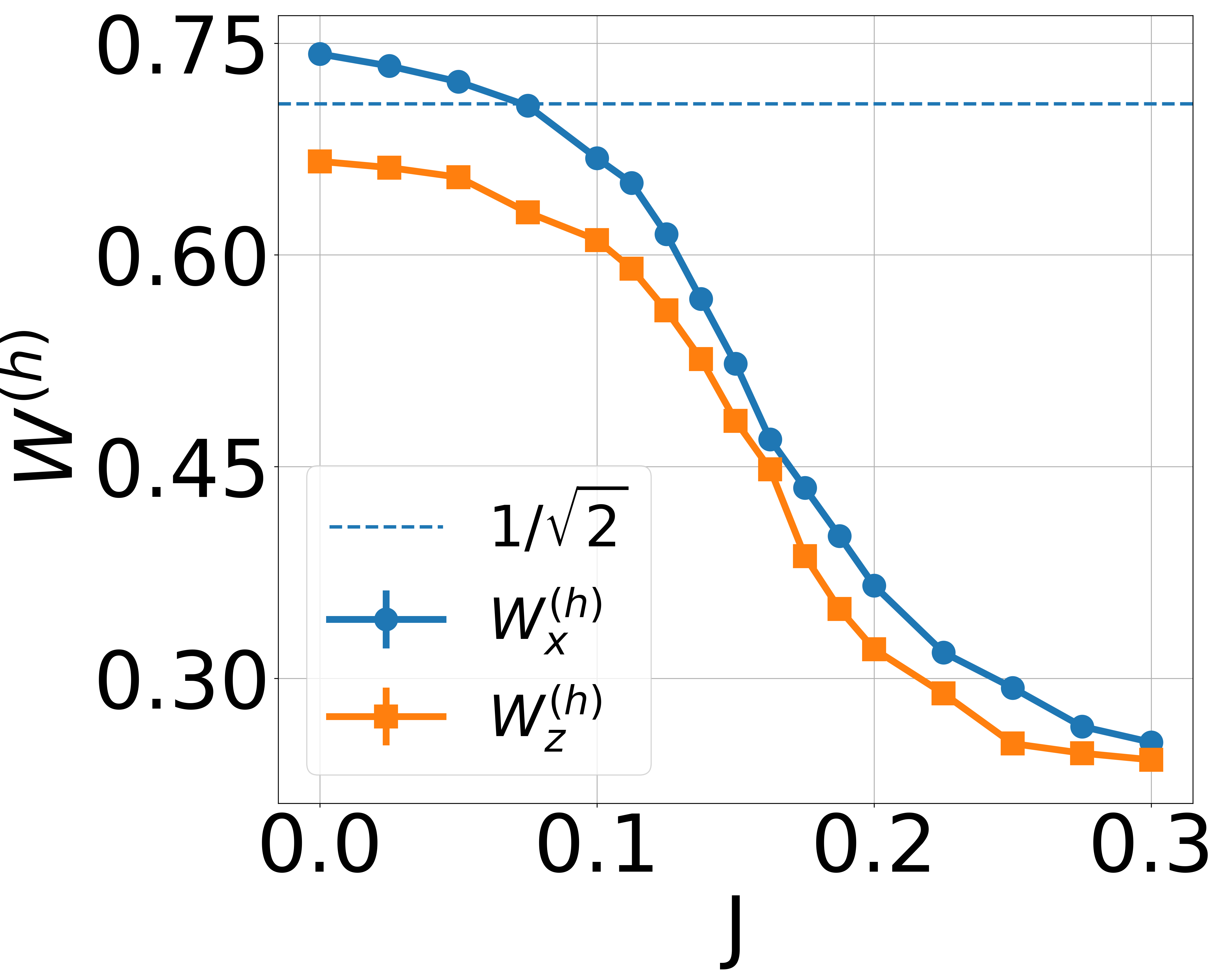}
 \put(0,82){\small(d)}
\end{overpic}
\caption{\label{fig:evenwilson}
Wilson-loop observables versus Heisenberg coupling $J$ for $L=4$ (a,b) and $L=6$ (c,d).
(a,c) Double-winding products $\langle W_x^{(h)}W_x^{(v)}\rangle_{\rm phys}$ and
$\langle W_z^{(h)}W_z^{(v)}\rangle_{\rm phys}$; (b,d) single-winding loops
$\langle W_x^{(h)}\rangle_{\rm phys}$ and $\langle W_z^{(h)}\rangle_{\rm phys}$.
A small-$J$ plateau near the perturbative reference values (dashed: $0.5$ and $1/\sqrt2$)
marks the topological regime, while the pronounced 
drop with $J$
signals the approach to a topological phase transition and increased logical error propensity.
Because of the near-degeneracy for even L discussed in Appendix~\ref{app:polarized_results}, they show slight deviation from the duality.}

\end{figure*}

\begin{figure}[t]
  \centering

\begin{minipage}{0.48\columnwidth}
  \centering
  \begin{overpic}[width=\linewidth,height=0.28\textheight,keepaspectratio]{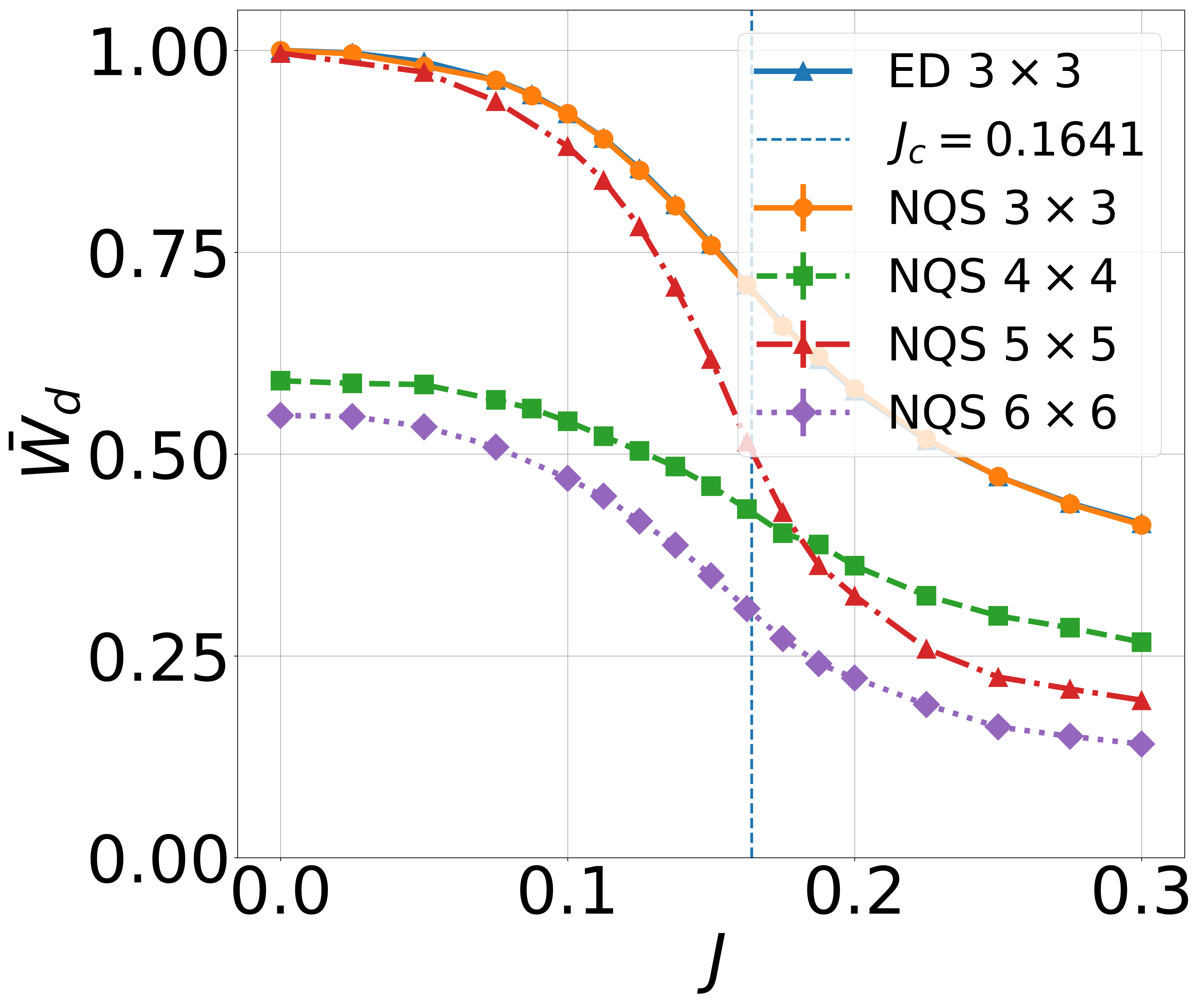}
    \put(-2,82){(a)}
  \end{overpic}
\end{minipage}\hfill
\begin{minipage}{0.48\columnwidth}
  \centering
  \begin{overpic}[width=\linewidth,height=0.28\textheight,keepaspectratio]{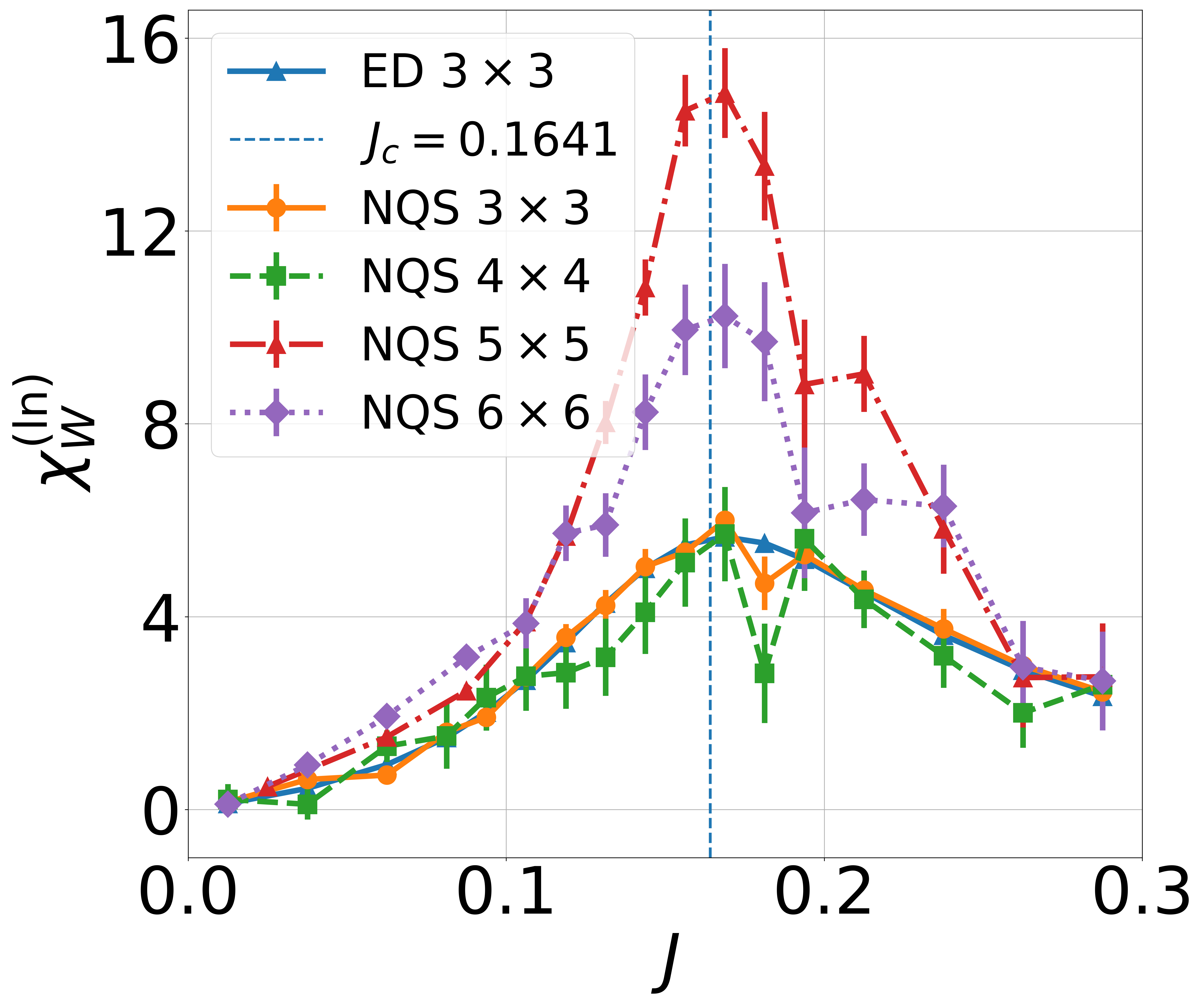}
    \put(-2,82){(b)}
  \end{overpic}
\end{minipage}
  
  \caption{\label{fig:Wd_bar}
  Composite non-contractible Wilson-loop diagnostic $\bar{W}_d$, defined in Eq.~\eqref{eq:Wd_bar}, for $L = 3, 4, 5, 6$ (ED for $3 \times 3$; NQS for all sizes). (a)~$\bar{W}_d(J, L)$ remains close to its baseline in the topological
phase and undergoes a pronounced suppression toward zero as $J$
increases, with the decay becoming steeper for larger system sizes within the same even- or odd-$L$ sector.
The even/odd-dependent baseline differs between odd $L$ ($\bar{W}_d \simeq 1$) and even $L$ ($\bar{W}_d \simeq 0.5$--$0.6$), consistent with the finite-size structure discussed in Sec.~\ref{sec:methods_sw_logical_summary}. The vertical dashed line marks the estimated critical coupling $J_c \approx 0.164$. (b)~The logarithmic susceptibility $\chi_W^{(\ln)}$ exhibits a peak that sharpens and grows with increasing $L$, signaling the breakdown of topological order. Because of the even/odd dependence on $L$, the peak heights follow separate even and odd branches---$\chi_W^{(\mathrm{ln},\mathrm{peak})}(L{=}3) < \chi_W^{(\mathrm{ln},\mathrm{peak})}(L{=}5)$ and $\chi_W^{(\mathrm{ln},\mathrm{peak})}(L{=}4) < \chi_W^{(\mathrm{ln},\mathrm{peak})}(L{=}6)$---while even and odd $L$ may alternate in magnitude at comparable system sizes. The peak positions are located near $J_c \simeq 0.164$, consistent with the fidelity susceptibility analysis.}
\end{figure}

\begin{figure}[t]
  \centering
  \includegraphics[width=0.32\textwidth]{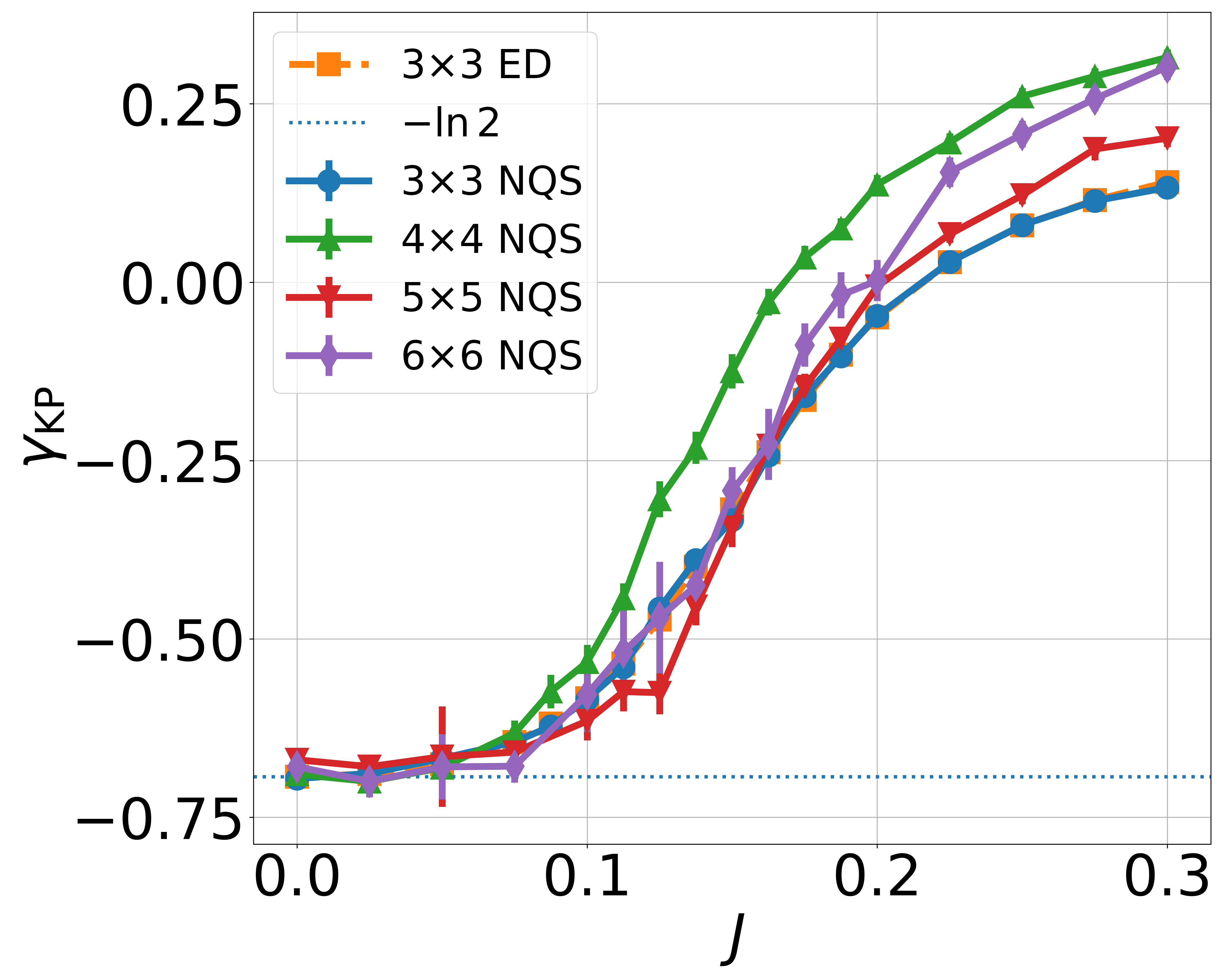}
  \caption{\label{fig:tee}Kitaev--Preskill topological entanglement indicator
\(\gamma_{\rm KP}\), extracted from the second R\'enyi entropies for the ground state of the toric code with AFM Heisenberg coupling \(J\).
Results are shown for system sizes $L=3-6$, obtained via NQS (We also include ED results for benchmarking).
We use the convention
\(\gamma_{\rm KP}\equiv S(A)+S(B)+S(C)-S(AB)-S(BC)-S(CA)+S(ABC)\). Thus, an ideal
\(\mathbb{Z}_2\) topological phase yields \(\gamma_{\rm KP}=-\ln 2\) (dotted line).
As \(J\) increases, \(\gamma_{\rm KP}\) rises toward zero, signaling the loss of
topological order. For larger \(J\), \(\gamma_{\rm KP}\) becomes positive
because of the contribution from cat state mixing and finite size effect; this induces a
Shannon-like mixing contribution to the R\'enyi entropies that is not removed by the KP subtraction. }
  
\end{figure}

In this section, we show our numerical results. All expectation values in this section are calculated in the physical basis, see Eq.~\eqref{eq:OM_definition}.

We begin with a benchmark of our methodologies in the perturbative regime and
present results that support the accuracy of our NQS ansatz and the validity of the SW expansion. 
Figure~\ref{fig:energy} compares the ground-state energies per spin for $L=3,4,5,6$ and the SW effective Hamiltonian up to second order. We furthermore include the ED result for $3\times3$. In the small-$J$ regime, the SW results agree quantitatively with the NQS energies, confirming the controlled nature of the expansion and the accuracy of the NQS. As $J$ increases, deviations gradually develop, and the variational NQS energy becomes systematically lower than the SW estimate, signaling the breakdown of the perturbative description.

Next, we show the fidelity susceptibility $\chi_F$ for $L=3,4,5,6$ in Fig.~\ref{fig:chiF_all}, which exhibits a pronounced peak that sharpens and drifts slightly 
toward larger $J$ as $L$ increases. 
By performing a finite-size scaling analysis of the peak heights and positions, see Eq.~\eqref{eq:chiF_peak_height} and Eq.~\eqref{eq:chiF_peak_shift}, we extrapolate the transition point in the 
thermodynamic limit, as shown in Fig.~\ref{fig:fidsus_scaling}. This analysis yields $J_c \simeq 0.164$ within the grid error.

We next present the numerical data for the non-contractible Wilson-loop expectation values.
For odd $L$, symmetry constraints enforce
\(\langle W_x\rangle_{\mathrm{phys}}=\langle W_z\rangle_{\mathrm{phys}}=0\),
as discussed in Sec.~\ref{Global Parity Symmetries and Logical Structure}, which implies strong mixing between topological sectors. 
On the other hand, the parity-even double-winding operators
\begin{equation}
W_z^{(h)}W_z^{(v)},\qquad
W_x^{(h)}W_x^{(v)} ,
\end{equation}
commute with both $P_x$ and $P_z$ even for odd $L$ and are thus not forbidden by symmetry.
The results for odd $L$ are presented in Fig.~\ref{fig:oddwilson}. 
The parity-even double-winding 
operators take negative expectation values, in excellent agreement with 
our SWT predictions. 
For odd system sizes and $J=0$, the global parity operators $P_x$ and $P_z$ coincide with the double-winding operators $W_x^{(h)}W_x^{(v)}$ and $W_z^{(h)}W_z^{(v)}$, respectively.  This follows because each parity operator can be decomposed into the product of double-winding operators and a set of contractible loops; the latter are products of local stabilizers $A_v$ or $B_p$, which evaluate to $+1$. 
To minimize the total energy for $J>0$, the ground state naturally resides in the 
$P_x = P_z = -1$ sector, which is in agreement with Eq.~\eqref{x-parity} and Eq.~\eqref{z-parity}. 
In this regime, the magnitudes of these loop expectation values increase gradually and remain well captured by the perturbative SW results. As the system approaches the transition region, however, we observe a markedly sharper increase together with substantial deviations from the SWT curves, signaling the breakdown of the low-order effective description.
Beyond the transition, the data suggest a gradual evolution toward smaller values, consistent with the loss of topological protection and the accompanying decay of non-local correlations.

Next, we examine the non-contractible Wilson loop data for even lattice 
sizes ($L=4, 6$) shown in Fig.~\ref{fig:evenwilson}. In the weak-coupling 
regime, the single-loop expectation values in the $x$ and $z$ channels 
remain close to the duality-symmetric value of 
$1/\sqrt{2}$. This confirms that the optimized finite-size ground state 
respects the Hamiltonian duality. Furthermore, the loop products 
systematically exceed the independent-qubit baseline of $1/2$, 
suggesting the presence of correlations between the logical sectors that 
are not captured by the simple mixing as predicted in Sec.~\ref{app:sw_higher_logical}.
We note that for very small $J$, the topological energy splitting
$\Delta E_{\rm topo}\sim J^{L}/\Delta^{L-1}$ becomes so small that the
variational optimization may converge to metastable, logically polarized
states rather than the duality-symmetric ground state. In such states, the
ansatz effectively polarizes into one Wilson-loop type, e.g. $\langle W_x\rangle\gg
\langle W_z\rangle$ or vice versa, thereby appearing to break the expected
duality at finite size, see Appendix~\ref{app:polarized_results}.

We can now use the non-contractible Wilson-loop data to build an indicator for the topological phase transition and to complement the fidelity-based estimate. We 
use the composite non-contractible Wilson-loop indicator $\bar{W}_d(J,L)$ defined in Eq.~(\ref{eq:Wd_bar}).
As shown in Fig.~\ref{fig:Wd_bar}(a), the suppression of $\bar{W}_d$ with increasing $J$ becomes progressively steeper for larger system sizes, consistent with a sharpening phase transition. To locate the coupling at which the topological order decays most rapidly, we monitor the logarithmic susceptibility $\chi_W^{(\ln)}= -d\ln\bar{W}_d/dJ$ [Eq.~(\eqref{eq:Wd_bar})], which measures the relative rate of change and is thereby insensitive to the even-odd-dependent baseline. As shown in Fig.~\ref{fig:Wd_bar}(b), $\chi_W^{(\ln)}$ develops a peak that sharpens and grows with $L$, with peak positions being located near $J_c \simeq 0.164$, providing a nonlocal estimate of the critical coupling that is fully consistent with the fidelity susceptibility analysis. We note that at comparable system sizes, even-$L$ systems exhibit systematically larger peak heights than odd-$L$ systems. This is attributed to the different baseline structures between even and odd $L$.

Having established the transition point and the role of 
even/odd L dependence of the non-contractible Wilson loop operators,
we next examine the TEE, $\gamma_{\mathrm{KP}}$ (Fig.~\ref{fig:tee}).
We find that the regime where the SW energy
predictions begin to deviate from our numerical results coincides
precisely with the rapid growth of the fidelity susceptibility and the
departure of $\gamma_{\mathrm{KP}}$ from its unperturbed topological value
of $-\ln 2$. The peak in $\chi_F$ and the simultaneous crossover
in $\gamma_{\mathrm{KP}}$ (Figs.~\ref{fig:chiF_all} and \ref{fig:tee}) thus
provide a consistent set of signatures for the breakdown of the
$\mathbb{Z}_2$ topological phase.

At larger $J$, the TEE remains distinctively above zero. We interpret this as a finite-size ``cat-state'' 
effect within the symmetry-preserving ordered regime, which makes a Shannon-like contribution to the Rényi entropies 
that does not cancel in the Kitaev--Preskill construction and a non-universal contribution depending on the correlation length, as discussed in Sec.~\ref{Detecting of quantum phase transition}.
\begin{figure*}[t]
\centering
\begin{overpic}[width=0.24\textwidth]{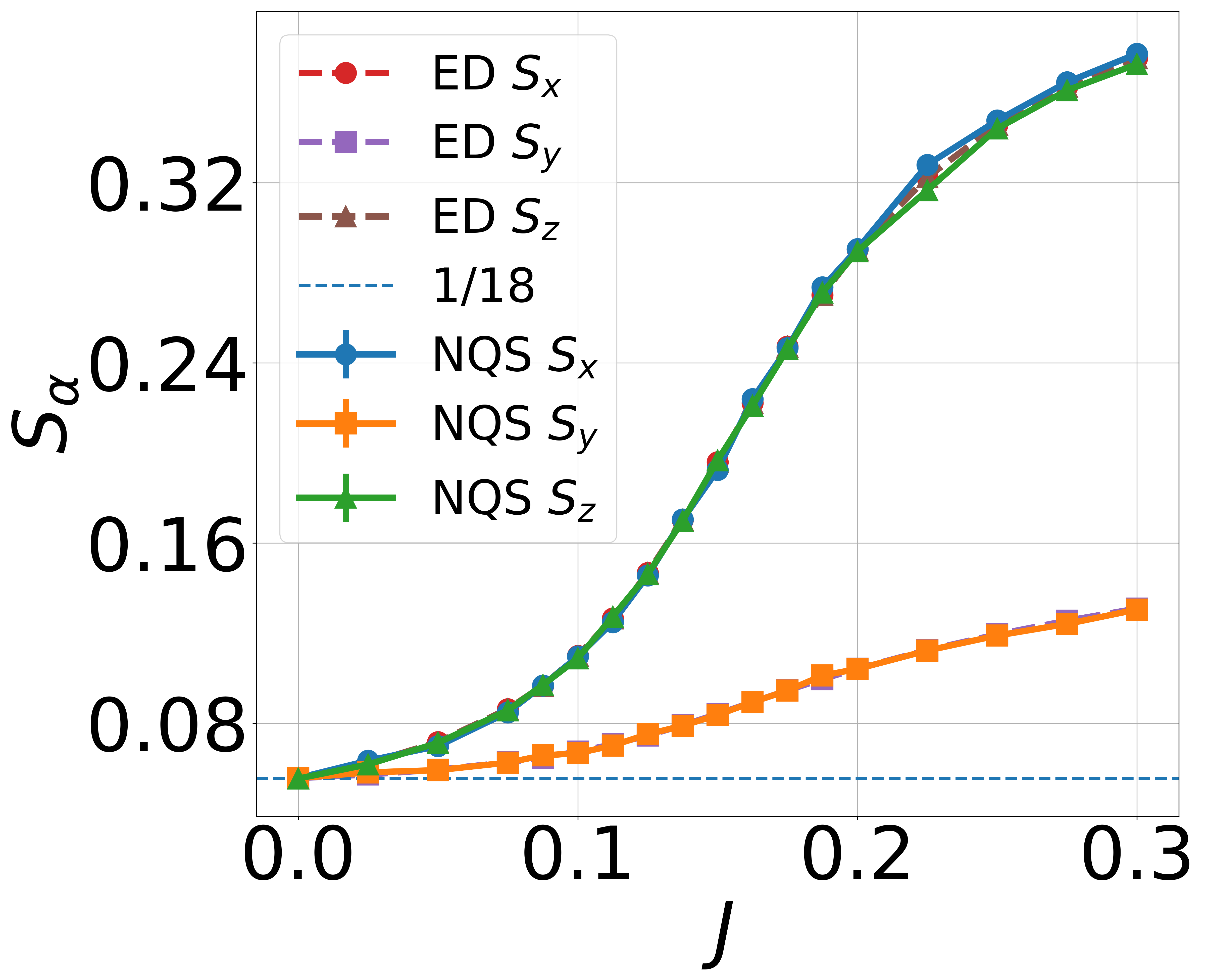}
 \put(2,82){\small(a)}
\end{overpic}
\begin{overpic}[width=0.24\textwidth]{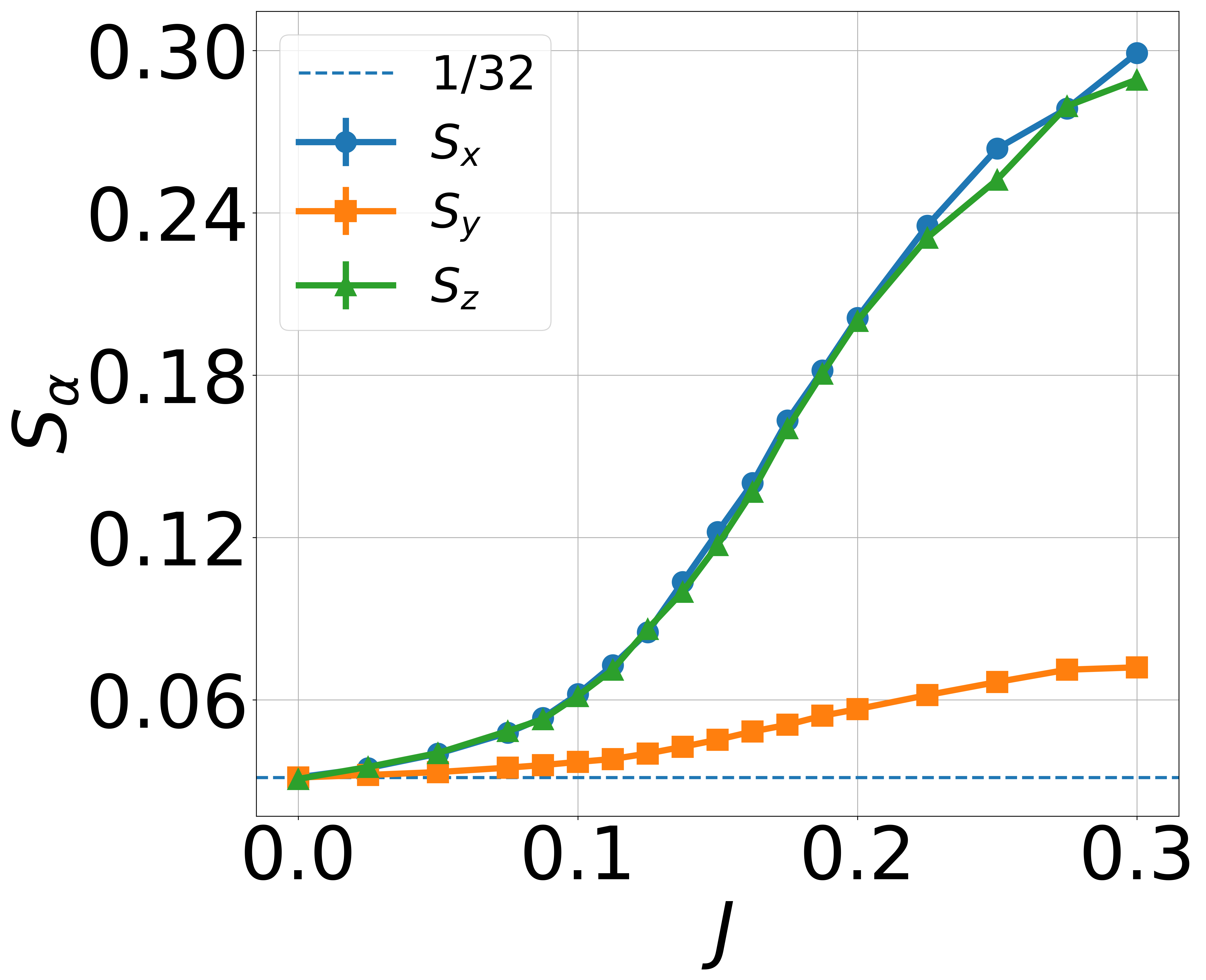}
 \put(2,82){\small(b)}
\end{overpic}
\begin{overpic}[width=0.24\textwidth]{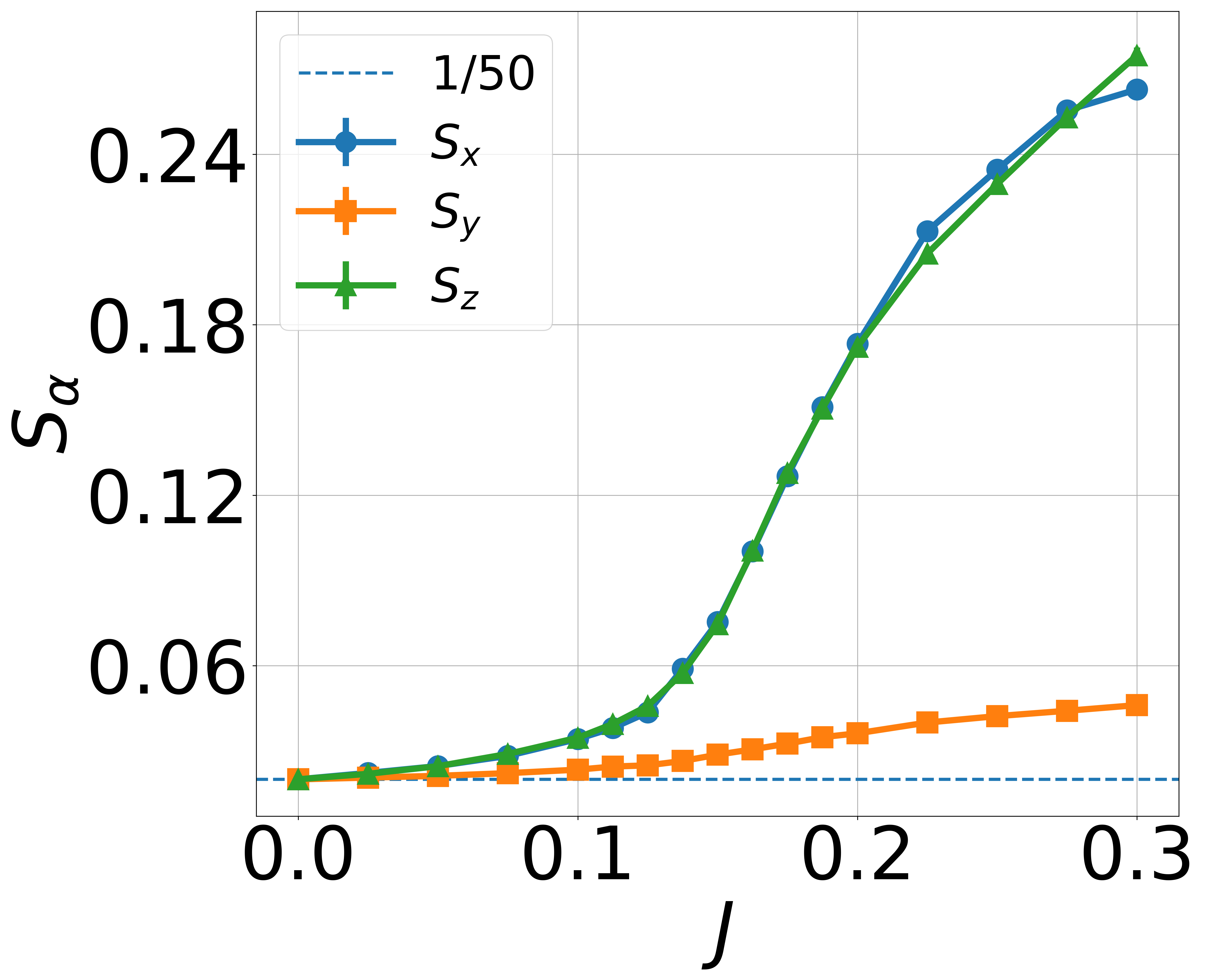}
 \put(2,82){\small(c)}
\end{overpic}
\begin{overpic}[width=0.24\textwidth]{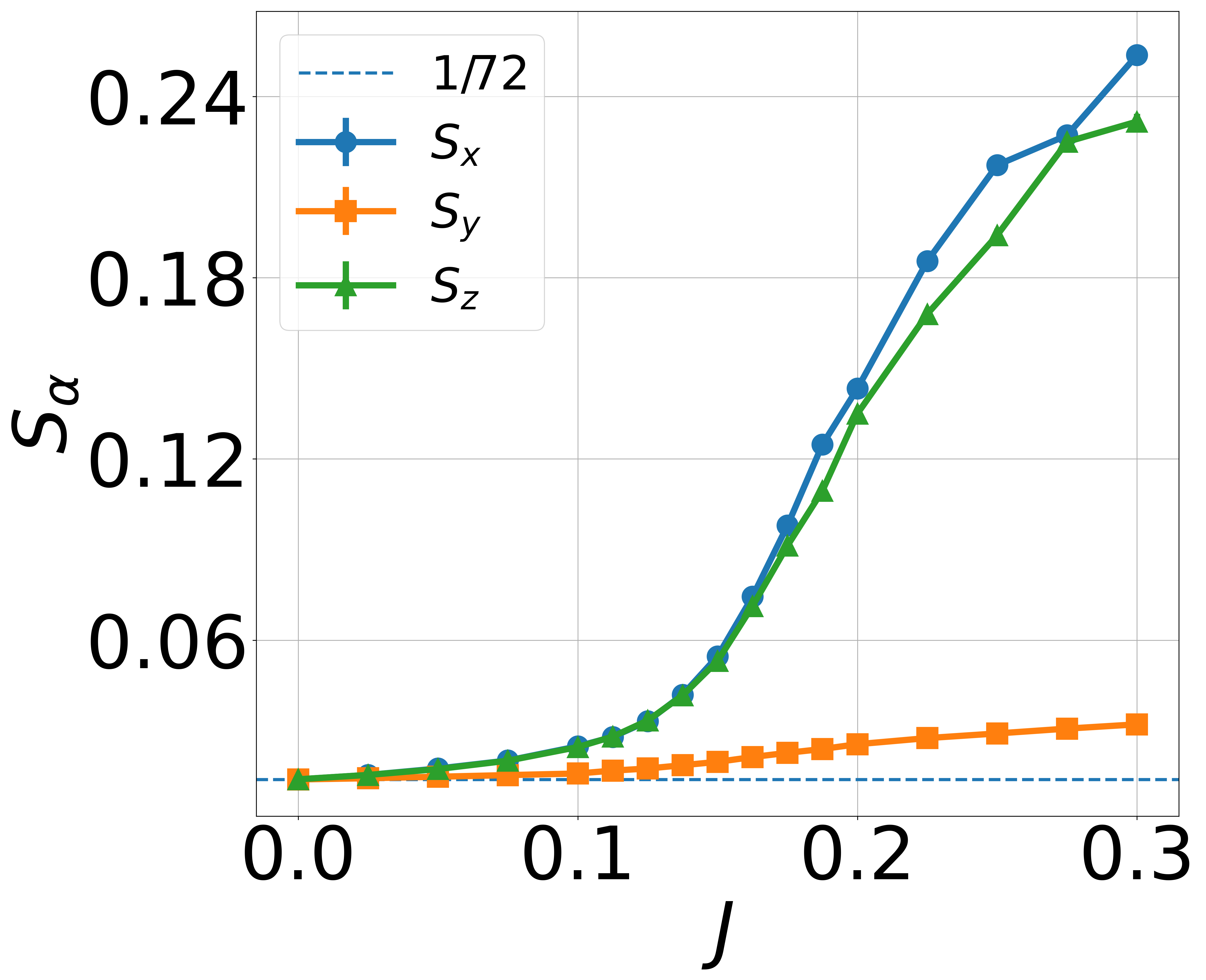}
 \put(2,82){\small(d)}
\end{overpic}
\caption{\label{fig:stag}Squared staggered magnetizations $S_\alpha = \langle(m^{\rm stag}_\alpha)\rangle^2$ ($\alpha=x,y,z$) versus $J$
for (a) $L{=}3$, (b) $L{=}4$, (c) $L{=}5$, and (d) $L{=}6$.
Horizontal dashed lines indicate the finite-size baseline $1/N$ of $N=2L^2$ spins. Because of the duality, $S_z \simeq S_x$ is expected. However, the
near-degeneracy for even $L$ discussed in Appendix~\ref{app:polarized_results}
can cause our variational ansatz to deviate from duality, as seen in panel~(d).}

\end{figure*}

Beyond the toric-code regime, the system enters a N\'eel-ordered phase.
To further analyze this phase, we monitor the staggered magnetizations defined in
Eq.~(\ref{eq:mstag}).
We note that  the ground state is necessarily an eigenstate of the
global parity operators, which forces the linear staggered order parameters
to vanish even deep in the N\'eel regime.
Indeed, we find
\begin{equation}
\begin{aligned}
P_x\, m_z^{\mathrm{stag}}\, P_x &= -\,m_z^{\mathrm{stag}}
\;\Rightarrow\;
\langle m_z^{\mathrm{stag}}\rangle_{\mathrm{phys}}=0,\\
P_z\, m_x^{\mathrm{stag}}\, P_z &= -\,m_x^{\mathrm{stag}}
\;\Rightarrow\;
\langle m_x^{\mathrm{stag}}\rangle_{\mathrm{phys}}=0.
\end{aligned}
\end{equation}

\begin{figure*}[t]
\centering
\begin{overpic}[width=0.48\textwidth]{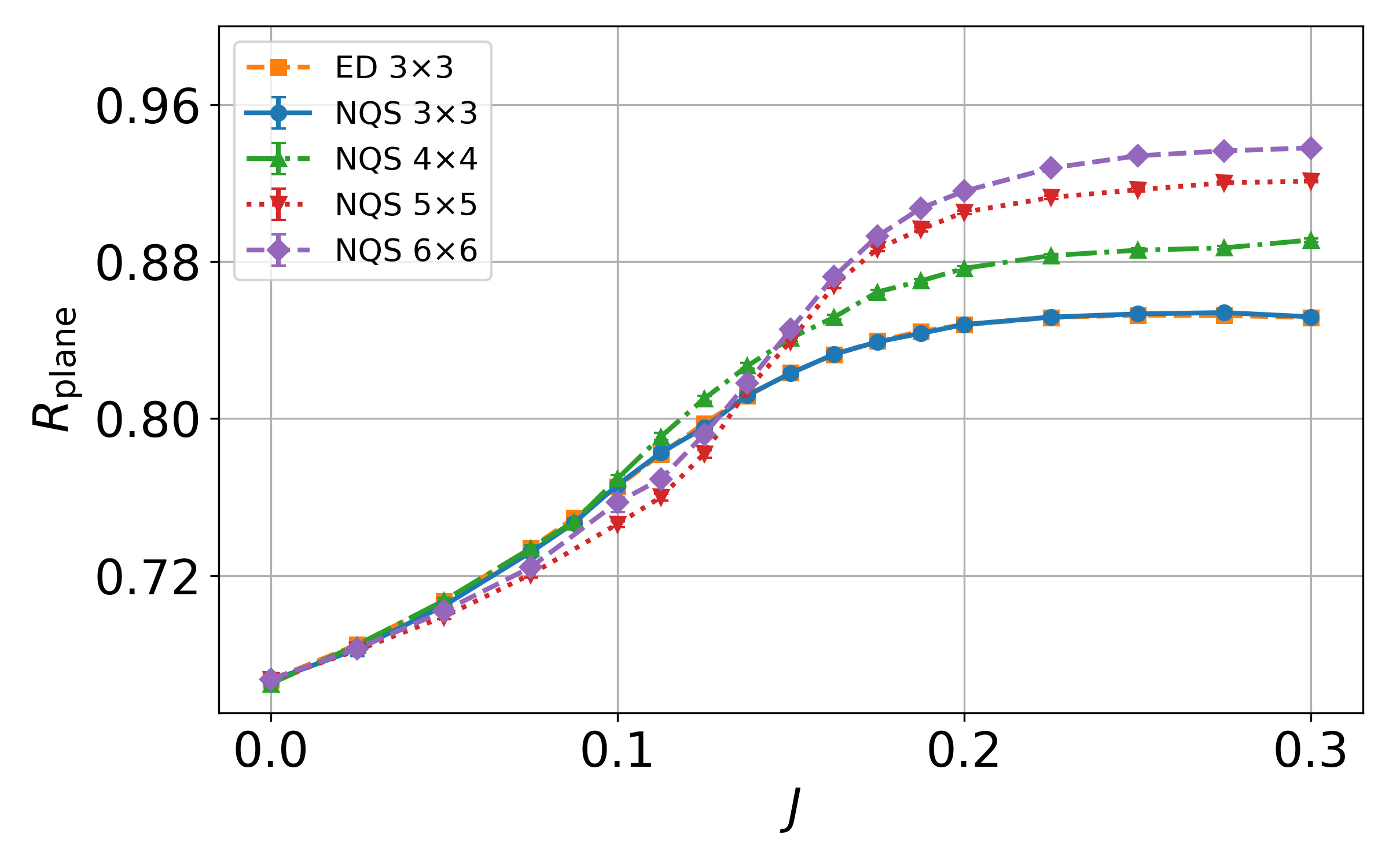}
  \put(10,62){(a)}
\end{overpic}
\begin{overpic}[width=0.48\textwidth]{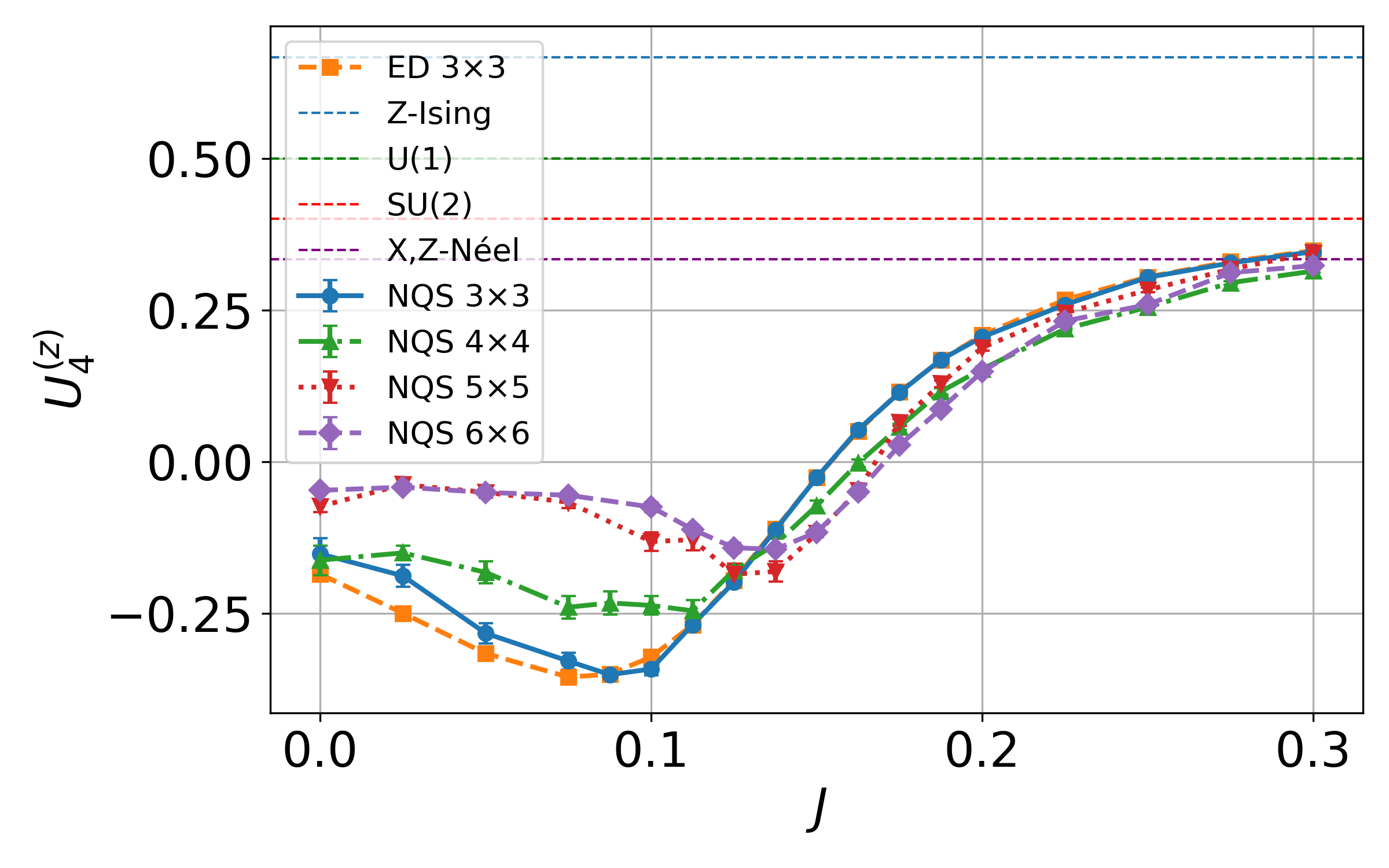}
  \put(10,62){(b)}
\end{overpic}
\caption{\label{fig:Rplane_Uz}
(a) Plane ratio $R_{\rm plane}=(S_x+S_z)/(S_x+S_y+S_z)$, with
$S_\alpha=\langle (m^{\rm stag}_\alpha)^2\rangle_{\mathrm{phys}}$.
$R_{\rm plane}=2/3$ would indicate comparable weight in all spin components, while
$R_{\rm plane}\to 1$ signals suppression of $S_y$ and easy-plane ($x$--$z$) order.
(b) We use the Binder cumulant
$U^{(z)}_4=1-\langle (m^{\rm stag}_z)^4\rangle_{\mathrm{phys}}/\bigl(3\langle (m^{\rm stag}_z)^2\rangle^2_{\mathrm{phys}}\bigr)$ as a dimensionless measure of the shape of the order-parameter distribution; it cleanly distinguishes discrete vs continuous symmetry realization and provides a finite-size–robust diagnostic.
Dashed lines mark the reference values for an $XZ$ rotor ($1/2$), an $SU(2)$ rotor ($0.4$),
and an equal-weight four-sector $X/Z$ cat distribution ($1/3$), see Appendix~\ref{app:binder} for derivation of these values.
ED results are shown for $3\times 3$, and NQS results for $L\times L$ with $L=3,4,5,6$.
}
\end{figure*}

\noindent To detect magnetic ordering, we therefore focus on even moments of the staggered
magnetization, which are proportional to the static spin structure factor at
$\mathbf Q=(\pi,\pi)$:
\begin{align}
&S_\alpha \equiv \Big\langle \big(m_{\alpha}^{\rm stag}\big)^2 \Big\rangle_{\rm phys},
\qquad \alpha\in\{x,y,z\},
\label{eq:Salpha_def} \\
&S_\alpha(\pi,\pi) = N\, S_\alpha .
\label{eq:Spi_pi_relation}
\end{align}
At finite size, the structure factor $S_\alpha$ contains a trivial self-correlation contribution (i.e., the product of the spin operator at the same site, which appears in Eq.~\eqref{eq:Salpha_def}) of 
order $1/N$, which vanishes in the thermodynamic limit.

As shown in Fig.~\ref{fig:stag}, the spin structure factors grow only weakly for small $J$ and increase rapidly once the topological phase melts.
We find that $S_y$ remains strongly suppressed, while $S_x$ and $S_z$ grow in a
duality-symmetric manner.
Correspondingly, the ratio $(S_x+S_z)/(S_x+S_y+S_z)$ approaches unity with
increasing $L$, indicating the development of $X/Z$ N\'eel order.

In addition, we employ Binder cumulants to further characterize the ordering
pattern and to distinguish between different symmetry realizations.
The motivation is that the exact global symmetries prohibit spontaneous symmetry breaking for finite system sizes with periodic boundary
conditions:
the ground state is a symmetric superposition of symmetry-related N\'eel
sectors, so that $\langle m^{\rm stag}\rangle_{\rm phys}=0$ even in an ordered
phase.
While the presence of order can be detected from the second moment, this does not uniquely determine how the symmetry is realized.
Binder cumulants provide a compact, scale-free probe of the shape of the
order-parameter distribution via the ratio of fourth and second moments, and
thus sharply discriminate between Ising-like single-sector order, continuous
rotor-like fluctuations, and multi-sector (cat) structures required by duality.
For a single spin component, we parametrize the staggered magnetization as
\[
m_z^{\rm stag} = m_0 \cos\theta ,
\]
and define the corresponding Binder cumulant
\begin{equation}
U_4^{(z)} = 1 -
\frac{\big\langle (m_z^{\rm stag})^4 \big\rangle_{\mathrm{phys}}}
{3\,\big\langle (m_z^{\rm stag})^2 \big\rangle_{\mathrm{phys}}^{\,2}} .
\end{equation}

\noindent To interpret the numerical Binder cumulants, we evaluate $U_4$ for several
representative symmetry scenarios using classical spin configurations, as
detailed in Appendix~\ref{app:binder}.
This classical analysis is appropriate in the large-$J$ regime, where the
antiferromagnetic Heisenberg interaction dominates and quantum fluctuations are
strongly suppressed, so that classical configurations capture the leading
symmetry structure of the ordered phase.
Characteristic limiting values,
$U_4=\tfrac12$ (U(1) rotor), $0.4$ (SU(2) rotor), $\tfrac13$ (four-sector $x$--$z$
cat), and $\tfrac23$ (single-sector Ising order), serve as benchmarks for
the underlying order-parameter distribution.
As shown in Fig.~\ref{fig:Rplane_Uz}, $U_4^{(z)}$ converges systematically towards the
value $\tfrac13$ with increasing system size.
This behavior identifies the ordered phase as an $X/Z$ N\'eel phase with duality
symmetry, corresponding to a four-sector cat structure in the $x$--$z$ plane.
The duality transformation acts on the staggered magnetizations as
\begin{equation}
\begin{aligned}
\mathcal D_{\rm dual}\, m_x^{\rm stag}\, \mathcal D_{\rm dual}^\dagger &= -\,m_z^{\rm stag},\\
\mathcal D_{\rm dual}\, m_z^{\rm stag}\, \mathcal D_{\rm dual}^\dagger &= -\,m_x^{\rm stag}.
\end{aligned}
\end{equation}
In finite systems, the ground state may be chosen as an eigenstate of
$\mathcal D_{\rm dual}$, which enforces dual equalities among diagnostics, such as
$U_4^{(x)}=U_4^{(z)}$ and $S_x(\pi,\pi)=S_z(\pi,\pi)$.

We next discuss the consequences for the thermodynamic limit.
As $L\to\infty$, spontaneous symmetry breaking selects a single ordered sector
out of the fourfold-degenerate ground space,
\begin{equation}
\begin{aligned}
\text{Z--N\'eel}^{\pm}:&\quad
(m_x^{\rm stag},\,m_z^{\rm stag}) \;\longrightarrow\; (0,\,\pm m_0),\\
\text{X--N\'eel}^{\pm}:&\quad
(m_x^{\rm stag},\,m_z^{\rm stag}) \;\longrightarrow\; (\pm m_0,\,0).
\end{aligned}
\end{equation}
Consequently,
\begin{equation}
U_4^{(z)}\xrightarrow[L\to\infty]{\text{Z--N\'eel}} \tfrac{2}{3}\qquad \text{or}
\qquad
U_4^{(z)}\xrightarrow[L\to\infty]{\text{X--N\'eel}} 0.
\end{equation}
And,
\begin{equation}
R_{\rm plane}\xrightarrow[L\to\infty]{}1,
\end{equation}
for both cases.

\section{Summary}
\label{summary}

We have investigated the stability of the $\mathbb{Z}_2$ topological 
phase in the square-lattice toric code under an isotropic 
antiferromagnetic Heisenberg perturbation. Our study aims to quantify 
the robustness of the topological order and identify the nature of 
the phase that emerges upon its breakdown. Our approach integrated
(i) 
an analytical SW expansion, providing a 
controlled low-energy description of the effective dynamics within 
the toric-code ground space, and
(ii) variational simulations using symmetry-adapted convolutional 
NQS, optimized within the Marshall gauge to exploit the stoquastic nature of the Hamiltonian.
The SW construction provides a microscopic understanding of the topological phase by clarifying how the Heisenberg perturbation renormalizes local stabilizers and dresses non-local observables, such as non-contractible Wilson loops, while leaving the topological protection intact, with a degeneracy splitting that is exponentially small in the system size. These analytical predictions are corroborated by our NQS calculations, which locate the phase transition using a set of complementary diagnostics. In particular, a pronounced peak in the fidelity susceptibility and the departure of the topological entanglement entropy from its quantized value $-\ln 2$ consistently signal the breakdown of topological order. Finite-size scaling of the fidelity-susceptibility peak provides an estimate of the critical exponent and, together with the peak positions of the logarithmic susceptibility of the composite non-contractible Wilson-loop diagnostic, yields an estimate of the thermodynamic critical coupling $J_c \approx 0.164$.

Beyond $J_c$, the topological order melts into a magnetically ordered 
regime. In our analysis, this phase is diagnosed primarily through 
the even-moment structure factors 
$S_\alpha(\pi,\pi) = N \langle (m_\alpha^{\rm stag})^2 \rangle_{\rm phys}$, 
which remain robust on finite tori where linear (odd-moment) 
magnetizations are suppressed by symmetry constraints. We find 
that the $y$-magnetization is strongly suppressed throughout the ordered 
regime, and we quantify this easy-plane character using the in-plane 
anisotropy ratio $R_{\rm plane}$ alongside Binder-cumulant analyzes of 
the order-parameter fluctuations. Consistent with the $x \leftrightarrow z$ 
duality and the constraints of the global parities, the finite-size 
ground state can maintain vanishing linear order while developing 
significant even-moment signatures. In the thermodynamic limit, 
spontaneous symmetry breaking will select a specific ordered sector, 
lifting the duality-symmetric finite-size behavior and rendering 
the $x$- and $z$-channel diagnostics inequivalent.
Finally, the SWT framework applies to a broad class of two-body spin perturbations, including ferromagnetic Heisenberg and XXZ-type anisotropies. The resulting analytic–numerical strategy, therefore, provides a general tool to study toric-code deformations, particularly those that remain stoquastic and are amenable to sign-problem–free quantum Monte Carlo simulations.

\begin{acknowledgments}
We thank A. Joshi, M. Imada for helpful discussions. R.P. is supported by JSPS KAKENHI Grant No. JP25H01535. T.P. acknowledges support by the Cluster of Excellence
‘Advanced Imaging of Matter’ (EXC 2056, project ID 390715994) and the European Union (ERC, QUANTWIST, Project number 101039098).
The views and opinions expressed
are however those of the authors only and do not
necessarily reflect those of the European Union or
the European Research Council, Executive Agency. 
Part of the calculations were performed on the supercomputer of the ISSP (University of Tokyo).
\end{acknowledgments}

\appendix
\section{Details of the neural network}
\label{appendix_A}
\begin{figure}
    \centering
    \includegraphics[width=\linewidth]{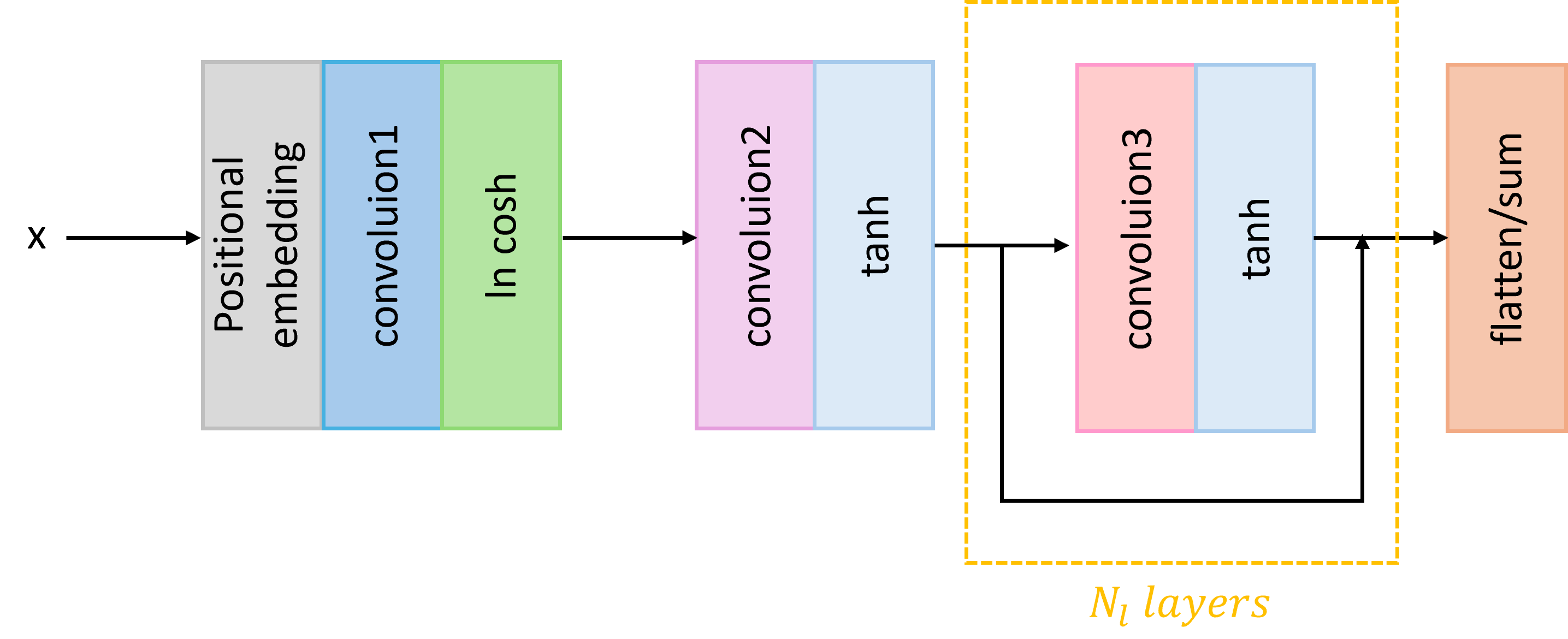}
	\caption{Convolutional neural network used as NQS. The inputs are the spin configurations in the $\sigma^z$ basis, and the output is the logarithm of the wave function. The architecture consists of: (i) a \textbf{positional embedding} stage that maps the edge spins onto a 2D grid; (ii) \textbf{convolution1}, a $1\times 2$ convolutional layer with stride $(1,2)$ and $\log\cosh$ activation; (iii) \textbf{convolution2}, a $2\times 2$ convolutional layer with $\tanh$ activation that compresses features into an $L\times L$ grid; and (iv) $N_l$ repeated layers (\textbf{convolution3}) of $2\times 2$ or $3\times 3$ convolutional layers with $\tanh$ activation, followed by a global \textbf{flatten/sum} aggregation. For $L \leq 4$, we use $12$--$24$ channels with $N_l = 2$ or $3$; for $L = 5$, $24$ channels with $N_l = 3$; and for $L = 6$, $56$--$75$ channels with cardinality $2$--$3$.}
	\label{CNN}
\end{figure}

\begin{figure}
    \centering
    \includegraphics[width=\linewidth]{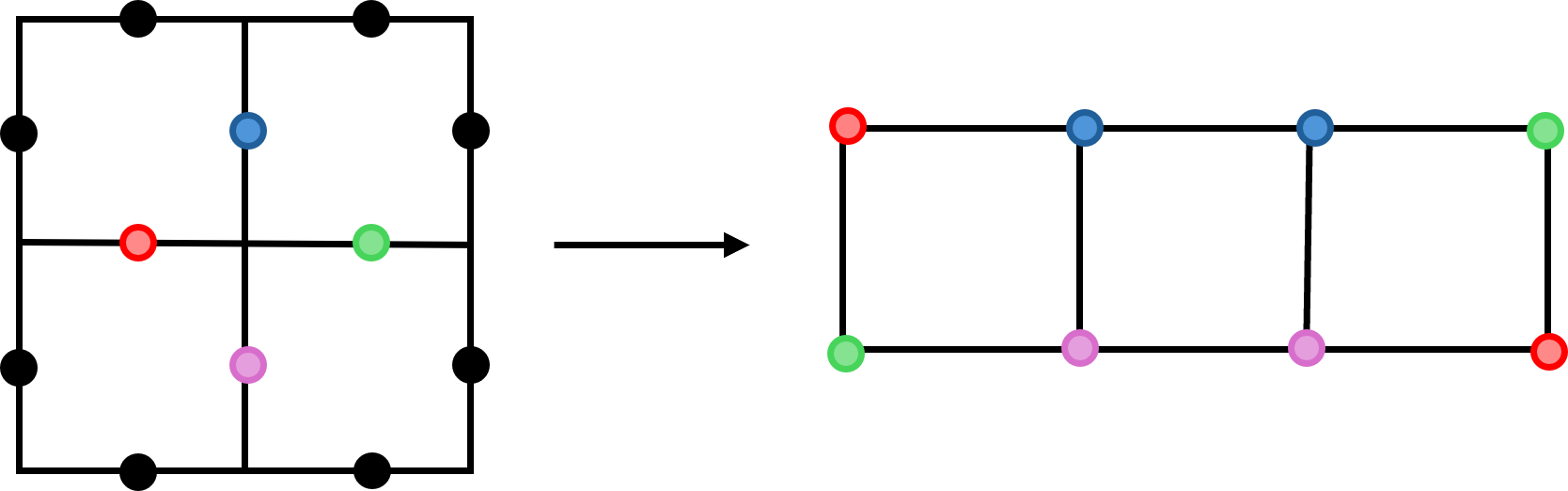}
	\caption{Symmetry-adapted input mapping and positional embedding. Physical edge spins of a star on the toric code are mapped onto a 2D grid using a locality-preserving positional embedding.}
	\label{construct}
\end{figure}

To leverage translational symmetry via convolutional weight sharing, 
we map the edge-spin configurations of the toric code onto a 
two-dimensional grid. 
We find that a naive unit-cell layout leads 
to optimization instabilities. This is attributed to the fact that 
the Hamiltonian couples spins through both nearest-neighbor bonds 
and four-spin star/plaquette operators; a naive mapping scatters 
these locally coupled physical degrees of freedom across distant 
grid coordinates, effectively increasing the complexity of the 
required convolutional kernels. 
This sensitivity to input encoding is consistent with previous 
benchmarks in the NQS literature \cite{corrRBM}. 
To ensure that the Hamiltonian terms remain local within the convolutional receptive fields, we implement a positional embedding stage.
The integrated global architecture of the symmetry-adapted NQS, 
including the positional embedding and the subsequent convolutional 
blocks, is illustrated in Fig.~\ref{CNN}.

The positional embedding is designed to prioritize the locality of the 
star operators ($A_v$), which are off-diagonal in the computational 
basis and empirically more challenging to optimize than the diagonal 
plaquette operators ($B_p$). On an $L \times L$ lattice, the $L^2$ 
stars are 
mapped into local $2 \times 4$ grid blocks, as illustrated in 
Fig.~\ref{construct}. Notably, this mapping involves a doubling of the 
spin degrees of freedom in the grid representation, where each 
physical spin is represented by two grid sites to maintain the 
required convolutional symmetry and local connectivity. This 
ensures that the off-diagonal flips are captured within the 
immediate receptive field of the initial convolutional layers.

The network architecture processes the embedded grid through a sequence of convolutional stages designed to respect the lattice symmetries. The first stage employs $1 \times 2$ kernels with stride $(1,2)$ and $\log\cosh$ activations, ensuring each kernel reads a single physical bond. In the second stage, $2 \times 2$ kernels with $\tanh$ activations aggregate these bond features into a compressed $L \times L$ lattice of ``condensed stars.'' Subsequent $N_l$ repeated layers consist of standard $2 \times 2$ or $3 \times 3$ convolutions with periodic boundary conditions. To improve expressivity without the optimization instabilities often associated with large channel dimensions, we utilize grouped convolutions (increasing cardinality) \cite{cardinality}. Where necessary, we incorporate skip connections \cite{resnet} to facilitate deep training.

The final log-amplitude $\ell_{\boldsymbol{\theta}}(\sigma)$ is obtained via a global sum over the spatial and channel axes. To prevent divergence in larger systems, we replace the raw sum with a normalized aggregation ($1/L$ or $1/\sqrt{L^2 C}$), ensuring the magnitude of the log-amplitude remains controlled across system sizes. By construction, the resulting ansatz is invariant under lattice translations.

To further constrain the variational search space, we explicitly enforce 
$C_4$ rotational invariance at the wavefunction level. For a spin 
configuration $\sigma$, let $\mathcal{R}$ denote the $90^\circ$ rotation 
operator. The $C_4$-symmetrized log-amplitude $\ell_{\theta}^{(C_4)}(\sigma)$ 
is constructed by averaging the network outputs over the four quarter turns:
\begin{equation}
\ell_\theta^{(C_4)}(\sigma) = \log \left[ \frac{1}{4} \sum_{k=0}^{3} 
\exp \left( \ell_\theta (\mathcal{R}^k \sigma) \right) \right].
\end{equation}
By construction, the resulting wavefunction 
$\psi_{\theta}^{(C_4)}(\sigma) = \exp[\ell_{\theta}^{(C_4)}(\sigma)]$ 
satisfies $\psi_{\theta}^{(C_4)}(\mathcal{R}\sigma) = \psi_{\theta}^{(C_4)}(\sigma)$. 
While this symmetrization increases the computational cost of each 
evaluation by a factor of four, it ensures that the NQS remains 
within the $C_4$-symmetric sector throughout the optimization process.

We find the intermediate coupling regime, $J \in [0.1, 0.2]$, to be 
the most numerically challenging, requiring higher network capacity 
compared to the deeper topological or magnetically ordered phases. For $L \le 4$, 
we employed $2 \times 2$ kernels with 12--24 channels and $N_l = 2$
or $3$ layers. Empirically, a parameter budget of approximately 
$10N^2$ (where $N=2L^2$) was adequate to obtain robust results across 
the transition.
For larger systems ($L \ge 5$), simply increasing channel width 
often led to optimization instabilities. We found it more effective 
to enlarge the convolutional receptive field to $3 \times 3$ and 
incorporate skip connections. For $L=5$, we utilized 24 channels 
with $N_l=3$. For $L=6$, we further enhanced expressivity while 
maintaining stability by increasing the convolutional cardinality to 
2 or 3 (excluding the first layer). This configuration utilized 
56--75 total channels per layer with a parameter budget of 
$\sim 10N^2$.

\section{OPTIMIZATION}
\label{appendix_B} 
Our training objective is the variational energy
\begin{equation}
E(\boldsymbol{\theta}) \equiv 
\frac{\langle \psi_{\boldsymbol{\theta}}|H'|\psi_{\boldsymbol{\theta}}\rangle}
     {\langle \psi_{\boldsymbol{\theta}}|\psi_{\boldsymbol{\theta}}\rangle}.
\end{equation}
The network outputs the log–amplitude 
$\ell_{\boldsymbol{\theta}}(\sigma)=\log \psi_{\boldsymbol{\theta}}(\sigma)$
(real in the Marshall gauge)hence 
$\psi_{\boldsymbol{\theta}}(\sigma)=\exp[\ell_{\boldsymbol{\theta}}(\sigma)]$ and 
we sample configurations $\sigma$ from $\pi_{\boldsymbol{\theta}}(\sigma)\propto|\psi_{\boldsymbol{\theta}}(\sigma)|^2$.
With importance sampling, the energy is estimated as
\begin{equation}
\begin{aligned}
\widehat{E}
&=\big\langle E_{\mathrm{loc}}\big\rangle_{\pi_{\boldsymbol{\theta}}},\\
E_{\mathrm{loc}}(\sigma)
&=\frac{(H'\psi_{\boldsymbol{\theta}})(\sigma)}{\psi_{\boldsymbol{\theta}}(\sigma)}\\
&=\sum_{\sigma'} H'_{\sigma,\sigma'}\,
\exp\!\Big[\ell_{\boldsymbol{\theta}}(\sigma')-\ell_{\boldsymbol{\theta}}(\sigma)\Big].
\end{aligned}
\end{equation}

For gradients, we use the log-derivative observables 
$O_i(\sigma)=\partial_{\theta_i}\ell_{\boldsymbol{\theta}}(\sigma)$ and define the
(force) statistics
\begin{equation}
F_i=\Big\langle O_i^\ast E_{\mathrm{loc}}\Big\rangle
     -\Big\langle O_i^\ast\Big\rangle\Big\langle E_{\mathrm{loc}}\Big\rangle,
\qquad
\nabla_{\theta_i}E(\boldsymbol{\theta})=2\,\mathrm{Re}\,F_i,
\end{equation}
where $\langle\cdot\rangle$ denotes Monte-Carlo averages over $\pi_{\boldsymbol{\theta}}$.
We minimize $E(\boldsymbol{\theta})$ using stochastic reconfiguration (SR) as a
preconditioner for stochastic gradient descent (SGD).
Let $g=\nabla_{\boldsymbol{\theta}}E$ be the Monte-Carlo energy gradient (computed from the
estimators above). At each iteration, we compute an SR direction $\mathbf{p}$ by solving
\begin{equation}
\label{eq:sr-precond}
\bigl(S + \lambda I \bigr)\,\mathbf{p}=\mathrm{Re}\,g,
\end{equation}
where $S$ is the quantum geometric tensor (QGT) and $\lambda$ is a small Tikhonov regularization. 
The linear
system~\eqref{eq:sr-precond} is solved by conjugate gradients (CG) to a prescribed
tolerance.
We then apply the preconditioned SGD update
\begin{equation}
\boldsymbol{\theta}\ \leftarrow\ \boldsymbol{\theta}\ -\ \eta\,\mathbf{p},
\end{equation}
with the learning rate $\eta$.

Unless otherwise stated, we use a fixed learning rate $\eta\in[0.005,\,0.01]$.
The SR regularization is also kept constant within each run,
with $\lambda\in[10^{-3},\,10^{-2}]$.
The specific $\lambda$ is chosen per run
based on the conditioning of the CG  and the variance of the gradient estimates. We employ 512 parallel Markov chains. Unless stated otherwise, we use \(2^{15}\) samples per VMC iteration, with the sweep length set to the number of spins (one full-lattice sweep per sample).

We next describe the Markov Chain Monte Carlo procedure used to  sample from the target distribution $|\psi_{\boldsymbol{\theta}}|^2$, 
focusing on the Metropolis--Hastings (MH) update schemes employed for 
ergodicity.
The proposal distribution is optimized for ergodicity and fast mixing. 
At each MH step, we select from the following moves 
with probabilities $\mathbf{p}=(0.50, 0.20, 0.15, 0.125, 0.025)$:
(i) vertex (star) 4-spin flip, 
(ii) non-contractible loop flip (Wilson-loop move), 
(iii) spin exchange up to next-nearest neighbors, 
(iv) two-spin flip up to next-nearest neighbors, 
(v) local single-spin flip.
The star and Wilson-loop moves facilitate transitions between 
topological sectors, while the exchange and local moves ensure 
granularity for local mixing. For odd linear sizes $L$, the 
physical-gauge structure in the Marshall gauge dictates that the 
low-energy state resides in the parity sector $(P_x, P_z) = (-1, -1)$. 
Because our real-valued ansatz inherently favors the $P_x$ sector, 
optimizing the $P_z$ sector can be slow for $J \in [0.1, 0.2]$. To 
accelerate convergence, we introduce a bias during the initial 
optimization stage:
\begin{equation}
H_{\rm guide} = H + h P_z, \quad h = 0.05.
\end{equation}
We gradually anneal $h \to 0$ and verify a posteriori that 
the resulting state is stable, with $\langle P_z \rangle \simeq -1$ 
and observables remaining consistent after the bias is removed. 
All models and procedures are implemented in \textsc{NetKet}
using a CNN ansatz with SR~\cite{netket}.

\section{Details of the Schrieffer--Wolff expansion} \label{appendix:SW-details}
In this section, we provide the technical details of the 
SW transformation used to derive the effective 
Hamiltonian within the toric-code ground space~\cite{bravyiSW}.
\subsection{Setup: projectors, resolvent, and off-diagonal gauge}
\label{sec:methods_sw_setup}
For the SW derivation, it is convenient to allow separate stabilizer couplings,
\begin{equation}
H_0^{\mathrm{SW}} = -J_e\sum_v A_v - J_m\sum_p B_p,
\qquad
H^{\mathrm{SW}} = H_0^{\mathrm{SW}} + V,
\label{eq:H0SW_methods}
\end{equation}
with $V$ the Heisenberg perturbation 
\begin{equation}
\begin{aligned}
\label{hei_SW2}
V &= J\sum_{b=\langle i,j\rangle}\bigl(X_b+Y_b+Z_b\bigr),\\
X_b &= \sigma^x_i\sigma^x_j,\quad
Y_b = \sigma^y_i\sigma^y_j,\quad
Z_b = \sigma^z_i\sigma^z_j .
\end{aligned}
\end{equation}
Let $P$ be the projector onto the unperturbed code space (all $A_v=B_p=+1$) and
$Q=\mathbf{1}-P$ its orthogonal complement. We define the reduced resolvent as
\begin{equation}
R
=
Q\left(E_P - H_0^{\mathrm{SW}}\right)^{-1}Q,
\label{eq:reduced_resolvent_methods}
\end{equation}
where $E_P$ is the ground-state energy of $H_0^{\mathrm{SW}}$.
The SW transformation is implemented by an anti-Hermitian generator
$S^\dagger=-S$ via
\begin{equation}
\tilde H
\equiv
e^{S} H^{\mathrm{SW}} e^{-S}.
\label{eq:Hbar_methods}
\end{equation}
We adopt the off-diagonal gauge for the generator $S$:
\begin{equation}
PSP = 0, \qquad QSQ = 0,
\label{eq:offdiag_gauge_methods}
\end{equation}
which constrains $S$ to only couple the code space $P$ and its complement $Q$, precluding rotations within the individual subspaces. The generator is expanded as a power series in $J$:
\begin{equation}
\begin{aligned}
S &= S_1 + S_2 + S_3 + \cdots,\\
S_k &= \mathcal{O}(J^k),\\
P S_k P &= Q S_k Q = 0.
\end{aligned}
\label{eq:S_series_methods}
\end{equation}
The defining SW condition is that $\tilde H = e^S H e^{-S}$ is block-diagonal at each order:
\begin{equation}
P\,\tilde H\,Q = 0,
\qquad
Q\,\tilde H\,P = 0,
\label{eq:blockdiag_methods}
\end{equation}
where the low-energy effective Hamiltonian is given by the projection:
\begin{equation}
H_{\mathrm{eff}} \equiv P\,\tilde H\,P.
\label{eq:Heff_def_methods}
\end{equation}
\subsection{First-order generator,  \texorpdfstring{$S_1$}{S1}}
\label{sec:methods_sw_S1}

Expanding $\tilde H$ to first order gives
$\tilde H=H_0^{\mathrm{SW}}+V-[H_0^{\mathrm{SW}},S_1]+\mathcal{O}(J^2)$.
Imposing the block-diagonal condition Eq.~\eqref{eq:blockdiag_methods} yields
\begin{equation}
0
=
P\bigl(V - [H_0^{\mathrm{SW}},S_1]\bigr)Q
=
Q\bigl(V - [H_0^{\mathrm{SW}},S_1]\bigr)P.
\label{eq:SW_first_order_condition}
\end{equation}
Using $PH_0^{\mathrm{SW}}P=E_PP$ and the identity
$(E_P-H_0^{\mathrm{SW}})R=R(E_P-H_0^{\mathrm{SW}})=Q$, one obtains the unique
off-diagonal solution
\begin{equation}
S_1
=
PVR - RVP,
\qquad
PS_1P=QS_1Q=0.
\label{eq:S1_methods}
\end{equation}
Physically, $S_1$ serves as the generator of virtual transitions $P \to Q$ and $Q \to P$ required to cancel the first-order off-diagonal coupling induced by $V$.
A key simplification in our model is that a single bond operator always
creates stabilizer violations (anyons), hence maps the code space to $Q$:
\begin{equation}
PVP=0.
\label{eq:PVP_zero_methods}
\end{equation}
Consequently, the first-order correction $H_{\mathrm{eff}}^{(1)} = PVP$ vanishes identically, and
the first nontrivial contribution to $H_{\mathrm{eff}}$ appears at
second order, $\mathcal{O}(J^2/\Delta)$.

\subsection{Energy denominators from single-bond excitations}
\label{sec:methods_sw_gaps}

To evaluate the expansion coefficients, we identify the excitation 
energies generated by a single bond operator acting on the ground space. The three components of the Heisenberg perturbation $V$ 
create distinct anyonic configurations with corresponding energy gaps:
\begin{itemize}
    \item $X_b$ generates a pair of $m$-anyons on adjacent plaquettes, 
    with gap $\Delta_x = 4J_m$.
    \item $Z_b$ generates a pair of $e$-anyons on adjacent vertices, 
    with gap $\Delta_z = 4J_e$.
    \item $Y_b \propto X_b Z_b$ generates both an $e$-pair and an 
    $m$-pair, with gap $\Delta_y = 4(J_e+J_m)$.
\end{itemize}
In this low-order regime, the resolvent operator $R$ acts as a scalar 
multiplier $1/\Delta_{x,y,z}$ determined by the specific 
intermediate anyon content of the state.

\subsection{Second-order expansion: Construction of \texorpdfstring{$H_{\mathrm{eff}}^{(2)}$}{Heff}}
\label{sec:methods_sw_second}

In the off-diagonal gauge, the second-order effective Hamiltonian takes the compact form
\begin{equation}
H_{\mathrm{eff}}^{(2)}
=
\,PVRVP.
\label{eq:Heff2_methods}
\end{equation}
Writing $V=J\sum_b V_b$ with $V_b\in\{X_b,Y_b,Z_b\}$,
each term $P V_b R V_{b'} P$ is nonzero only if:
(i) $V_b$ creates a definite two-anyon excitation out of the code space,
(ii) $R$ contributes the corresponding denominator, and
(iii) $V_{b'}$ annihilates the excitations and returns the state to the code space.
On the square lattice, the only two-bond connected processes that return
to $P$ and build an operator are:
opposite-corner $XX$ pairs around a vertex, which multiply to a star $A_v$, and
opposite-corner $ZZ$ pairs around a plaquette, which multiply to a plaquette
$B_p$.  All other two-bond strings either fail to annihilate the anyons and
therefore vanish after projection, or reduce to same-bond ``backtracking''
($V_b^2=\mathbf{1}$) and build only a constant term.
As a result,
\begin{equation}
H_{\mathrm{eff}}^{(2)}
=P \bigl(
-\frac{J^2}{J_m}\sum_v A_v
-\frac{J^2}{J_e}\sum_p B_p
+
E_{\mathrm{const}}^{(2)} \bigr) P .
\label{eq:Heff2_final_methods}
\end{equation}
The geometry-dependent constant $E_{\mathrm{const}}^{(2)}$ arises from same-bond
backtracking in the $XX$, $YY$, and $ZZ$ channels. Explicitly, for an $L\times L$
torus,
\begin{equation}
E_{\mathrm{const}}^{(2)}
= -\,L^{2}J^{2}\!\left(\frac{1}{J_m}+\frac{1}{J_e+J_m}+\frac{1}{J_e}\right).
\label{eq:Econst2_methods}
\end{equation}
Moreover, within the code space $P$, we have $A_v=B_p=+1$, so
$P\sum_v A_v P=L^2P$ and $P\sum_p B_p P=L^2P$, and therefore the entire
second-order contribution reduces to a constant energy shift,
\begin{equation}
H_{\mathrm{eff}}^{(2)}
=\Bigl[E_{\mathrm{const}}^{(2)}-L^2J^2\!\left(\frac{1}{J_m}+\frac{1}{J_e}\right)\Bigr]\,P.
\label{eq:Heff2_constant_methods}
\end{equation}

\subsection{Vanishing of the third-order corrections}
\label{sec:methods_sw_third}

At third order, the effective Hamiltonian consists of terms with the 
schematic structure $PVRVRVP$ (plus convention-dependent subtraction 
terms). For a contribution to be non-vanishing, the product of three 
single-bond operators must act as an identity or a logical operator 
within the code space.

However, on the square lattice, any product $\mathcal{O}$ of three 
local Heisenberg bond operators necessarily leaves at least one 
unpaired anyonic excitation. 
This is formally demonstrated by the fact that for any 
such three-bond product $O_3$, there exists at least one 
stabilizer $O_{st} \in \{A_v, B_p\}$ that anticommutes with it: 
$\{ O_3, O_{st} \} = 0$. Using the property that $PSP = P$ 
for any stabilizer, we have:
\begin{equation}
PO_3 P = P (O_{st}O_3 O_{st}) P = -PO_3 P \implies PO_3P = 0.
\end{equation}
Consequently, the local effective Hamiltonian vanishes at this order:
\begin{equation}\label{3rd_Heff}
H_{\mathrm{eff}}^{(3)} = 0.
\end{equation}
We note that on a torus, non-contractible contributions appear at order $L$; thus, for $L=3$ such terms can coincide with the third order, 
while for $L \ge 4$ they occur only at higher orders.

\subsection{Higher-order expansions: A logical-operator formalism}
\label{app:sw_higher_logical}

At order $n \ge 4$, the effective Hamiltonian $H_{\mathrm{eff}}^{(n)}$ 
consists of operator structures generated by sequences of the 
perturbation $V$ and the reduced resolvent $R$, schematically 
represented as $P V R V R \cdots R V P$ with $n$ insertions of $V$ 
and $(n-1)$ resolvents $R$. These contributions fall into two categories:
\begin{itemize}
    \item Connected bond clusters: These generate new 
    gauge-invariant structures, including multi-stabilizer couplings 
    and contractible loop operators.
    \item Reducible contributions: Convention-dependent 
    subtraction terms remove these processes, primarily serving to 
    renormalize the coefficients of operators established at lower order.
\end{itemize}
Following the convention for the reduced resolvent $R$ introduced in 
Eq.~\eqref{eq:reduced_resolvent_methods}, we define the excitation gap 
for any intermediate state $|\alpha\rangle \in Q$ with energy $E_\alpha$ 
as the positive quantity:
\begin{equation}
\Delta_\alpha \equiv E_\alpha - E_P > 0.
\label{eq:gap_positive_def}
\end{equation}
The reduced resolvent is defined relative to the unperturbed 
ground-state energy $E_P$, allowing it to be expressed directly 
in terms of these spectral gaps:
\begin{equation}
R = Q (E_P - H^{SW}_0)^{-1} Q = -\sum_{\alpha \in Q} \frac{|\alpha\rangle \langle \alpha|}{\Delta_\alpha}.
\label{eq:resolvent_positive_gap}
\end{equation}
Under this definition, each of the $(n-1)$ resolvents contributes an 
explicit minus sign. Consequently, the amplitude of a typical $n$th-order 
connected process scales as:
\begin{equation}
\text{Coeff.} \sim (-1)^{n-1} \frac{J^n}{\bar{\Delta}^{n-1}},
\label{eq:connected_scaling_positive_gap}
\end{equation}
where $\bar{\Delta}$ represents the characteristic positive gap of the 
intermediate anyonic excitations. While the final sign of a specific 
operator coefficient also depends on the microscopic signs within $V$ 
and the operator algebra of the bond cluster, 
Eq.~\eqref{eq:connected_scaling_positive_gap} fixes the sign convention 
for the denominators.

For the perturbations considered in this model, a single insertion of $V$ 
creates anyonic defects and leaves the code space, such that 
$PVP=0$. In this case, the first few orders of the effective Hamiltonian take the following explicit form:
\begin{align}
    H_{\rm eff}^{(2)} 
        &= P\,V R V\,P, 
        \label{eq:Heff2} \\
    H_{\rm eff}^{(3)} 
        &= P\,V R V R V\,P, 
        \label{eq:Heff3} \\
    H_{\rm eff}^{(4)} 
        &= P\,V R V R V R V\,P \notag\\
        &\quad - P\,V R^{2} V\,P\,V R V\,P \notag\\
        &\quad - P\,V R V\,P\,V R^{2} V\,P,
        \label{eq:Heff4}
\end{align}
where the first term is the connected string and the remaining terms are subtraction
terms that cancel reducible (disconnected) contributions.  This linked-cluster
structure ensures that new operator content is generated by connected bond clusters,
while subtraction terms predominantly renormalize lower-order structures.

After projection, any operator that creates stabilizer defects is annihilated
by the ground-state projector.  Thus, the only non-vanishing operators in the projected theory
are those that preserve all stabilizer eigenvalues, i.e., those that commute with
every $A_v$ and $B_p$.  
In the toric 
code, this commutant is generated by:
\begin{enumerate}
    \item Local stabilizers, which generate all 
    contractible loop operators.
    \item Non-contractible Wilson loops, which wind around 
    the cycles of the torus.
\end{enumerate}
Consequently, any gauge-invariant operator acting within the code space can be
expanded in the generator set
\begin{equation}
\{A_v\},\ \{B_p\},\ \{W_x^{(h)},W_x^{(v)},W_z^{(h)},W_z^{(v)}\}.
\label{eq:code_space_generators}
\end{equation}
In particular, SW-generated terms generally appear as decorated winding
operators: a non-contractible Wilson loop multiplied by an arbitrary product of
local/contractible factors.  Because $A_v$ and $B_p$ act as $+1$ within $P$, all
such decorations reduce to the same winding operator with a renormalized
coefficient.  
Thus, the projected Hamiltonian may be organized as a sum of products of
the four winding generators, with coefficients that absorb all contractible-loop
dressings.

\begin{equation}
\begin{aligned}
H_{\rm eff}
&= c_0\,P
+ P\Bigg[
\sum_{\gamma\in\{h,v\}}\sum_{\alpha\in\{x,z\}}
t_{\alpha,\gamma}\sum_{r} W_\alpha^{(\gamma)}(r)
\\
&\qquad
+ \sum_{\alpha\in\{x,z\}} u_\alpha\,
\smashoperator{\sum_{\mathclap{r,r'}}}
W_\alpha^{(h)}(r)\,W_\alpha^{(v)}(r')
\\
&\qquad
+ \sum_{\gamma,\gamma'\in\{h,v\}}
u_{xz}^{\gamma\gamma'}\,
\smashoperator{\sum_{\mathclap{r,r'}}}
W_x^{(\gamma)}(r)\,W_z^{(\gamma')}(r')
\\
&\qquad
+ \sum_{\gamma\in\{h,v\}} r_{xxz}^{\gamma}\,
\smashoperator{\sum_{\mathclap{r_1,r_2,r_3}}}
i\,W_x^{(h)}(r_1)\,W_x^{(v)}(r_2)\,W_z^{(\gamma)}(r_3)
\\
&\qquad
+ \sum_{\gamma\in\{h,v\}} r_{zzx}^{\gamma}\,
\smashoperator{\sum_{\mathclap{r_1,r_2,r_3}}}
i\,W_z^{(h)}(r_1)\,W_z^{(v)}(r_2)\,W_x^{(\gamma)}(r_3)
\\
&\qquad
+ y\,
\smashoperator{\sum_{\mathclap{r_1,r_2,r_3,r_4}}}
W_x^{(h)}(r_1)\,W_x^{(v)}(r_2)\,\allowbreak
W_z^{(h)}(r_3)\,W_z^{(v)}(r_4)
\\
&\qquad
+ H_{\rm loc}
\Bigg]P \, ,
\end{aligned}
\label{eq:Heff_taxonomy_all_nc}
\end{equation}
where
\begin{equation}
\begin{aligned}
H_{\rm loc}\equiv\;
&\sum_v a_v A_v
+\sum_p b_p B_p
+\sum_{v<v'} a_{vv'} A_v A_{v'} \\
&+\sum_{p<p'} b_{pp'} B_p B_{p'}
+\sum_{v,p} c_{vp}\,A_v B_p
+\cdots .
\end{aligned}
\label{eq:Heff_local_block}
\end{equation}
We note that because a non-contractible Wilson loop $W$ requires a sequence of perturbations to span the lattice, such terms only appear at order $n \ge L$. In the thermodynamic limit ($L \to \infty$), these logical operators vanish to all finite orders of the expansion, leaving an effective Hamiltonian composed entirely of local, contractible loop structures. However, for the finite-size systems considered in our numerical calculations, these expansion terms are essential as they represent the physical mechanism for topological degeneracy splitting. This splitting is 
exponentially suppressed with system size as $\mathcal{O}((J/\bar{\Delta})^L)$.

Having established the general form of the expansion, we next relate the taxonomy in Eq.~\eqref{eq:Heff_taxonomy_all_nc} to the non-contractible Wilson loops, thereby mapping the effective Hamiltonian onto a logical-qubit description. Within the toric-code ground space on a torus, the non-contractible Wilson 
loops act as logical Pauli operators for two encoded qubits. Adopting 
our horizontal/vertical loop convention, we fix the identification:
\begin{equation}
\begin{aligned}
\bar Z_1 &\equiv W_z^{(h)},\qquad
\bar X_1 \equiv W_x^{(v)},\\
\bar Z_2 &\equiv W_z^{(v)},\qquad
\bar X_2 \equiv W_x^{(h)}.
\end{aligned}
\label{eq:logical_mapping}
\end{equation}
We further define the logical $Y$ operators as:
\begin{equation}
\bar Y_j \equiv i\,\bar X_j\bar Z_j,
\qquad (j=1,2).
\label{eq:logical_y_def}
\end{equation}
With this choice, the winding operators satisfy the standard two-qubit 
Pauli algebra within the code space: operators on different logical 
qubits commute, $[\bar\mu_1, \bar\nu_2] = 0$, while on the same 
qubit they obey $\{\bar\mu_j, \bar\nu_j\} = 0$ for 
$\bar\mu_j \neq \bar\nu_j$ and $\bar\mu_j^2 = \mathbf{1}$ for 
$\bar\mu_j,\bar\nu_j  \in \{\bar X_j, \bar Y_j, \bar Z_j\}$.

The most general code-space Hamiltonian can next be expressed in the 
two-qubit logical Pauli basis as:
\begin{equation}
H_{\rm eff}
=
c_0\,P
+
P\left[
\sum_{\mu,\nu\in\{I,X,Y,Z\}}
c_{\mu\nu}\,\bar\mu_1\,\bar\nu_2
\right]P,
\label{eq:Heff_Pauli_basis_sym}
\end{equation}
where the $H_{\rm loc}$ is projected and absorbed to the $c_0$.
Because the SW transformation is constructed from $H^{SW}_0$ and $V$ using only
projectors and resolvents, the effective Hamiltonian $H_{\rm eff}$ inherits all
symmetries of the microscopic Hamiltonian that preserve the low-energy space.

The symmetry group of the model at the $J_e = J_m$ point significantly 
reduces the number of independent parameters in $H_{\mathrm{eff}}$. 
First, the $C_4$ rotational symmetry of the torus maps the two 
fundamental cycles onto each other, effectively swapping the logical 
qubits:
\begin{equation}
C_4:\quad \bar X_1 \leftrightarrow \bar X_2, \qquad \bar Z_1 \leftrightarrow \bar Z_2.
\label{eq:C4_action_logical_sym}
\end{equation}
This symmetry enforces parity between the logical qubits, ensuring that corresponding field and coupling coefficients are identical. 
Furthermore, at the dual point $J_e = J_m$, the Hamiltonian is 
invariant under the duality transformation exchanging $e \leftrightarrow m$ 
anyons. In the logical basis, this maps $x$-type operators to $z$-type 
operators, thereby equating the longitudinal and transverse field 
strengths as well as the related logical interactions. Combined with 
 the requirement of a real-valued matrix representation, these 
constraints lead to the following symmetric effective Hamiltonian

\begin{equation}
\begin{aligned}
H_{\rm eff}^{\mathrm{(real)}+C_4+\mathrm{(dual)}}
&=
c_0\,P
+
P\Big[
t(\bar X_1+\bar X_2+\bar Z_1+\bar Z_2)
\\
&\qquad
+u(\bar X_1\bar X_2+\bar Z_1\bar Z_2)
\\
&\qquad
+v(\bar X_1\bar Z_2+\bar Z_1\bar X_2)
+w\,\bar Y_1\bar Y_2
\Big]P 
\end{aligned}
\label{eq:Heff_sym_reduced_compact}
\end{equation}
with real coefficients $t,u,v,w$ generated by winding and higher-order connected
processes. 
In the subsequent even--odd analysis, we show how global parities further constrain these operators and which operators appear at leading order for a given $L$.

First, we consider the systems where $L$ is odd.
Within a fixed $(P_x, P_z)$ parity sector, any operator that is odd 
under either parity is forbidden. Since a non-contractible loop 
contains $L$ single-spin Pauli factors, conjugation gives:
\begin{equation}
\begin{aligned}
P_x W_z^{(h/v)} P_x &= (-1)^L W_z^{(h/v)}, \\
P_z W_x^{(h/v)} P_z &= (-1)^L W_x^{(h/v)}.
\end{aligned}
\label{eq:parity_action_loops_odd_app3}
\end{equation}
Thus, for odd $L$, the elementary loops $W_z^{(h/v)}$ and $W_x^{(h/v)}$ 
are parity-odd, and all terms linear in $\bar X_i$ or $\bar Z_i$ 
vanish within a fixed parity sector. The mixed cross-terms 
$\bar X_1 \bar Z_2$ and $\bar Z_1 \bar X_2$ are also parity-odd 
for odd $L$ and are forbidden as well.

Thus, the leading symmetry-allowed winding operators are parity-even loop products,
\begin{equation}
\bar X_1\bar X_2 = W_x^{(h)}W_x^{(v)},\qquad
\bar Z_1\bar Z_2 = W_z^{(h)}W_z^{(v)},
\label{eq:odd_allowed_products_app3}
\end{equation}
which can already be generated at order $n=L$ by length-$L$ winding bond sequences (each bond insertion contributes two single-spin Pauli operators).  
In addition,  
$\bar Y_1 \bar Y_2$ is allowed. This term is parity-even because each logical 
operator $\bar Y_j \propto i \bar X_j \bar Z_j$ is parity-odd for 
odd $L$; their product therefore transforms trivially under both 
global parities.

At the dual point $J_e=J_m=1$ with $C_4$ symmetry, the odd-$L$ code-space Hamiltonian
therefore takes the form
\begin{equation}
H_{\rm eff}^{\rm odd}
=
c_0\,P
+
P\Big[
u\big(\bar X_1\bar X_2+\bar Z_1\bar Z_2\big)
+w\,\bar Y_1\bar Y_2
\Big]P
\label{eq:Heff_odd_withYY_app3}
\end{equation}
with real couplings $u,w\sim J^L/\bar{\Delta}^{L-1}$ in the topological regime.  For the odd $L$ cases considered here, we find $u>0$
at leading winding order (See, Eq.~\eqref{eq:connected_scaling_positive_gap}).  The $\bar Y_1\bar Y_2$ term is typically smaller because
the intermediate $YY$ defects carry the larger positive gap
$\Delta_y=4(J_e+J_m)=8$, whereas $XX$ and $ZZ$ processes involve $\Delta_x=\Delta_z=4$.
Parametrically, this yields the suppression estimate
\begin{equation}
\frac{|w|}{|u|}
\sim
\left(\frac{\Delta_x}{\Delta_y}\right)^{L-1}
=
2^{-(L-1)}
\qquad (J_e=J_m=1),
\label{eq:w_over_u_suppression_app3}
\end{equation}
up to cluster-dependent prefactors and possible interference between distinct paths.

The three operators $\bar X_1\bar X_2$, $\bar Y_1\bar Y_2$, and $\bar Z_1\bar Z_2$
mutually commute and have eigenvalues $\pm1$.  It is therefore convenient to use the
Bell basis in the $\bar Z$-sector basis $\{|z_1,z_2\rangle\}$,
\begin{equation}
|\Phi^\pm\rangle=\frac{|+,+\rangle\pm|-,-\rangle}{\sqrt2},\qquad
|\Psi^\pm\rangle=\frac{|+,-\rangle\pm|-,+\rangle}{\sqrt2}.
\label{eq:bell_basis_app3}
\end{equation}
In this basis one finds the eigenvalue table
\begin{equation}
\begin{array}{c|ccc}
\text{state} & \bar X_1\bar X_2 & \bar Y_1\bar Y_2 & \bar Z_1\bar Z_2\\ \hline
|\Phi^+\rangle & +1 & -1 & +1\\
|\Phi^-\rangle & -1 & +1 & +1\\
|\Psi^+\rangle & +1 & +1 & -1\\
|\Psi^-\rangle & -1 & -1 & -1
\end{array}
\label{eq:bell_eigenvalues_app3}
\end{equation}
and thus the code space energies
\begin{equation}
\begin{aligned}
E_{\Phi^+}&=E_0+2u-w,\qquad &E_{\Phi^-}&=E_0+w,\\
E_{\Psi^+}&=E_0+w,\qquad &E_{\Psi^-}&=E_0-2u-w.
\end{aligned}
\label{eq:bell_energies_app3}
\end{equation}

For $u>0$, the state $|\Psi^-\rangle$ is the ground state whenever
$|w|<u$.  In the perturbative regime this condition is generically satisfied,
since Eq.~\eqref{eq:w_over_u_suppression_app3} implies $|w|/|u|\sim 2^{-(L-1)}\ll1$
for odd $L$ at $J_e=J_m=1$.
 In this ground state,
\begin{equation}
\langle \bar X_1\bar X_2\rangle=\langle \bar Z_1\bar Z_2\rangle=\langle \bar Y_1\bar Y_2\rangle=-1,
\label{eq:odd_products_in_gs_app3}
\end{equation}
while the single-loop expectation values vanish by parity,
\begin{equation}
\langle \bar X_1\rangle=\langle \bar X_2\rangle=
\langle \bar Z_1\rangle=\langle \bar Z_2\rangle
=0,
\qquad (L\ \text{odd}).
\label{eq:odd_singleloops_zero_app3}
\end{equation}
In terms of Wilson loop operators this implies
\begin{equation}
\begin{aligned}
\langle W_x^{(h)}W_x^{(v)}\rangle
=\langle W_z^{(h)}W_z^{(v)}\rangle &= -1,\\
\langle W_x^{(h)}\rangle
=\langle W_x^{(v)}\rangle
=\langle W_z^{(h)}\rangle
=\langle W_z^{(v)}\rangle &= 0,
\end{aligned}
\label{eq:odd_wilson_predictions_app3}
\end{equation}
which agrees with our numerical results for small $J$.

For even $L$, the elementary non-contractible loops are parity-even. 
Length-$L$ winding bond strings
can therefore project to single logical loops, and the effective Hamiltonian in the
code space may contain linear (single-qubit) terms at the leading winding order.

The effective code-space Hamiltonian
takes the form
\begin{equation}
\begin{aligned}
H_{\rm eff}^{\rm even}
&=
c_0\,P
+
P\Big[
t(\bar X_1+\bar X_2+\bar Z_1+\bar Z_2)
\\
&\qquad
+u(\bar X_1\bar X_2+\bar Z_1\bar Z_2)
\\
&\qquad
+v(\bar X_1\bar Z_2+\bar Z_1\bar X_2)
\\
&\qquad
+w\,\bar Y_1\bar Y_2
\Big]P .
\end{aligned}
\label{eq:Heff_even_withYY_clean}
\end{equation}

Next, we consider the coefficients of each logical operator.
At leading order, a winding process around a length-$L$ cycle requires $L$
insertions of $V$ separated by $L{-}1$ resolvents, so the induced logical
couplings scale as \(\sim J^{L}/\bar{\Delta}^{\,L-1}\).
\begin{equation}
t,\ u,\ v,\ w \;=\; \mathcal{O}\!\left(\frac{J^L}{\bar{\Delta}^{\,L-1}}\right),
\qquad (J\ll \bar{\Delta}),
\label{eq:even_winding_order}
\end{equation}
with an overall denominator sign factor $(-1)^{L-1}=-1$ for even $L$.
At $J_e=J_m=1$ the relevant 
gap values are
\begin{equation}
\Delta_x=\Delta_z=4,\qquad \Delta_y=8,
\label{eq:gaps_dual_point}
\end{equation}
corresponding to intermediate configurations with pure $m$-type, pure $e$-type, and
combined ($e{+}m$) defects, respectively.
A useful way to organize magnitudes is to separate the leading $XX/ZZ$ winding channels
from channels that require combined defects.  Writing the leading winding coefficients as
\begin{equation}
t \sim -A_t\,\frac{J^L}{4^{\,L-1}},\qquad
u \sim -A_u\,\frac{J^L}{4^{\,L-1}},
\label{eq:t_u_scale_even}
\end{equation}
and the mixed/combined-defect channels as
\begin{equation}
v \sim -A_v\,\frac{J^L}{8^{\,L-1}},\qquad
w \sim -A_w\,\frac{J^L}{8^{\,L-1}},
\label{eq:v_w_scale_even}
\end{equation}
illustrates the suppression of the mixed-species couplings
\begin{equation}
\frac{|v|}{|t|}\sim \frac{|w|}{|t|}
\sim
\left(\frac{4}{8}\right)^{L-1}
=
2^{-(L-1)},
(J_e=J_m=1),
\label{eq:vw_suppression_even}
\end{equation}
up to cluster-dependent prefactors $A_{t,u,v,w}>0$ and possible interference between
distinct winding paths. While the amplitudes $t$ and $u$ are of the same perturbative 
order in $J$, their combinatorial prefactors exhibit markedly different 
scalings with system size. We find that the coefficient $A_t$ scales 
exponentially, $A_t \sim 2L\times2^{L/2}$, whereas the prefactor $A_u$ increases only linearly, $A_u \sim 8L$. For the 
lattice sizes accessible to our numerics ($L=4, 6$), these competing 
scalings render $t$ and $u$ comparable in practice. For larger 
even $L$, by contrast, the exponential enhancement of $A_t$ implies 
a clear separation, with $|t|$ dominating over $|u|$.
Therefore, one expects the typical hierarchy
\begin{equation}
\begin{aligned}
|t| &\gtrsim |u| \gg |v|,\,|w| \qquad &(\text{intermediate even }L),\\
|t| &\gg |u| \gg |v|,\,|w| \qquad &(\text{larger even }L),
\end{aligned}
\label{eq:hierarchy_even_typical}
\end{equation}
in the perturbative regime (small $J$) for even $L$ at the dual point.
This hierarchy is not enforced by symmetry but follows from the minimal winding
construction (single-loop terms for $t$) and the larger gap $\Delta_y$ associated with
combined-defect channels.

If $|t|$ is the dominant logical coupling and $t<0$, the leading physics is set by the single-qubit
field $t(\bar X+\bar Z)$ on each logical qubit.  The corresponding dominant-field ground
state is the $+1$ eigenstate of $(\bar X+\bar Z)/\sqrt2$, which yields
\begin{equation}
\langle \bar X\rangle=\langle \bar Z\rangle=\frac{1}{\sqrt2},\qquad \langle \bar Y\rangle=0.
\label{eq:singlequbit_bisector_even_clean}
\end{equation}
Applying this to both logical qubits gives
\begin{equation}
\langle \bar X_1\rangle\simeq \langle \bar X_2\rangle
\simeq
\langle \bar Z_1\rangle\simeq \langle \bar Z_2\rangle
\simeq \frac{1}{\sqrt2}.
\label{eq:even_expect_bisector_clean}
\end{equation}
In Wilson-loop language this corresponds to
\begin{equation}
\langle W_x^{(h)}\rangle\simeq \langle W_x^{(v)}\rangle
\simeq
\langle W_z^{(h)}\rangle\simeq \langle W_z^{(v)}\rangle
\simeq \frac{1}{\sqrt2}.
\label{eq:even_wilson_bisector_clean}
\end{equation}
The remaining couplings $u,v,w$ act as subleading interactions within the same
four-dimensional ground space and primarily (i) correlate the two logical qubits and
(ii) shift the bisector expectations away from the ideal value $1/\sqrt2$.
Specifically, $u(\bar X_1\bar X_2+\bar Z_1\bar Z_2)$ favors aligned logical responses in
both $\bar X$ and $\bar Z$ (enhancing $\langle \bar X_1\bar X_2\rangle$ and
$\langle \bar Z_1\bar Z_2\rangle$ relative to the product-state estimate, 0.5),
$v(\bar X_1\bar Z_2+\bar Z_1\bar X_2)$ locks $\bar X$ of one qubit to $\bar Z$ of the
other (producing nonzero cross-correlators $\langle \bar X_1\bar Z_2\rangle$ and
$\langle \bar Z_1\bar X_2\rangle$), and $w\,\bar Y_1\bar Y_2$ generates additional
two-qubit correlations while remaining consistent with $\langle \bar Y_i\rangle=0$.
However, as reflected in the hierarchy above, these terms are exponentially suppressed with increasing system size $L$.

In practice, when $J/\Delta$ is very small, the overall tunneling scale
$\Delta E_{\rm topo}(L,J)\sim \mathcal{O}(J^L/\Delta^{L-1})$ can become so small that it
falls below the variational/numerical resolution (set by sampling noise, optimizer
tolerances, and the expressivity/initialization of the ansatz).  In this regime, VMC may converge to a metastable logical polarization that is close to an eigenstate of a single
loop (for example, a state with $\langle W_x^{(h)}\rangle_{phys}\simeq \langle W_x^{(v)}\rangle_{phys}$
large while $\langle W_z^{(h)}\rangle_{phys},\langle W_z^{(v)}\rangle_{phys}$ are strongly suppressed, or vice versa) depending on the initialized sector. 
This is shown in Appendix~\ref{app:polarized_results}.
This reflects the near degeneracy of the ground space rather than the
unique finite-size eigenstate selected by the nonzero tunneling terms.
Strictly speaking, for any fixed finite $L$ and any $J>0$ the exact ground state is the
lowest eigenstate of $H_{\rm eff}^{\rm even}$ and is generically a sector-mixed state,
but the energy difference to nearby logically polarized states can be exponentially small
in $L$.  In the thermodynamic limit $L\to\infty$ (at fixed $J$ in the topological phase), the tunneling amplitudes vanish, and the four torus sectors become exactly degenerate; in
that limit any of these logically polarized states may be chosen as a legitimate ground
state.

\subsection{Dressing and mixing of contractible loop operators}
\label{app:stabilizer_dressing_contractible}

In the toric code, the plaquette and star stabilizers are the minimal
contractible $\mathbb{Z}_2$ Wilson loops: $B_p$ is the smallest magnetic
(``$m$'') loop encircling a single plaquette, while $A_v$ is the smallest
electric (``$e$'') loop on the dual lattice.  Any larger contractible Wilson loop can therefore be written as a product of stabilizers over the
enclosed region.  For a $Z$-type contractible loop $C$ bounding a simply
connected set of plaquettes $S(C)$,
\begin{equation}
W_z(C)\equiv \prod_{e\in C}\sigma^z_e
\;=\;\prod_{p\in S(C)} B_p,
\label{eq:contractible_loop_product_Bp}
\end{equation}
and similarly an $X$-type contractible loop is a product of star operators over
vertices in the enclosed dual region,
\begin{equation}
W_x(C)\equiv \prod_{e\in C}\sigma^x_e
\;=\;\prod_{v\in S^\star(C)} A_v.
\label{eq:contractible_loop_product_Av}
\end{equation}
This representation is convenient: once $A_v^{\rm eff}$ and $B_p^{\rm eff}$ are
known, dressed contractible-loop observables follow immediately by multiplying
the corresponding dressed stabilizers.

For the operators of interest here (\(A_v\), \(B_p\), and \(W_{x/z}\)) we have
\([O,H^{\rm SW}_0]=0\).  Thus \(O\) is block diagonal in the \(P\oplus Q\)
decomposition and, in particular, commutes with both \(P\) and the reduced
resolvent \(R\).  With the leading SW generator \(S_1=PVR-RVP\)
[Eq.~\eqref{eq:S1_methods}], the first commutator simplifies to
\begin{equation}
[S_1,O]
=
P[V,O]R
-
R[V,O]P.
\label{eq:S1O_commutator_PVOR}
\end{equation}
We next evaluate \([V,O]\) using the bond decomposition of the Heisenberg
perturbation in Eq.~\eqref{hei_SW}.  In particular, for \(O=B_p\) only those
bonds with \([V_b,B_p]\neq 0\) can contribute to \([S_1,B_p]\) and hence to the
second-order correction.

A convenient selection rule follows from a simple parity count.  For a bond
\(b=\langle i,j\rangle\) and channel \(V_b\in\{X_b,Y_b\}\), define
\begin{equation}
n_p(b)\equiv \bigl|\langle i,j\rangle\cap\partial p\bigr|\in\{0,1,2\}.
\end{equation}
Using \(\sigma^{x/y}\sigma^z=-\sigma^z\sigma^{x/y}\) on the same spin (and
commutation otherwise), we obtain
\begin{equation}
V_b\,B_p
=
(-1)^{n_p(b)}\,B_p\,V_b,
\qquad (V_b\in\{X_b,Y_b\}),
\label{eq:bond_Bp_parity_rule}
\end{equation}
so only bonds with \(n_p(b)=1\) (i.e., bonds that touch \(\partial p\) at a
single endpoint) anticommute with \(B_p\) and therefore contribute.  The
\(Z_b\) channel always commutes with \(B_p\) and drops out at this order.

The projected second-order term $\tfrac12 P[S_1,[S_1,B_p]]P$ can be interpreted
as a sum over two-bond virtual processes: the first bond takes a code-space
state out to $Q$ (creating defects), and the second must remove those defects in
order to return to the code space.  For block-diagonal $O$ (hence
$[O,P]=[O,R]=0$), it is convenient to use the equivalent ``two-$V$''
representation
\begin{equation}
\begin{aligned}
\frac12\,P[S_1,[S_1,O]]P
&=
P V R\, O\, R V P
\\
&\quad
-\frac12\,P\Bigl(VR^2V\, O + O\, VR^2V\Bigr)P ,
\end{aligned}
\label{eq:SW_doublecomm_twoV}
\end{equation}
which separates the virtual-process contributions term by term.
To see the mechanism explicitly, consider a single active $XX$ bond
$b=\langle i,j\rangle$ with $n_p(b)=1$, and write
\begin{equation}
V_b = J\,X_b,
\qquad
X_b\equiv \sigma_i^x\sigma_j^x,
\qquad
X_bB_p=-B_p X_b.
\end{equation}
Acting on the code space, $X_b$ creates a virtual $m$-pair with energy cost
$\Delta_x$, so the reduced resolvent obeys
\begin{equation}
\begin{gathered}
R\,X_b P = -\frac{1}{\Delta_x}\,X_b P, \qquad
R^2\,X_b P = \frac{1}{\Delta_x^2}\,X_b P, \\[4pt]
\Rightarrow\quad P\,X_b R^2 X_b\,P = \frac{1}{\Delta_x^2}\,P.
\end{gathered}
\label{eq:RvP_rules}
\end{equation}
Using Eq.~\eqref{eq:RvP_rules} together with $X_b^2=\mathbf{1}$ and
$X_bB_p=-B_pX_b$ (hence $X_bB_pX_b=-B_p$), the two contributions in
Eq.~\eqref{eq:SW_doublecomm_twoV} reduce to
\begin{equation}
P\,V_b R\,B_p\,R V_b\,P
= -\Bigl(\frac{J}{\Delta_x}\Bigr)^2\,P B_p P,
\label{eq:term_sandwich_eval}
\end{equation}
and
\begin{equation}
\begin{split}
-\frac{1}{2}\,P\Bigl(V_b R^2 V_b\,B_p 
  + B_p\,V_b R^2 V_b\Bigr)P
  = -\Bigl(\frac{J}{\Delta_x}\Bigr)^2&\,P B_p P.
\end{split}
\label{eq:term_subtract_eval}
\end{equation}
Adding Eqs.~\eqref{eq:term_sandwich_eval} and \eqref{eq:term_subtract_eval}
gives the repeat-process correction associated with this single bond,
\begin{equation}
\frac12\,P[S_{1,b},[S_{1,b},B_p]]P
=
-2\Bigl(\frac{J}{\Delta_x}\Bigr)^2\,P B_p P.
\label{eq:repeat_singlebond_result}
\end{equation}
An analogous higher-order process arises in the $YY$ channel, with intermediate
energy cost $\Delta_y$.

For a fixed plaquette $p$, there are eight active bonds with $n_p(b)=1$ (two
emanating from each corner).  
Adding up these contributions
therefore gives $-16(J/\Delta_x)^2$ from $XX$ and $-16(J/\Delta_y)^2$ from $YY$.
In the $XX$ channel, there is an additional class of nonvanishing two-bond
processes.  Namely, let $b_1\neq b_2$ be the two active bonds emanating from
the same plaquette corner (equivalently, incident on the same corner
vertex $v$), and write $V_{b_\ell}=J\,X_{b_\ell}$ $(\ell=1,2)$.  Since
$n_p(b_1)=n_p(b_2)=1$, each bond anticommutes with $B_p$,
\begin{equation}
X_{b_\ell}B_p=-B_p X_{b_\ell}\qquad(\ell=1,2).
\label{eq:cornercross_each_anticommutes}
\end{equation}
However, the product of the two distinct corner bonds is parity-even with
respect to $B_p$ and, for the allowed corner-cross geometry, equals the local
star stabilizer at that corner vertex,
\begin{equation}
X_{b_2}X_{b_1}=A_v,
\qquad\Rightarrow\qquad
[X_{b_2}X_{b_1},\,B_p]=0.
\label{eq:cornercross_Av_identity}
\end{equation}
Consequently the two-bond operator appearing in the second-order dressing can be
reordered as
\begin{equation}
X_{b_2}\,B_p\,X_{b_1}
=
-\;B_p\,X_{b_2}X_{b_1}
=
-\;B_p\,A_v,
\end{equation}
and projection onto the code space removes the stabilizer because $A_v=+1$ in
$P$,
\begin{equation}
P\,X_{b_2}\,B_p\,X_{b_1}\,P
=
-\,P\,B_p A_v\,P
=
-\,P\,B_p\,P.
\label{eq:cornercross_reduction}
\end{equation}
Thus corner-cross sequences return to the code space even though $b_1\neq b_2$.
There are four plaquette corners and two orderings $(b_1,b_2)$ per corner,
giving $8$ such sequences.  Each carries the same denominator $\Delta_x$ as the
$XX$ repeat process and contributes $-2(J/\Delta_x)^2\,PB_pP$, yielding an
additional $-16(J/\Delta_x)^2$.

Collecting all
contributions, the $\mathcal{O}(J^2)$ dressing
of $B_p$ is purely multiplicative,
\begin{equation}
B_p^{\mathrm{eff}}
=
\Bigl[
1
-32\Bigl(\frac{J}{\Delta_x}\Bigr)^2
-16\Bigl(\frac{J}{\Delta_y}\Bigr)^2
\Bigr]\,
P B_p P
+\mathcal{O}(J^3).
\label{eq:Bp_eff_short}
\end{equation}
The corresponding result for $A_v$ follows by the $x\leftrightarrow z$ duality
of the toric code (exchanging plaquettes and stars).  Concretely, the $X_b$
channel commutes with $A_v$ and drops out, while $Z_b$ and $Y_b$ create virtual
electric $(e)$ and composite $(e,m)$ pairs with denominators $\Delta_z$ and
$\Delta_y$, yielding
\begin{equation}
A_v^{\mathrm{eff}}
=
\Bigl[
1
-32\Bigl(\frac{J}{\Delta_z}\Bigr)^2
-16\Bigl(\frac{J}{\Delta_y}\Bigr)^2
\Bigr]\,
P A_v P
+\mathcal{O}(J^3).
\label{eq:Av_eff_short}
\end{equation}
Equations~\eqref{eq:contractible_loop_product_Bp}--\eqref{eq:contractible_loop_product_Av}
imply that for any contractible loop $C$,
\begin{equation}
\begin{aligned}
W_z^{\mathrm{eff}}(C) &\equiv \prod_{p\in S(C)} B_p^{\mathrm{eff}}, \\
W_x^{\mathrm{eff}}(C) &\equiv \prod_{v\in S^\star(C)} A_v^{\mathrm{eff}}.
\end{aligned}
\label{eq:contractible_loops_from_dressed_stabilizers}
\end{equation}
Since $B_p^{\mathrm{eff}}$ and $A_v^{\mathrm{eff}}$ are renormalized
multiplicatively at $\mathcal{O}(J^2)$, any fixed-size contractible loop built
from a finite number of stabilizers inherits the same structure: At this order, it acquires only an overall prefactor, with no operator mixing inside the code space.

For $O\in\{A_v,B_p\}$ (and similarly for contractible loops built from their
products), cubic contributions are eliminated by the projection $P(\cdots)P$.
Any $\mathcal{O}(J^3)$ term corresponds to a product of three local bonds, which
cannot form a closed contractible loop on the square lattice (the smallest contractible cycle requires four bonds).  The resulting operator has
open-string endpoints (it creates defects) and is therefore annihilated by the
projection,
\begin{equation}
P\,(\text{three-bond open-string operator})\,P=0.
\end{equation}
Thus, for local stabilizers and contractible loops, the first corrections beyond
the multiplicative $\mathcal{O}(J^2)$ dressing arise at $\mathcal{O}(J^4)$, when
four-bond virtual processes can generate closed contractible structures and
induce genuine operator mixing within $P$, which is consistent with the case of the Hamiltonian in Eq.~\eqref{3rd_Heff}.

An important exception is provided by non-contractible winding operators
on a finite torus.  The shortest winding cycle has length $L$, so at order $J^L$
a virtual process can generate a closed string that winds around the system and
projects to a logical Wilson loop.  In particular, for $L=3$ the winding length
equals three, so $\mathcal{O}(J^3)$ processes can already produce
non-contractible loop operators (and hence survive for winding observables),
even though $\mathcal{O}(J^3)$ terms remain absent for $A_v$, $B_p$, and all
contractible loops.

\subsection{Dressing of non-contractible Wilson loops}
\label{app:wilson_dressing_short}

We define dressed non-contractible Wilson loops using the same SW unitary that
generates \(H_{\rm eff}\) and the dressed stabilizers,
\begin{equation}
W^{\rm eff}_{z}\equiv P\,e^{S} W_{z}\,e^{-S}P,
\qquad
W^{\rm eff}_{x}\equiv P\,e^{S} W_{x}\,e^{-S}P,
\label{eq:W_eff_def}
\end{equation}
with \(S=S_1+S_2+\cdots\).  Here \(S_1\) is the leading Schrieffer--Wolff
generator defined in Eq.~\eqref{eq:S1_methods}, constructed from the Heisenberg
perturbation \(V\) written in bond form in Eq.~\eqref{hei_SW}.

Since $W_{x/z}$ commute with $H_0$ (and are therefore block-diagonal in the
$P\oplus Q$ decomposition), the first nontrivial projected correction again
starts at second order, in direct analogy to $B_p^{\rm eff}$ and $A_v^{\rm eff}$.
Let $C_z$ be a straight non-contractible cycle of length $L$ and define
$W_z\equiv W_z(C_z)=\prod_{e\in C_z}\sigma^z_e$.
As in the stabilizer case, the structure of the second-order term implies a
simple selection rule: A term can contribute after the projection $P(\cdots)P$
only if the two perturbing bond operators create and then annihilate the
same virtual anyon pair, returning the net action to the code space.  In practice, this requires that each active bond $b$ intersect the loop support in
the minimal way---it must touch $C_z$ on exactly one edge spin.  Bonds
touching $C_z$ on zero edges commute trivially with $W_z$, while bonds touching it on two edges produce two sign flips that cancel; both cases drop out of the
commutators and do not contribute.

The channel structure follows from the same commutation logic.  For $W_z$ the
$Z_b$ channel commutes and does not contribute at this order, whereas the
$X_b$ and $Y_b$ channels enter with denominators $\Delta_x$ and $\Delta_y$,
corresponding to the virtual $m$-pair and $(e,m)$-pair gaps.  For a straight
loop, each of the $L$ loop vertices admits four nearest-neighbor bonds that
touch the loop on exactly one edge, giving $4L$ contributing bonds in total.

The nonvanishing $\mathcal{O}(J^2)$ contributions fall into two classes:
(i) repeated processes, in which the same contributing bond acts twice, and
(ii) vertex-cross processes, in which two distinct contributing bonds sharing a loop vertex act in sequence and collectively return to the code space.

For the repeated processes, each of the $4L$ contributing bonds produces a
multiplicative correction proportional to $P W_z P$, giving
$-8L(J/\Delta_x)^2$ from the $X_b$ channel and $-8L(J/\Delta_y)^2$ from the
$Y_b$ channel.  In addition, at each loop vertex there is one nonvanishing
cross term in the $X_b$ channel: The product of the two bonds reduces to $W_z$
times a local stabilizer, which evaluates to $+1$ after projection and supplies
an additional $-8L(J/\Delta_x)^2$.  By contrast, the analogous $Y_b$ cross terms
do not return to the code space and vanish under the projection.

Collecting all contributions yields, again, a purely multiplicative
renormalization.
\begin{multline}
P W^{\rm eff}_z P = \Bigl[ 1 -16L\Bigl(\frac{J}{\Delta_x}\Bigr)^2 
  - 8L\Bigl(\frac{J}{\Delta_y}\Bigr)^2 \Bigr] P W_z P \\
  + \mathcal{O}(J^3).
\label{eq:Wz_eff_renorm}
\end{multline}

The derivation for $W_x\equiv W_x(C_x)=\prod_{e\in C_x}\sigma^x_e$ is identical
upon exchanging $x\leftrightarrow z$.  For $W_x$, the $X_b$ channel commutes and
drops out, so only $Z_b$ and $Y_b$ contribute with denominators $\Delta_z$ and
$\Delta_y$.  The same counting gives
\begin{multline}
P W^{\rm eff}_x P
= \Bigl[
1 -16L\Bigl(\frac{J}{\Delta_z}\Bigr)^2
- 8L\Bigl(\frac{J}{\Delta_y}\Bigr)^2
\Bigr] P W_x P \\
+ \mathcal{O}(J^3).
\label{eq:Wx_eff_renorm}
\end{multline}

As for the stabilizers, cubic terms in $P e^S W_{x/z} e^{-S} P$ 
cancel 
generically: Any $\mathcal{O}(J^3)$ contribution involves a product of three
local bond operators $V_{b_3}V_{b_2}V_{b_1}$, which cannot form a closed string
on the square lattice and therefore leaves open-string endpoints (defects).
Such terms are annihilated by the projection, $P(\cdots)P=0$.  Thus, beyond the
multiplicative $\mathcal{O}(J^2)$ renormalization, the first genuinely new local
structures in $W^{\rm eff}_{x/z}$ appear at $\mathcal{O}(J^4)$.

On a finite torus, however, winding loops provide an important exception.  A
product of $L$ local bond operators can form a non-contractible closed
string that winds around the system and projects to a logical Wilson loop.
Consequently, for $L=3$ the minimal winding length equals three and
$\mathcal{O}(J^3)$ processes can, in principle, contribute to the dressing of
winding loops through terms proportional to logical operators, with amplitudes
scaling as $\sim J^3/\Delta^{\,2}$.  For $L>3$ the same mechanism first appears
at order $J^L$ (or higher, depending on lattice-parity constraints), so cubic
contributions remain absent.

Moreover, composite loop products are obtained simply as products of the
corresponding non-contractible Wilson loops.  For example, within the projected
theory one may form $W_{x}^{\rm eff}(C_1)\,W_{x}^{\rm eff}(C_2)$ (or mixed
$x/z$ products) to represent the effective operator associated with the
combined winding along both cycles.

Finally, any non-straight non-contractible Wilson loop can be reduced, within
the code space, to a product of the corresponding straight dressed Wilson loop
and local dressed stabilizers (plaquettes and stars), which project to products of $B_p^{\rm eff}$ and $A_v^{\rm eff}$. Hence, we can obtain an effective non-straight non-contractible Wilson loop from a product of effective plaquettes, stars, and a straight non-contractible Wilson loop.

\section{Metastable logically-polarized states}
\label{app:polarized_results}
\begin{figure}[t]
  \centering

  \begin{minipage}{0.48\columnwidth}
    \raggedright
    \textbf{(a)}\\[-2pt]
    \centering
    \includegraphics[width=\linewidth,height=0.28\textheight,keepaspectratio]{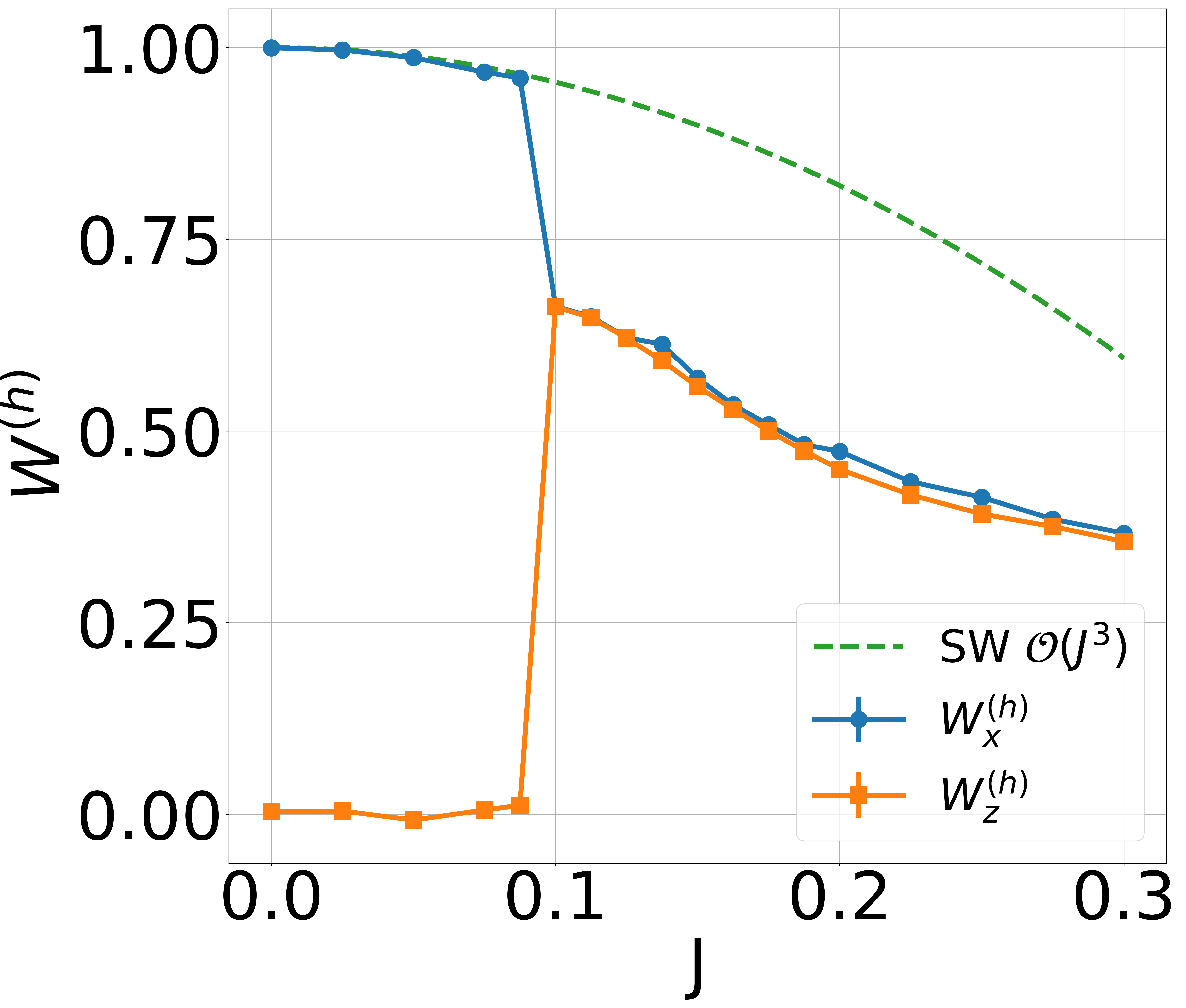}
    \label{fig:XhZh_L4}
  \end{minipage}\hfill
  \begin{minipage}{0.48\columnwidth}
    \raggedright
    \textbf{(b)}\\[-2pt]
    \centering
    \includegraphics[width=\linewidth,height=0.28\textheight,keepaspectratio]{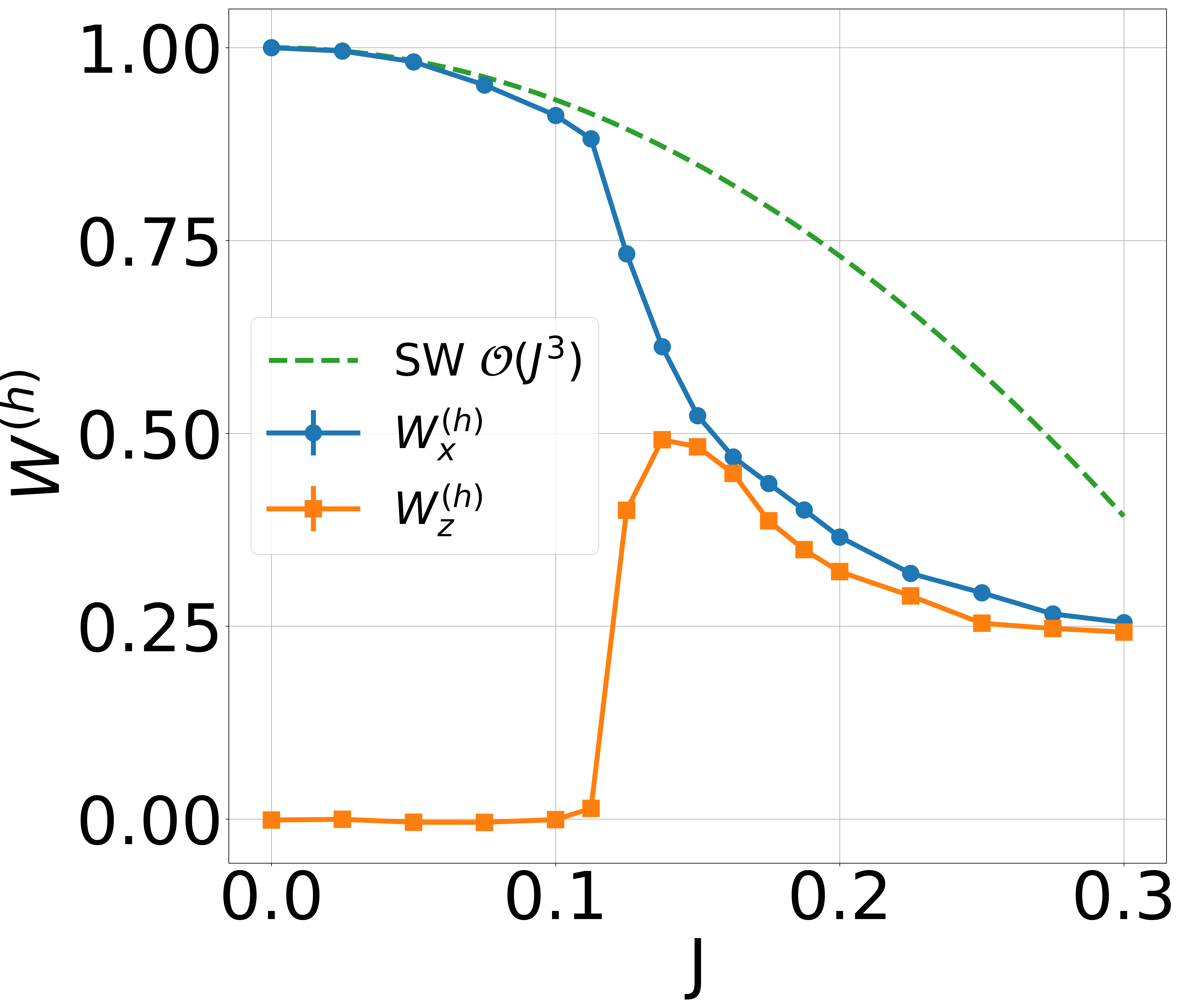}
    \label{fig:XhZh_L6}
  \end{minipage}

  \caption{\label{fig:evenwilson_single}
  Single non-contractible Wilson loops along the horizontal cycle,
  \(W_x^{(h)}\) and \(W_z^{(h)}\), for even system sizes (a) \(L{=}4\) and (b) \(L{=}6\) in a duality-broken metastable  state.
  Symbols with error bars are NQS measurements, while the green dashed curve shows the
  \(O(J^2)\) Schrieffer--Wolff (SW) prediction for the renormalized loop amplitude.}
\end{figure}

\begin{figure}[t]
  \centering

  \begin{minipage}{0.48\columnwidth}
    \raggedright
    \textbf{(a)}\\[-2pt]
    \centering
    \includegraphics[width=\linewidth,height=0.28\textheight,keepaspectratio]{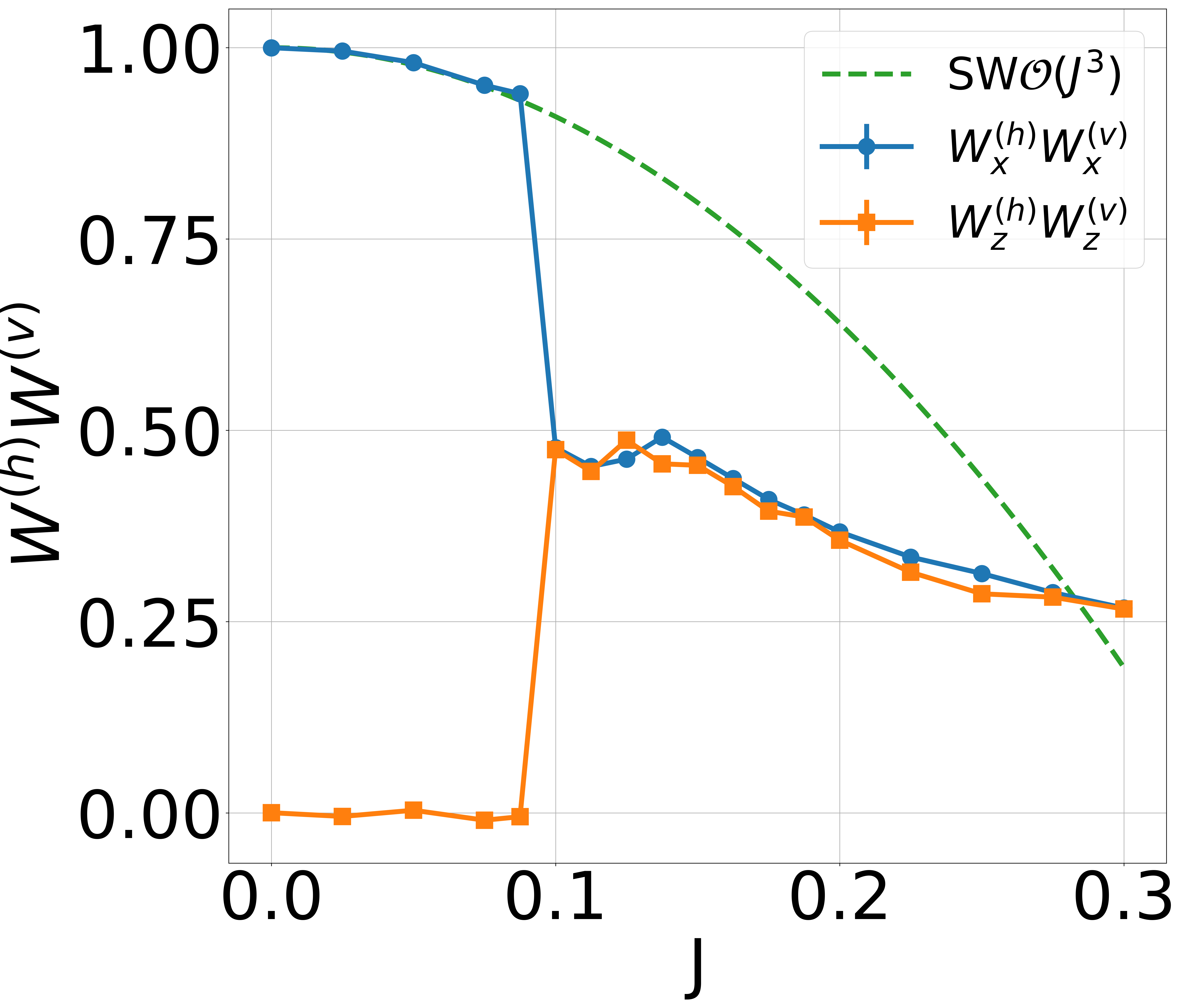}
    \label{fig:XdZd_L4}
  \end{minipage}\hfill
  \begin{minipage}{0.48\columnwidth}
    \raggedright
    \textbf{(b)}\\[-2pt]
    \centering
    \includegraphics[width=\linewidth,height=0.28\textheight,keepaspectratio]{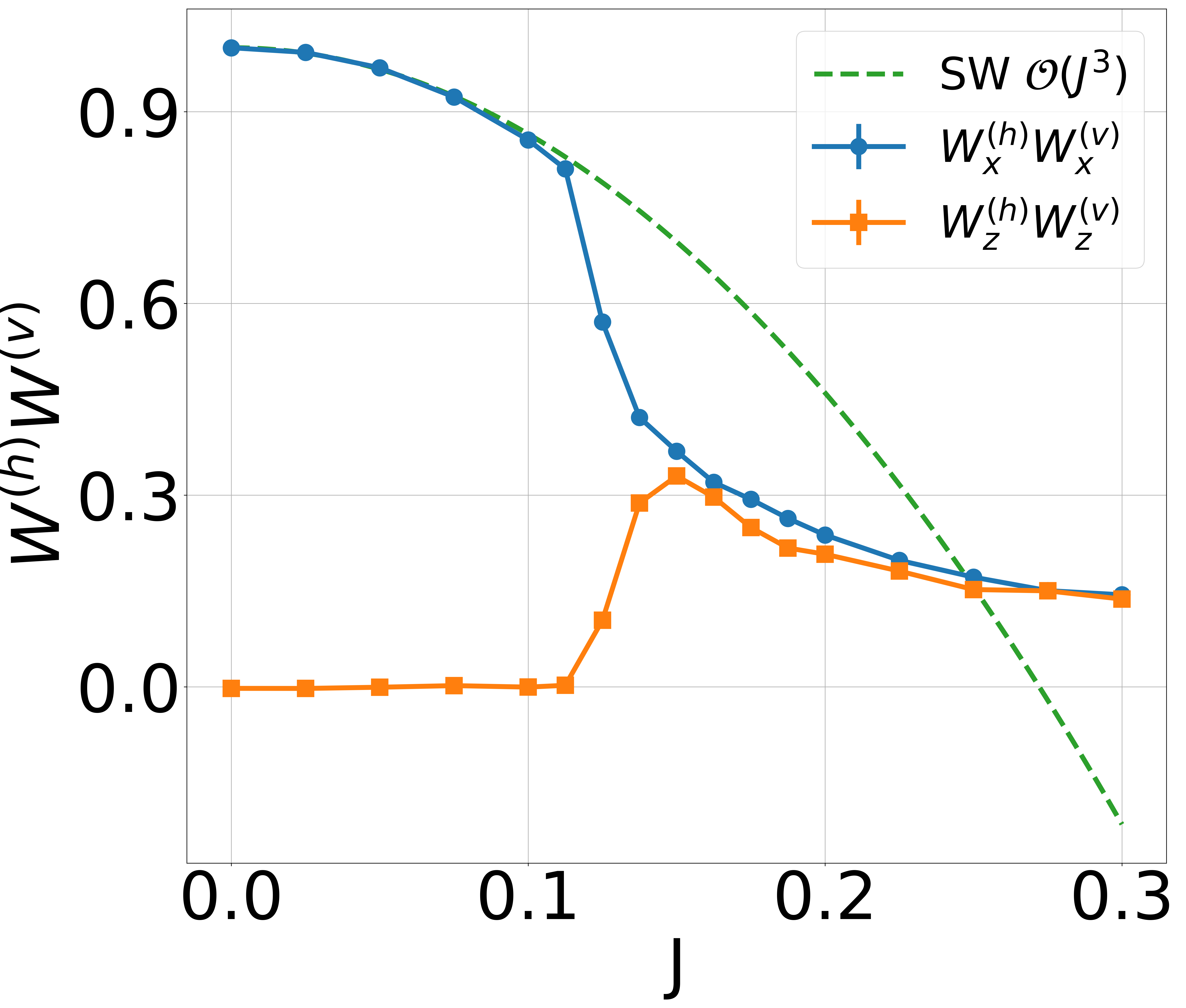}
    \label{fig:XdZd_L6}
  \end{minipage}

  \caption{\label{fig:evenwilson_products}
  Composite non-contractible Wilson-loop products
  \(W_x^{(h)}W_x^{(v)}\) and \(W_z^{(h)}W_z^{(v)}\) for even system sizes
  (a) \(L{=}4\) and (b) \(L{=}6\) in a duality-broken metastable  state.
   Symbols with error bars are NQS results, and the green dashed curve is the
  \(O(J^2)\) SW prediction.}
\end{figure}
At very small $J$, the overall tunneling scale that splits the near-degenerate torus sectors
becomes exponentially small, $\Delta E_{\rm topo}(L,J)\sim J^L/\Delta^{L-1}$. In this regime, the energy differences between distinct logical polarizations can fall below the practical
resolution of the variational optimization (sampling noise, SR regularization, finite ansatz
capacity). As a result, unconstrained runs initialized far from the mixed-sector solution can
converge to a metastable, logically polarized state that strongly favors one loop
channel, e.g.\ $\langle W_x\rangle_{\rm phys}\gg \langle W_z\rangle_{\rm phys}$,
thereby appearing to ``break'' duality at finite size. 
(See, Fig.~\ref{fig:evenwilson_single} and Fig.~\ref{fig:evenwilson_products} .)

Strictly speaking, at any fixed finite $L$ the true ground state is unique and must respect all exact symmetries of the Hamiltonian; any apparent duality breaking at very
small $J$ therefore reflects variational metastability rather than a distinct thermodynamic
phase. In the thermodynamic limit, however, the tunneling-induced splitting between the
near-degenerate logical sectors vanishes, the sector space becomes exactly degenerate,
and a duality-broken (sector-polarized) state can be selected as a legitimate ground state.

To obtain the duality-preserving finite-size ground state, we adopt the following
strategy: We first optimize at an intermediate coupling where the logical
sectors are already appreciably mixed and then warm-start (adiabatically retrain) toward
smaller $J$. This procedure strongly suppresses trapping in polarized metastable solutions
and yields loop data close to the duality preserved sector, while a noticeable deviation still remains, as shown in Fig.~\ref{fig:evenwilson}.
In practice, within our numerical resolution, we find essentially no energy difference between
(i) a $W_x$-dominant solution obtained from random initialization that remains trapped in a
single-loop sector (favored by the real, parity-even ansatz) and (ii) the duality-symmetric
solution obtained by warm-starting from intermediate $J$. Other observables in this section are calculated from the near-duality preserved ground state.

Furthermore, in both cases the optimized state
lies in the $(P_x,P_z)=(+1,+1)$ sector.
As discussed in Sec.~II, when the ground state is non-degenerate and $L$ is even, the
duality-preserving branch should lie in the $P_x = P_z=+1$ sector. 
One might still worry that, at very small $J$, the near-degeneracy of
the ground space could complicate this sector assignment. Nevertheless, for even
$L$ both $P_x$ and $P_z$ can be written as global products of local star or plaquette
operators, which are positive deep in the topological regime. This supports the expectation
that the physically relevant duality-preserving state remains in $(P_x,P_z)=(+1,+1)$ in the
parameter range considered here.

\section{Binder cumulants for representative symmetry classes}
\label{app:binder}

In this appendix, we evaluate the Binder cumulant
\[
U_4^{(z)} = 1 -
\frac{\langle (m_z^{\rm stag})^4\rangle_{phys}}
{3\,\langle (m_z^{\rm stag})^2\rangle_{phys}^2}
\]
for several idealized symmetry scenarios.
Note that this classical approximation is appropriate deep in the ordered
phase, where quantum fluctuations are strongly suppressed.
We parametrize the staggered magnetization as
$m_z^{\rm stag}=m_0\cos\theta$ and compute the moments assuming classical
distributions $p(\theta)$ appropriate to each case.
These results provide useful reference values for interpreting the numerical
Binder cumulants in the main text.

\medskip
\noindent U(1) / $XZ$ rotor (direction uniform on a circle):
Here the angle $\theta$ is uniformly distributed on $[0,2\pi)$,
$p(\theta)=1/(2\pi)$. One finds
\begin{align}
\langle \cos^2\theta\rangle
&=\frac{1}{2\pi}\int_0^{2\pi}\cos^2\theta\,d\theta
=\frac{1}{2},\\
\langle \cos^4\theta\rangle
&=\frac{1}{2\pi}\int_0^{2\pi}\cos^4\theta\,d\theta
=\frac{3}{8}.
\end{align}
The Binder cumulant is therefore
\[
U_4^{(z)}=1-\frac{(3/8)}{3(1/2)^2}
=\frac{1}{2}.
\]

\medskip
\noindent SU(2) / Heisenberg rotor (direction uniform on a sphere):
The polar angle obeys $p(\theta)=\tfrac12\sin\theta$ on $[0,\pi]$.
Using $u=\cos\theta$ yields
\begin{align}
\langle \cos^2\theta\rangle
&=\int_0^\pi \cos^2\theta\,\tfrac12\sin\theta\,d\theta
=\int_{-1}^{1} u^2\,\tfrac12\,du
=\frac{1}{3},\\
\langle \cos^4\theta\rangle
&=\int_0^\pi \cos^4\theta\,\tfrac12\sin\theta\,d\theta
=\int_{-1}^{1} u^4\,\tfrac12\,du
=\frac{1}{5}.
\end{align}
Hence
\[
U_4^{(z)}=1-\frac{(1/5)}{3(1/3)^2}
=0.4.
\]

\medskip
\noindent Four-sector cat state in the $x$--$z$ plane:
Equal weight is assigned to
$\theta\in\{0,\pi,\tfrac{\pi}{2},\tfrac{3\pi}{2}\}$.
Thus $\cos\theta\in\{+1,-1,0,0\}$ with probability $1/4$ each, giving
\begin{align}
\langle \cos^2\theta\rangle &= \tfrac12,\\
\langle \cos^4\theta\rangle &= \tfrac12.
\end{align}
The corresponding Binder cumulant is
\[
U_4^{(z)}=1-\frac{(1/2)}{3(1/2)^2}
=\frac{1}{3}.
\]

\medskip
\noindent Ising single-sector state:
For two angles $\theta\in\{0,\pi\}$ with equal weight,
$\langle \cos^2\theta\rangle=1$ and $\langle \cos^4\theta\rangle=1$, yielding
\[
U_4^{(z)}=1-\frac{1}{3}=\frac{2}{3}.
\]

\bibliography{references}

\end{document}